\documentclass[sigconf]{acmart}
\pdfoutput=1

\makeatletter%
\providecommand\authornotemark{}%
\renewcommand\authornotemark[1][\relax]{%
  \ifx#1\relax\relax\relax
  \g@addto@macro\addresses{\@authornotemark}%
  \else
  \g@addto@macro\addresses{\@@authornotemark{#1}}%
  \fi}
\def\@authornotemark{\g@addto@macro\@currentauthors{\footnotemark\relax}}
\def\@@authornotemark#1{\g@addto@macro\@currentauthors{\footnotemark[#1]}}
\makeatother

\usepackage[utf8]{inputenc}
\usepackage{graphicx}
\usepackage{booktabs}
\usepackage{mathtools}
\usepackage{amsmath}
\usepackage{amssymb}
\usepackage{amsthm}
\usepackage{bm}
\usepackage{enumitem}
\usepackage{ifthen}
\usepackage{style/algorithm}
\usepackage{array}
\usepackage{subfigure}
\usepackage{style/algorithmic}

\newcommand{\systemA}{\textsc{ByzSGD}}
\newcommand{\systemS}{\textsc{ByzSGD}}
\newcommand{\papertitle}{Genuinely Distributed Byzantine Machine Learning}
\newcommand{\paragraphspaceheight}{0.2cm}
\newcommand{\paragraphspace}{\vspace{\paragraphspaceheight}}
\newcommand{\mathsep}{,~}
\newcommand{\median}{\emph{Median}}

\newcommand{\modelsf}{\emph{Outliers}}
\newcommand{\lipschitzf}{\emph{Lipschitz}}
\newcommand{\brute}{\emph{MDA}}
\newcommand{\brutelong}{\emph{Minimum--Diameter Averaging}}
\newcommand{\mkrum}{\emph{Multi--Krum}}

\newcommand{\ldotexp}{\ldots{}}
\newcommand{\setr}{\mathbb{R}}
\newcommand{\setn}{\mathbb{N}}
\newcommand{\expect}{\mathop{{}\mathbb{E}}}
\newcommand{\alphaf}{\ensuremath{\left( \alpha, f \right)}}
\newcommand{\suchthat}{\ensuremath{~\middle|~}}
\newcommand{\card}[1]{\left\lvert{#1}\right\rvert}
\newcommand{\absv}[1]{\card{#1}}
\newcommand{\normone}[1]{\left\lVert{#1}\right\rVert_{1}}
\newcommand{\normtwo}[1]{\left\lVert{#1}\right\rVert_{2}}
\newcommand{\norminf}[1]{\left\lVert{#1}\right\rVert_{\infty}}
\newcommand{\norm}[1]{\normtwo{#1}}
\newcommand{\floor}[1]{\left\lfloor{#1}\right\rfloor}

\newcommand{\realgrad}[1]{\nabla{}L\left({#1}\right)}
\newcommand{\stocgrad}[1]{\widehat{\nabla{}L}\left({#1}\right)}
\newcommand{\indexed}[1]{\!\left[{#1}\right]}
\newcommand{\indexvar}[3]{{#3}^{\ifthenelse{\equal{#1}{}}{}{\left({#1}\right)}}_{#2}}
\newcommand{\range}[2]{\left[{#1} \,..\, {#2}\right]}
\newcommand{\locals}[2]{\indexvar{#1}{#2}{\overline{\theta}}}
\newcommand{\params}[2]{\indexvar{#1}{#2}{\theta}}
\newcommand{\gradvar}[2]{\indexvar{#1}{#2}{g}}
\newcommand{\graddist}[2]{\indexvar{#1}{#2}{\mathcal{G}}}
\newcommand{\gradaggr}[2]{\indexvar{#1}{#2}{G}}
\newcommand{\argsfor}[2]{{\color{gray}\,\times\!\left({#2}\right)^{\left({#1}\right)}}}

\DeclareMathOperator*{\argmin}{arg\,min}

\renewcommand{\paragraph}[1]{\textbf{#1}~}

\newcolumntype{L}[1]{>{\raggedright\let\newline\\\arraybackslash\hspace{0pt}}m{#1}}

\AtBeginDocument{%
  \providecommand\BibTeX{{%
    \normalfont B\kern-0.5em{\scshape i\kern-0.25em b}\kern-0.8em\TeX}}}

\setcopyright{none}
\renewcommand\footnotetextcopyrightpermission[1]{}
\begin{document}
\title{\papertitle}

\author{El-Mahdi El-Mhamdi}
\authornote{Equal contribution. Authors are listed alphabetically.}
\author{Rachid Guerraoui}
\authornotemark[1]
\author{Arsany Guirguis}
\authornotemark[1]
\author{Lê Nguyên Hoang}
\authornotemark[1]
\author{Sébastien Rouault}
\authornotemark[1]
\email{firstname.lastname@epfl.ch}
\affiliation{%
  \institution{EPFL}
  \streetaddress{Route Cantonale}
  \city{Lausanne (VD)}
  \state{Switzerland}
  \postcode{1015}
}

\renewcommand{\shortauthors}{EGGHR}

\begin{abstract}
Machine Learning (ML) solutions are nowadays distributed, according to the so-called \emph{server/worker} architecture.
One \emph{server} holds the model parameters while several \emph{workers} train the model.
Clearly, such architecture is prone to various types of component failures, which can be all encompassed within the spectrum of a Byzantine behavior.
Several approaches have been proposed recently to tolerate Byzantine workers.
Yet all require trusting a central parameter server.
We initiate in this paper the study of the \emph{``general'' Byzantine-resilient} distributed machine learning problem where no individual component is trusted.
In particular, we distribute the parameter server computation on several nodes.

We show that this problem can be solved in an asynchronous system, despite the presence of $\frac{1}{3}$ Byzantine parameter servers and $\frac{1}{3}$ Byzantine workers (which is optimal).
We present a new algorithm, \systemA{}, which solves the general Byzantine-resilient distributed machine learning problem by relying on three major schemes.
The first, \emph{Scatter/Gather}, is a communication scheme whose goal is to bound the maximum drift among models on correct servers.
The second, \emph{Distributed Median Contraction} (DMC), leverages the geometric properties of the median in high dimensional spaces to bring parameters within the correct servers back close to each other, ensuring learning convergence.
The third, \brutelong{} (\brute{}),
is a statistically--robust gradient aggregation rule whose goal is to tolerate Byzantine workers.
\brute{} requires loose bound on the variance of non--Byzantine gradient estimates, compared to existing alternatives (e.g.,\ Krum~\cite{krum}).
Interestingly, \systemA{} ensures Byzantine resilience without adding communication rounds (on a normal path), compared to vanilla non-Byzantine alternatives. \systemA{} requires, however, a larger number of messages which, we show, can be reduced if we assume synchrony.

We implemented \systemA{} on top of TensorFlow, and we report on our evaluation results.
In particular, we show that \systemA{} achieves convergence in Byzantine settings with around 32\% overhead compared to vanilla TensorFlow.
Furthermore, we show that \systemS{}'s throughput overhead is 24--176\% in the synchronous case and 28--220\% in the asynchronous case.
\end{abstract}

\maketitle

\section{Introduction}
\label{sec:intro}

\paragraph{Distributed ML is fragile.}
Distributing Machine Learning (ML) tasks seems to be the only way to cope with ever-growing datasets \cite{kim1many,meng2016mllib,chilimbi2014project}. A common way to distribute the learning task is through the now classical  \emph{parameter server} architecture~\cite{li2013parameter,li2014scaling}. In short, a central  \emph{server} holds the model parameters (e.g.,\ weights of a neural network) whereas a set of \emph{workers} perform the backpropagation computation~\cite{hecht1992theory}, typically following the standard optimization algorithm: \emph{stochastic gradient descent} (SGD)~\cite{rumelhart1986learning}, on their local data, using the latest model they pull from the server. This server in turn gathers the updates from the workers, in the form of \emph{gradients}, and aggregates them, usually through averaging ~\cite{konevcny2015federated}. This scheme is, however, very fragile because averaging does not tolerate a single corrupted input~\cite{krum}, whilst the multiplicity of machines increases the probability of a misbehavior somewhere in the network.

This fragility is problematic because ML is expected to play a central role in safety-critical tasks such as driving and flying people, diagnosing their diseases, and recommending treatments to their doctors~\cite{lecun2015deep, esteva2017dermatologist}.
Little room should be left for the routinely reported~\cite{papernot2017practical, biggio2017wild, gilmer2018adversarial} vulnerabilities of today's ML solutions.

\paragraph{Byzantine-resilient ML.}
Over the past three years, a growing body of work, e.g.,\ \cite{krum, alistarh2018byzantine, chen2018draco, xie2018zeno, bernstein2018signsgd, moitra2018, lilisu2018, lilisu2019}, took up the challenge of  \emph{Byzantine-resilient ML}.
The Byzantine failure model, as originally introduced in distributed computing~\cite{lamport1982Byzantine}, encompasses crashes, software bugs, hardware defects, message omissions, corrupted data, and even worse, hacked machines~\cite{biggio2012poisoning,xiao2015feature}.
So far, all the work on Byzantine-resilient ML assumed that
a fraction of workers could be Byzantine. But all assumed  the central parameter server to be always \emph{honest} and \emph{failure-free}.
In other words, none of the previous approaches considered a genuinely Byzantine--resilient distributed ML system, in the distributed computing sense, where no component is trusted.
Consider a multi--branch organization with sensitive data, e.g.,\ a hospital or a bank, that would like to train an ML model among its branches.
In such a situation, the worker machines, as well as the central server, should be robust to the worst: the adversarial attacks.

A natural way to prevent the parameter server from being a  \emph{single point--of--failure} is to \emph{replicate} it. But this
poses the problem of how to synchronize the replicas.
The classical synchronization technique, \emph{state machine replication} \cite{schneider1990implementing,rachidbouquin} (SMR), enforces a \emph{total order for updates} on all replicas through \emph{consensus},  providing the abstraction of a single parameter server, while benefiting from the resilience of the multiplicity of underlying replicas.
Applying SMR to distributed SGD would however lead to a potentially
huge \emph{overhead}. In order to maintain the same state, replicas would need to agree on a total order of the model updates, inducing frequent exchanges (and retransmissions) of gradients and parameter vectors, that can be several hundreds of MB large~\cite{kim1many}.
Given that a distributed ML setup is network bound~\cite{poseidon,hsieh2017gaia},  SMR is impractical in this context.

The key insight underlying our paper is that the \emph{general Byzantine SGD problem}, even when neither the workers nor the servers are trusted, is easier than consensus, and total ordering of updates is not required in the context of ML applications; only convergence to a good final cross-accuracy is needed.
We thus follow a different route where we do not require all the servers to maintain the same state. Instead, we allow mildly diverging parameters (which have proven beneficial in other contexts~\cite{zhang2015deep,alistarh2016qsgd}) and present new ways to \emph{contract} them in a \emph{distributed} manner.

\paragraph{Contributions.}
In short, we consider, for the first time in the context of distributed SGD, the general Byzantine ML problem, where no individual component is trusted.
First, we show that this problem can be solved in asynchronous settings, and then we show how we can utilize synchrony to boost the performance of our asynchronous solution.

Our main algorithm, \systemA{}, achieves \emph{general} resilience without assuming bounds on communication delays, nor inducing additional communication rounds (on a normal path), when compared to a non-Byzantine resilient scheme.
We prove that \systemA{} tolerates $\frac{1}{3}$ Byzantine servers and $\frac{1}{3}$ Byzantine workers, which is optimal in an asynchronous setting.
\systemA{} employs a novel communication scheme, \emph{Scatter/Gather}, that bounds the maximum drift between models on correct servers.
In the \emph{scatter} phase, servers work independently (they do not communicate among themselves) and hence,
their views of the model could drift away from each other.
In the \emph{gather} phase, correct servers communicate and
apply collectively a \emph{Distributed Median-based Contraction} (DMC) module.
This module is crucial for it
brings the diverging parameter vectors back closer to each other, despite each parameter server being only able to gather a fraction of the parameter vectors.
Interestingly, \systemA{} ensures Byzantine resilience without adding communication rounds (on a normal path), compared to non-Byzantine alternatives. \systemA{} requires, however, a larger number of messages
which we show we can reduce if we assume synchrony.
Essentially, in a synchronous setting, workers can use a novel filtering mechanism we introduce to eliminate replies from Byzantine servers without requiring to communicate with all servers.

\systemS{}\footnote{We use the same name for our asynchronous and synchronous variants of the algorithm when there is no ambiguity.}  uses a \emph{statistically--robust gradient aggregation rule} (GAR), which we call \brutelong{} (\brute{}) to tolerate Byzantine workers.
Such a choice of a GAR has
two advantages compared to  previously--used GARs in the literature to tolerate Byzantine workers, e.g.,\ \cite{krum,xie2018generalized}.
First, \brute{} requires a loose bound on the variance of the non--Byzantine gradient estimates, which makes it practical\footnote{In Appendix~D, we report on our experimental validation of such a requirement. We compare the bound on the variance require by \brute{} to the one required by \mkrum{}~\cite{krum}, a state--of--the--art GAR. For instance, we show that with a batch--size of $100$ and assuming $1$ Byzantine failure, the requirement of \brute{} is satisfied in our experiments, while that of \mkrum{} is not.}.
Second, \brute{} makes the best use of the variance reduction resulting from the gradient estimates on multiple workers, unlike Krum~\cite{krum} and Median~\cite{xie2018generalized}.

We prove that \systemS{} guarantees convergence despite the presence of Byzantine machines, be they workers or servers.
We implemented \systemS{} on top of TensorFlow~\cite{abadi2016tensorflow}, while achieving transparency: applications implemented with TensorFlow need not to change their interfaces to be made Byzantine--resilient.
We report on our evaluation of \systemS{}.
We show that \systemS{} tolerates Byzantine failures with a reasonable convergence overhead ($\sim 32\%$) compared to the non Byzantine-resilient vanilla TensorFlow deployment.
Moreover, we show that the throughput overhead of \systemS{} ranges from 24\% to 220\% compared to vanilla TensorFlow.

The paper is organized as follows.
Section~\ref{sec:background} provides some background on SGD, the problem settings and the threat model.
Section~\ref{sec:systemA} describes our \systemA{} algorithm, while Section~\ref{sec:proof} sketches its correctness proof.
Section~\ref{sec:systemS} discusses how \systemS{} leverages synchrony to boost performance.
Section~\ref{sec:eval} reports on our empirical evaluation of \systemS{}.
Section~\ref{sec:conc} concludes the paper by discussing related work and highlighting open questions.
In the Appendix to the main paper, we give the convergence proof of \systemA{} (Appendix~C), we empirically validate our assumptions (Appendix~D), and
we assess \systemA{}'s performance in a variety of cases (Appendix~E).
\section{Background and Model}
\label{sec:background}

\subsection{Stochastic Gradient Descent}
\label{sec:background-sgd}
Stochastic Gradient Descent (SGD)~\cite{rumelhart1986learning} is a widely-used \emph{optimization} algorithm in ML applications~\cite{chilimbi2014project,li2014scaling,abadi2016tensorflow}.
Typically, SGD is used to minimize a \emph{loss function}
$L\left( \theta \right) \in \setr$, which measures how accurate the model $\theta$ is when classifying an input.
Formally, SGD addresses the following optimization problem:
\begin{equation}
\label{eq:opt}
\argmin_{\theta{} \in{} \mathbb{R}^d} L(\theta)
\end{equation}

SGD works iteratively and consists, in each step $t$, of:
\begin{enumerate}[noitemsep,nolistsep]
    \item{Estimating the gradient $G\left(\theta^{(t)}, \xi\right)$, with a subset $\xi$ of size $b$ of the training set, called \emph{mini--batch}.
    Such a gradient is a \emph{stochastic} estimation of the real, uncomputable one $\nabla{}L\left(\theta^{(t)}\right)$.}
    \item{Updating the parameters following the estimated gradient:
    \begin{equation}
        \label{eq:opt-step}
        \theta^{(t+1)} = \theta^{(t)} - \eta_t \cdot G\left( \theta^{(t)}, \xi \right)
    \end{equation}
    The sequence $\{\eta_t \in \setr_{>0}\}$ is called the \emph{learning rate}.}
\end{enumerate}

\subsection{The Parameter Server Architecture}
\label{subsec:ps_model}

Estimating one gradient is computationally expensive,
as it consists in computing $b$ estimates of $G\left(\theta, \xi_i \right)$, where $\xi_i$ is the $i^\text{th}$ pair (input, label) from the mini--batch, and where
each $G\left(\theta, \xi_i\right)$ involves one \emph{backpropagation} computation~\cite{hecht1992theory}.
Hence, the amount of arithmetic operations to carry out to estimate $G\left(\theta, \xi\right) \approx \nabla{}L\left( \theta^{(t)} \right)$ is $\mathcal{O}\left( b\!\cdot\!d \right)$.

However, this gradient estimation can be easily distributed:
the $b$ computations of $G\left( \theta^{(t)}, \xi_i \right)$
can be executed in parallel on $n$ machines,
where the aggregate of such computations gives $G\left( \theta^{(t)}, \xi \right)$.
This corresponds to the now standard \emph{parameter server} architecture~\cite{li2013parameter}, where a central server holds the parameters $\theta$.

Each training step includes two communication rounds: the server first broadcasts the parameters to workers,
which then estimate the gradient $G\left( \theta^{(t)}, \xi_i \right)$ (i.e.,\ each with a mini--batch of $\frac{b}{n}$).
When a worker completes its estimation, it sends it back to the parameter server, which in turn \emph{averages} these estimations and updates the parameters $\theta$, as in Equation~\ref{eq:opt-step}.

\subsection{Byzantine Machine Learning}
\label{subsec:byzml}
The Byzantine failure abstraction~\cite{lamport1982Byzantine} models any arbitrary behavior and encompasses software bugs, hardware defects, message omissions, or even hacked machines. We then typically assume that a subset of the machines can be Byzantine and controlled by an adversary whose sole purpose is to defeat the computation.

A Byzantine machine can, for instance, send a biased estimate of a gradient to another machine, which leads to a corrupted learning model accordingly or even to learning divergence~\cite{baruch2019little}. 
Byzantine failures also abstract the \emph{data poisoning} problem~\cite{biggio2012poisoning}, which happens when a machine owns maliciously--labeled data. This may result in learning a corrupted model, especially--crafted by the adversary.
Clearly, assuming a central machine controlling the learning process (as with the standard parameter server architecture~\cite{li2013parameter}) is problematic if such a machine is controlled by an adversary, for 
this machine can write whatever it wants to the final model, jeopardizing the learning process.
\begin{table}[!ht]
\centering
\caption{The notations used throughout this paper.}
\vspace{-4mm}
\begin{tabular}{c L{6.9cm}}
$n_{ps}$ & Total number of parameter servers \\[0.5mm]
$f_{ps}$ & Maximal number of Byzantine parameter servers \\[0.5mm]
$q_{ps}$ & Number of parameter vectors a node waits for from servers, $2 f_{ps} + 2 \le q_{ps} \le n_{ps} - f_{ps}$ \\[0.5mm]
$n_w$ & Total number of workers \\[0.5mm]
$f_w$ & Declared, maximal number of Byzantine workers \\[0.5mm]
$q_w$ & Number of gradients a node waits for from workers, $2 f_w + 1 \le q_w \le n_w - f_w$ \\[0.5mm]
$d$ & Dimension of the parameter space $\setr^d$ \\[0.5mm]
$L$ & Loss function we aim to minimize \\[0.5mm]
$l$ & Lipschitz constant of the loss function \\[0.5mm]
$\params{i}{t}$ & Parameter vector (i.e.\ model) at the parameter server $i$ at step $t$ \\[0.5mm]
$\graddist{i}{t}$ & Gradient distribution at the worker $i$ at step $t$ \\[0.5mm]
$\gradvar{i}{t}$ & Stochastic gradient estimation of worker $i$ at step $t$ \\[0.5mm]
$\realgrad{\params{}{}}$ & Real gradient of the loss function $L$ at $\params{}{}$ \\[0.5mm]
$\stocgrad{\params{}{}}$ & A stochastic estimation of $\nabla{}L\left( \params{}{} \right)$ \\[0.5mm]
$\eta_t$ & Learning rate at step $t$
\end{tabular}
\parbox[b][11pt]{\linewidth}{Without loss of generality in the analysis, we note:}
\begin{tabular}{c L{4.5cm}}
$\range{1}{n_{ps} - f_{ps}}$ & the indexes of the correct servers \\[0.5mm]
$\range{1}{n_w - f_w}$ & the indexes of the correct workers
\end{tabular}
\label{table:notations}
\vspace{-4mm}
\end{table}

\subsection{System Model}

\begin{figure}
\centering
\centerline{\includegraphics[width=0.75\linewidth]{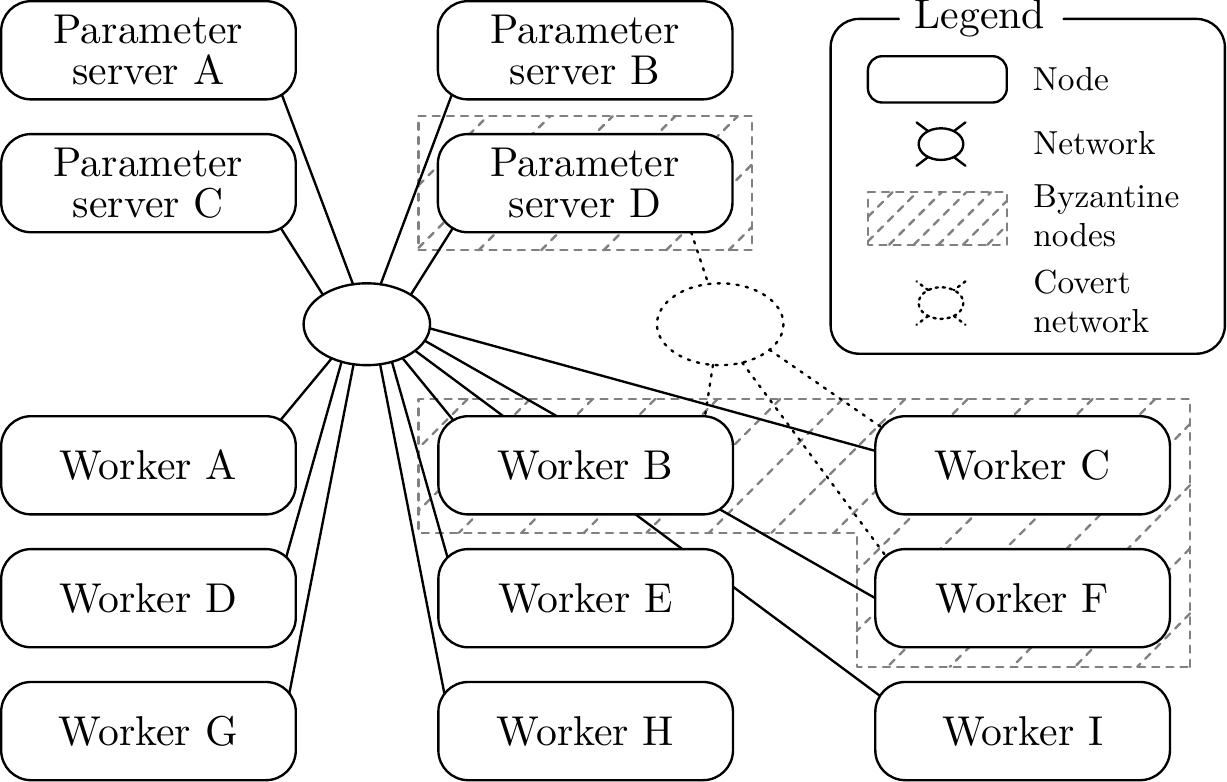}}
\caption{A distributed ML setup with 4 parameter servers and 9 workers, including respectively 1 and 3 Byzantine nodes, which all
can be viewed as a single adversary.}
\label{fig:model-overview}
\vspace{-5mm}
\end{figure}

\paragraph{Computation and communication models.}
We build on the standard parameter server model,
with two main variations (Fig.\ \ref{fig:model-overview}).
\newline
{\bf 1.}  We assume a subset of the nodes (i.e.,\ machines) involved in the distributed SGD process to be \emph{adversarial}, or \emph{Byzantine}, in the parlance of \cite{krum, chen2018draco, bulyanPaper}.
The other nodes are said to be \emph{correct}. Clearly, which node is correct and which is Byzantine is not known ahead of time.
{\bf 2.} We consider several \emph{replicas} of the parameter server (we call them \emph{servers}), instead of a single one,
preventing it from being a single \emph{point--of--failure}, unlike in classical ML approaches.

In our new context, workers send their gradients to all servers (instead of one in the standard case), which in turn send their parameters to all workers.
Periodically, servers communicate with each other, as we describe later in Section~\ref{sec:systemA}.
We consider \emph{bulk--synchronous training}: only the gradients computed at a learning step $t$
are used for the parameters update at the same step $t$.

\paragraph{Adversary capabilities.}
The adversary is an entity that controls all Byzantine nodes (Figure \ref{fig:model-overview}), and whose goal is to prevent the  SGD process from converging to a state that could have been achieved if there was no adversary.
As in \cite{krum, chen2018draco, bulyanPaper}, we assume an omniscient adversary that can see the full training datasets as well as the packets being transferred over the network (i.e.,\ gradients and models).
However, the adversary is not omnipotent: it can only send arbitrary messages from the nodes it controls,
or force them to remain silent.
We assume nodes can authenticate the source of a message, so no Byzantine node can forge its identity or create multiple fake ones~\cite{castro1999practical}.

\subsection{Convergence Conditions}
\label{sec:system-assumptions}

We assume the classical conditions\footnote{The exhaustive list, with their formal formulations, is available in Appendix~C.1.} for convergence in non--convex optimization~\cite{bottou1998online}.
For instance, we assume that the training data is identically and independently distributed (\emph{i.i.d}) over the workers, and
the sequence of learning rates $\eta_t$ is monotonically decreasing (i.e.,\ $\lim\limits_{t\rightarrow +\infty} \eta_t = 0$).
We also assume that correct workers compute unbiased estimates of the true gradient with sufficiently low variance, namely (see Table~\ref{table:notations} for notations):
\begin{align}
    &\exists \kappa \in \left] 1, +\infty \right[ \mathsep
    \forall \left( i, t, \params{}{} \right) \in \range{1}{n_{w} - f_{w}} \times \setn \times \setr^d \mathsep \nonumber \\
    &\kappa \, \frac{2 f_{w}}{n_{w} - f_{w}} \, \sqrt{\expect\left( \norm{\gradvar{i}{t} - \expect \gradvar{i}{t}}^2 \right)} \le \norm{\realgrad{\params{}{}}}.
    \label{eqn:main-variance}
\end{align}

Such an equation bounds the ratio of the standard deviation of the stochastic gradient estimations to the norm of the real gradient with $\frac{n_w - f_w}{2 f_w}$.
This assumption is now classical in the Byzantine ML literature~\cite{krum,bulyanPaper}. We also empirically verify this assumption in Appendix~D in our experimental setup.

In addition, we assume:
(1) $L$ is $l$-Lipschitz continuous, and (2) no (pattern of) network partitioning among the correct servers lasts forever.
The standard $l$-Lipschitz continuity assumption~\cite{rosasco2004loss,lipschitzNIPS2017,virmaux2018lipschitz} acts as the only bridge between the parameter vectors $\params{i}{t}$ and the stochastic gradients $\gradvar{i}{t}$ at a given step $t$.
This is a \emph{liveness} assumption: the value of $l$ can be arbitrarily high and is only used to bound the expected maximal distance between any two correct parameter vectors.
The second assumption ensures that eventually the correct parameter servers can communicate with each other in order to pull their views of the model back close to each other; this is crucial to achieve Byzantine resilience.

Denote by $S_j \triangleq \left\lbrace s \subset \range{1}{n_{ps}\!-\!f_{ps}}\!-\!\left\lbrace j \right\rbrace \suchthat \card{s} \in \range{q\!-\!f\!-\!1}{q\!-\!1} \right\rbrace$ the subset of correct parameter server indexes that parameter server $j$ can deliver at a given step.
We then call $S \triangleq \prod_{j \in \range{1}{n_{ps}-f_{ps}}} S_j$ the set of all correct parameter server indexes the $n_{ps}-f_{ps}$ correct parameter servers $i \in \range{1}{n_{ps}-f_{ps}}$ can deliver.
We further assume $\exists \rho > 0 \mathsep \forall s \in S \mathsep P\left( X = s \right) \ge \rho$.
For any server $j$,  $\normtwo{\realgrad{\params{j}{t}}} \geq 9 \left( n_w - f_w \right) \sigma' + \frac{C d l^2}{\rho}$, for some constant $C$, and where $\sigma'$ is the upper bound on the distance between the estimated and the true gradients (see Appendix~C.1).
Finally, \systemA{} requires $n_w \ge 3f_w+1$ and $n_{ps} \ge 3f_{ps} + 2$.
\section{\systemA{}: General Byzantine Resilience}
\label{sec:systemA}

We present here  \systemA{}, the first algorithm to 
tolerate Byzantine workers and servers without making any assumptions on node relative speeds and communication delays. 
\systemA{} does not add, on average, any communication rounds compared to the standard parameter server communication model (Section~\ref{subsec:ps_model}).
However, periodically,
\systemA{} adds a communication round between servers to enforce contraction and convergence, as we show in this section.

We first describe the fundamental technique to tolerate Byzantine servers: \emph{Distributed Median-based Contraction} (DMC). 
Then, we describe the Byzantine--resilient \emph{gradient aggregation rule} we use to tolerate Byzantine workers.
Finally, we explain the overall functioning of \systemA{}, highlighting our novel \emph{Scatter/Gather} communication scheme.

\subsection{\emph{Distributed Median--based Contraction}}

The fundamental problem addressed here is induced by the multiplicity of servers and consists of 
bounding the drift among correct parameter vectors $\params{1}{t} \ldotexp \params{n_{ps} - f_{ps}}{t}$, as $t$ grows.
The problem is particularly challenging because of the combination of three constraints: 
(a) we consider a Byzantine environment, (b) we assume an asynchronous network, and (c) 
we do not want to add communication rounds, compared to those done by vanilla non--Byzantine deployments, 
given the expensive cost of communication in distributed ML applications~\cite{poseidon,hsieh2017gaia}.
The challenging question can then be formulated as follows: 
given that the correct parameter servers should not expect to receive more than $n\!-\!f\!-\!1$ messages per round, how to keep the correct parameters close to each other, knowing that a fraction of the received messages could be Byzantine?

\paragraphspace
Our solution to this issue is, what we call, \emph{Distributed Median--based Contraction (DMC)}, 
whose goal is to decrease the expected maximum distance between any two honest parameter vectors (i.e.,\ \emph{contract} them).
DMC
is a combination of (1) the application of \emph{coordinate--wise \median{}} (which is Byzantine--resilient as soon as $q_{ps} \ge 2 f_{ps} + 1$) on the parameter vectors 
and (2) the over--provisioning of $1$ more correct parameter server (i.e.,\ $q_{ps} \ge 2 f_{ps} + 2$); both constitute the root of what we call the \emph{contraction effect}.
Assuming each honest parameter server can deliver a subset of $q_{ps} - f_{ps} - 1$ \emph{honest} parameters, the expected median of the gathered parameters is then both (1) \emph{bounded} between the gathered honest parameters and (2) \emph{different} from any extremum among the gathered honest parameters (as $1$ correct parameter server was over-provisioned).
Since each subset of the gathered honest parameters contains a subset of all the $n_{ps} - f_{ps}$ honest parameters, the expected maximum distance between two honest parameters is thus decreased after applying DMC.

\subsection{\brutelong{}}
\label{subsec:mda}
To tolerate Byzantine workers, we use 
a statistically--robust \emph{Gradient Aggregation Rule} (GAR).
A GAR is merely a function of $\left( \setr^d \right)^n \rightarrow \setr^d$.
Reminiscent of the \emph{Minimum Volume Ellipsoid}~\cite{rousseeuw1985multivariate},
we use \brutelong{} (\brute{}) as our Byzantine--resilient GAR.
Given $n$ input gradients (with $f$ possible Byzantine ones), \brute{} averages only a subset of $n - f$ gradients.
The selected subset must have the \emph{minimum diameter} among all the subsets of size $n - f$ taken from the $n$ input gradients; hence the name.
The \emph{diameter} of a subset is defined as the maximum $\ell_2$ distance between any two gradients of this subset.

\begin{figure}[!t]
\vspace{-3mm}
\caption{\systemA{}: worker and parameter server logic.}
\vspace{-5mm}
\begin{minipage}[t]{0.48\linewidth}
\begin{algorithm}[H]
    \centering
    \caption{Worker}\label{alg:wrk}
    \begin{algorithmic}[1]
        \STATE $t \gets 0$
        \REPEAT
            \STATE \text{ms} $\gets$ \text{read\_models()}
            \STATE \text{m}.set(\text{\median{}(ms)})
            \STATE \text{g $\gets$ compute\_grad()}
        \STATE $t \gets t+1$
        \UNTIL $t > $ \text{max\_steps}
    \end{algorithmic}
\end{algorithm}
\end{minipage}
\begin{minipage}[t]{0.48\linewidth}
\begin{algorithm}[H]
    \centering
    \caption{Parameter Server}\label{alg:ps}
    \begin{algorithmic}[1]
        \STATE \text{Calculate $T$ \& \emph{seed}}
        \STATE \text{m $\gets$ init\_model(seed)}
        \STATE $t \gets 0$
        \REPEAT
        \STATE \text{gs} $\gets$ \text{read\_gradients()}
        \STATE \text{m.update(\brute{}(gs))}
        \IF{$t\text{ mod }T = 0$}
            \STATE \text{m $\gets$ read\_models()}
            \STATE \text{m $\gets$ \median{}(ms)}
        \ENDIF
        \STATE $t \gets t+1$
        \UNTIL $t > $ \text{max\_steps}
    \end{algorithmic}
\end{algorithm}
\end{minipage}
\vspace{-3mm}
\end{figure}

\paragraphspace
\brute{} can be formally described as follows.
Let $\left( x_1 \ldotexp x_{q} \right) \in \left( \mathbb{R}^d \right)^q$,
and $\mathcal{Q} \triangleq \left\lbrace\, x_1 \ldotexp x_q \,\right\rbrace$ the set of all the input gradients.
Let $\mathcal{R} \triangleq \left\lbrace\, \mathcal{X} \mid \mathcal{X} \subset \mathcal{Q} \mathsep \card{\mathcal{X}} = q - f \,\right\rbrace$ the set of subsets of $\mathcal{Q}$ with cardinality $q - f$.
Let $\mathcal{S} \triangleq \argmin_{\mathcal{X} \in \mathcal{R}}\left( \max_{\left( x_i, x_j \right) \in \mathcal{X}^2}\left( \norm{x_i - x_j} \right) \right)$.
Then, the aggregated gradient is $\brute{}\left( x_1 \ldotexp x_q \right) \triangleq \frac{1}{q - f} \sum_{x \in \mathcal{S}}{x}$.

\subsection{The \systemA{} Algorithm}
\label{sec:system-algorithm}

\systemS{} operates iteratively in two phases: \emph{scatter} and \emph{gather}.
One \emph{gather} phase is entered every $T$ steps (lines 8 to 11 in Algorithm~\ref{alg:ps}); we call the whole $T$ steps the \emph{scatter} phase.

Algorithms~\ref{alg:wrk} and~\ref{alg:ps} depict the training loop applied by workers and servers respectively.
As an initialization step, correct servers initialize the model with the same random values, i.e.,\ using the same \emph{seed}.
Moreover, the servers compute the value of $T$.

The subsequent steps $t \in \setn{}$ work as follows.
The algorithm starts with the \emph{scatter} phase, which includes doing a few learning steps. In each step, each server $i$ broadcasts its current parameter vector $\params{i}{t}$ to every worker.
Each worker $j$ then (1) \emph{aggregates} with \emph{coordinate--wise \median{}} (hereafter simply, \median{}) the first $q_{ps}$ received $\params{i}{t}$ (line~4 in Algorithm~\ref{alg:wrk}),
and (2) computes an estimate $\gradvar{j}{t}$ of the gradient at the aggregated parameter vector, i.e.,\ model.
Then, each  worker $j$ broadcasts its computed gradient estimation $\gradvar{j}{t}$ to all parameter servers.
Each parameter server $i$ in turn \emph{aggregates} with \brute{} the first $q_w$ received $\gradvar{j}{t}$ (line~6 in Algorithm~\ref{alg:ps}) and then performs a local parameter update with the aggregated gradient, resulting in $\locals{i}{t}$.
In normal steps (in the \emph{scatter} phase), $\params{i}{t+1}=\locals{i}{t}$.

Every $T$ steps (i.e.,\ in the \emph{gather} phase), correct servers apply DMC:
each parameter server $i$ broadcasts $\locals{i}{t}$ to every other server and then aggregates with \median{} the first $q_{ps}$ received $\locals{k}{t}$; this aggregated parameter vector is~$\params{i}{t+1}$.
\section{Correctness of \systemA{}}
\label{sec:proof}
This section sketches the convergence proof \systemA{}.
The full proof is detailed in Appendix~C.

\begin{theorem}[Correctness of \systemA{}]
Under the assumptions of Section~\ref{sec:system-assumptions}, \systemA{} guarantees the convergence of all correct parameter servers, i.e.
\begin{equation}
    \forall i \in \range{1}{n_{ps} - f_{ps}}, \lim\limits_{t \rightarrow +\infty}{\norm{\realgrad{\params{i}{t}}}} = 0.
    \label{eqn:convergence}
\end{equation}
\label{th:correctness}
\end{theorem}

Equation~\ref{eqn:convergence} is the commonly admitted termination criteria for non--convex optimization tasks~\cite{bottou1998online}.
We provide here the key intuitions and steps of the proof,
which relies on two key observations:
(1) assuming that the learning rate $\eta_t$ converges to 0, \median{} guarantees that all correct servers' parameters $\params{}{t}$ all eventually reach the same values (i.e.,\ \emph{contract}) and
(2) any given server $i$'s parameters $\params{i}{t}$ eventually essentially follows a stochastic gradient descent dynamics, which guarantees its convergence as described by Equation~\ref{eqn:convergence}.
In the sequel, we detail these two points.

\subsection{Contraction by \median{}}

The first key element of the proof is to prove that \median{} contracts the servers' parameters in expectation, despite the Byzantines' attacks. This turns out to be nontrivial. In fact, the diameter of servers' parameters does not decrease monotonically. The trick is to follow the evolution of another measure of the spread of the servers' parameters, namely the sum of coordinate-wise diameters. We prove that, using \median{}, Byzantines can never increase this quantity.

\begin{lemma}[Safety of \median{}]
Let $\Delta_{\params{}{t}}$ the sum of coordinate-wise diameters, i.e.:
$\Delta_{\params{}{t}} = \sum_{i=1}^d \max_{j,k \in \range{1}{n_{ps}-f_{ps}}} \absv{\params{j}{t}[i] - \params{k}{t}[i]}$.
Then, no matter which gradients are delivered to which servers, and no matter how Byzantines attack, we have $\Delta_{\params{}{t+1}} \leq \Delta_{\params{}{t}}$.
\end{lemma}

\begin{proof}[Sketch of proof]
On each dimension $j$, since $q_{ps} \geq 2 f_{ps} +2$, for any server $i$, there is a majority of correct servers whose parameters have been delivered to $i$. Thus, the median computed by server $i$ must belong to the interval containing the $j$-coordinates of all correct servers. Since this holds for all servers $i$, the diameter along dimension $j$ cannot increase.
\end{proof}

But interestingly, assuming that all delivering configurations can occur with a positive probability, we can show that there will be contraction in expectation.

\begin{lemma}[Expected contraction by \median{}]
There is a constant $m<1$ such that $\mathbb E \left[ \Delta_{\params{}{t+1}} \right] \leq m \Delta_{\params{}{t}}$. Note that the expectation is taken over all delivering configurations, and given Byzantines' attacks for a delivering configuration.
\end{lemma}

\begin{proof}[Sketch of proof]
Using the assumption $q_{ps} \geq 2 f_{ps} +2$, for any coordinate, we show the existence of a delivering configuration such that the diameter along this coordinate is guaranteed to decrease by a factor $3/4$.
The trick to identify this configuration is to divide, for each coordinate $i$, the set of servers into those that have a high coordinate $i$ and those that have a low coordinate $i$. One of these two subsets, call it $X^*$, must contain a majority of correct servers.

Given that $q_{ps} \leq \lfloor \frac{n_{ps}-f_{ps}}{2} \rfloor$, we can thus guarantee the existence of a delivering configuration such that all parameter servers all contain a majority of inputs that come from $X^*$. If this happens, then all servers' parameters after application of \median{} must then have their $i$'s coordinates closer to the subset $X^*$. In fact, we show that their new diameters along the coordinate $i$ is at most $3/4$ of their previous diameter, which proves the contraction along coordinate $i$ for at least one delivering configuration.
Taking the expectation implies a strict expected contraction along this coordinate.
Summing up over all coordinates yields the lemma.
\end{proof}

Unfortunately, this is still not sufficient to guarantee the convergence of the servers' parameters, because of a potential drift during learning. However, as the learning rate decreases, we show that contraction eventually becomes inevitable, which guarantees eventual convergence.

\begin{lemma}[Bound on the drift]
There exists constants $A$ and $B$ such that $\mathbb E\left[\Delta_{\params{}{t+1}}\right] \leq (m+A \eta_t) \Delta_{\params{}{t}} + B \eta_t$. Here, the expectation is over delivering configurations, from parameter servers to workers, from workers to parameter servers and in-between parameter servers. The expectation is also taken over stochastic gradient estimates by workers, and for any attacks by the Byzantines.
\end{lemma}

\begin{proof}[Sketch of proof]
This bound requires to control the drift despite Byzantine attacks on two different levels, namely when parameter servers deliver their parameters to workers, and when workers deliver their gradients to servers.

In the former case, we exploit the fact that the spread of all correct parameters along, measured in terms of sum of coordinate-wise diameters, is at most $\Delta_{\params{}{t}}$, to guarantee that the coordinate-wise medians of the parameters computed by workers are also close to one another.

Using Lipschitz continuity, this then implies that the gradients they compute are also close to one another, in the sense that the diameter of gradients is at most $l\Delta_{\params{}{t}}$ plus some deviation $2 h_w \sigma'$ due to the noise in the stochastic gradient estimates.

Then, using the Byzantine resilience property of \brute{}, we show that the diameter of servers' aggregated gradients is at most that of workers' estimated gradients. Combining it all yields the lemma.

Note that the constant $A$ depends on the Lipschitz continuity of the gradient of the loss function, while the constant $B$ comes from the randomness of the gradient estimates by workers.
See more details in Appendix~C.4.
\end{proof}

\begin{lemma}[Eventual contraction of \median{}]
As a corollary, assuming $\eta_t \rightarrow 0$, we have $\mathbb E[\Delta_{\params{}{t}}] \rightarrow 0$.
\end{lemma}

\begin{proof}[Sketch of proof]
This comes from the eventual contraction of $\Delta_{\params{}{t}}$. Note that it implies that some other measures of the spread of servers' parameters, like their diameter measured in $\ell_2$, also converge to zero.
\end{proof}

\subsection{Liveness of Server Parameters}

The previous section showed that eventually, the parameters will all have nearly identical values. In this section, we show that, as a result, and thanks to a Byzantine--resilient GAR like \brute{}, any parameter's update is nearly an update that would be obtained by vanilla Byzantine-free SGD, for which the guarantee of Theorem \ref{th:correctness} has been proven.

First, we note that \brute{} ensures that its output gradient lies within the correct set of gradients submitted to a correct server, as stated by the following lemma.
\begin{lemma}[\brute{} bounded deviation from majority]
The distance between the output of \brute{} and (at least) one of the correct gradients is bounded below the diameter of the set of correct gradients.

Formally, let $\left( d, f \right) \in \left( \setn - \left\lbrace 0 \right\rbrace \right)^2$ and let $q \in \setn$ such that $q \ge 2 \, f + 1$.
Let note $H \triangleq \range{1}{q - f}$.

It holds that, $\forall \left( g_1 \ldotexp g_q \right) \in \left( \setr^d \right)^q \mathsep \exists k \in H \mathsep$:
\begin{equation}
    \norm{\brute{}\left( g_1 \ldotexp g_q \right) - g_k} \le \max\limits_{\left( i, j \right) \in H^2}{\norm{g_i - g_j}}
     \label{eqn:mda_bound}
\end{equation}
\end{lemma}

Hence, SGD, with \brute{}, alone would converge.

Now, we show that a server $j$'s update can be written $\params{j}{t+1} = \params{j}{t} - \eta_t \hat G^{(j)}_{t}$, where $\hat G^{(j)}_t$ satisfies the classical assumptions of Byzantine machine learning. Namely, we show that $\expect \hat G^{(j)}_t \cdot \realgrad{\params{j}{t}}$ is positive, under our assumptions.

\begin{theorem}[Liveness and safety of parameter servers]
Under our assumptions for any parameter server $j$, $\expect\hat G^{(j)}_t \realgrad{\params{j}{t}} > 0$.
\end{theorem}

\begin{proof}[Sketch of proof]
To guarantee this condition, we first prove that the expected maximum distance between an estimate $\gradaggr{j}{t}$ by worker $j$ and server $k$'s parameters $\realgrad{\params{k}{t}}$ is upper-bounded by the spread of all servers' parameters and a term due to the randomness of the worker's gradient estimates.

We then show that the servers' updates are all similar, which allows to say that after applying \median{}, all servers are guaranteed to move along the direction of the true gradient, given our assumptions.
\end{proof}

\paragraph{Complexity.}
The communication complexity of \systemS{} is $\mathcal{O}(dn_wn_{ps})$ when $T$ is very large.
With small values for $T$, the communication complexity is $\mathcal{O}(dn_{ps}(n_w+n_{ps}))$.
The computation complexities of \median{} and \brute{} are $\mathcal{O}(n_{ps}d)$ and $\mathcal{O}\!\left( \binom{n_w}{f_w} + n_w^2 d \right)$.
Notably, distributed ML problem is network--bound~\cite{poseidon,hsieh2017gaia} and hence, the (possibly) exponential complexity of \brute{} does not aggressively harm the performance, as we show in Section~\ref{sec:eval}.
\section{\systemS{}: Reducing Messages With Synchrony}
\label{sec:systemS}
We show here that assuming network synchrony, we can
boost \systemS{}'s performance while keeping the same resilience guarantees.
In particular, the number of communicated messages can be reduced as follows:
instead of pulling an updated model from $q_{ps}$ servers (line~3 in Algorithm~\ref{alg:wrk}), each worker pulls only one model and then checks its legitimacy using two filters: \lipschitzf{} and \modelsf{} filters.
In this case, \systemS{} requires $n_w \ge 2f_w+1$ while keeping $n_{ps} \ge 3f_{ps}+2$.
The full proof of \systemS{}, while considering synchronous networks, is in Appendix~C.

\subsection{\lipschitzf{} Filter}
Based on the standard Lipschitz continuity of the loss function assumption~\cite{bottou1998online,krum}, previous work used empirical estimations for the Lipschitz coefficient to filter out gradients from Byzantine workers in asynchronous learning~\cite{kardam}.
We use a similar idea, but we now apply it to filter out \emph{models} from Byzantine \emph{servers}.
The filter works as follows: consider a worker $j$ that owns a model $\params{j}{t}$ and a gradient it computed $\gradvar{j}{t}$ based on that model at some step $t$.
A correct server $i$ should include $\gradvar{j}{t}$ while updating its model $\params{i}{t}$, given network synchrony.
Worker $j$ then: (1) estimates the updated model locally $\params{j(l)}{t+1}$ based on its own gradient and
(2) pulls a model $\params{i}{t+1}$ from a parameter server $i$.
If server $i$ is correct then the growth of the pulled model $\params{i}{t+1}$ (with respect to the local gradient $\gradvar{j}{t}$) should be close to that of the estimated local model $\params{j(l)}{t+1}$, based on the guarantees given by \brute{} (see Appendix~C.2.1).
Such a growth rate is encapsulated in the \emph{Lipschitz coefficient} of the loss function.
If the pulled model is correct then, the worker expects that the Lipschitz coefficient computed based on that model be close to those of the other correct models received before by the worker.
Concretely, a worker computes an empirical estimation of the Lipschitz coefficient $k = \frac{\norm{ \gradvar{j}{t+1} - \gradvar{j}{t}}}{\norm{ \params{j(l)}{t+1} - \params{j}{t}}}$
and then, ensures that it follows the condition  $k \le K_p \triangleq \text{quantile}_\frac{n_{ps} - f_{ps}}{n_{ps}} \{K\}$, where $K$ is the list of all previous Lipschitz coefficients $k$ (i.e.,\ with $t_{prev} < t$).
Note the Lipschitz filter requires $n_{ps} > 3f_{ps}$ (see Appendix~C.2.3).

\subsection{\modelsf{} Filter}
Although the Lipschitz filter can bound the model growth with respect to gradients, a server can still trick this filter by sending a well-crafted model that is arbitrarily far from the other correct models~\cite{baruch2019little}.
To overcome this problem, we use another filter, which we call \emph{\modelsf{} filter}, to bound the distance between models in any two successive steps.
In short, a worker $j$ assumes the distance between a local estimate of a model $\params{j(l)}{t+1}$ and a pulled model $\params{i}{t+1}$ to be upper--bounded as follows: $\norm{ \params{j(l)}{t+1} - \params{i}{t+1}} < \eta_{T \cdot (t\text{ mod }T)} \norm{ g_{T \cdot (t\text{ mod }T)}} \bigg( \frac{(3 T + 2)(n_w - f_w)}{4 f_w} + 2 \big((t-1)\text{ mod }T\big) \bigg)$.

\begin{proof}[Sketch of proof]
Each worker $j$ at time $t$ computes a local updated model $\params{j(l)}{t}$ as follows:
\begin{equation*}
    \params{j(l)}{t} = \params{j}{t-1} - \eta_{t}\gradvar{j}{t}.
\end{equation*}
Simultaneously, a correct server $i$ computes an update as follows:
\begin{equation*}
    \params{i}{t} = \params{i}{t-1} - \eta_{t}\brute{}\left(\gradvar{1}{t} \ldots \gradvar{n_w}{t}\right).
\end{equation*}
Based on the guarantees given by \brute{}~\cite{bulyanPaper}, the following holds:
\begin{equation*}
    \norm{\brute{}\left(\gradvar{1}{t} \ldots \gradvar{n_w}{t}\right) - \gradvar{j}{t}} \le \frac{n_w - f_w}{2 f_w} \norm{\gradvar{j}{t}}.
\end{equation*}
Hence, based on the Lipschitzness of the loss function, we have:
\begin{align}
    \norm{\params{j(l)}{t} - \params{i}{t}}
    &\le \eta_{(T \cdot (t\text{ mod }T))} \norm{\gradvar{}{(T \cdot (t\text{ mod }T))}} \nonumber \\
    &\cdot \left( 2 \bigg( (t\text{ mod }T) - 1\bigg) + \frac{(n_w - f_w)(3 T + 2)}{4 f_w} \right)
    \label{eqn:model-bound}
\end{align}
\end{proof}
Such a bound is also based on the \emph{scatter}/\emph{gather} scheme we are using. The details of deriving this term is in Appendix~C.2.3.

\begin{figure}[!t]
\vspace{-5mm}
\begin{minipage}{0.9\linewidth}
\begin{algorithm}[H]
    \centering
    \caption{\systemS{}: worker logic (synchronous)}\label{alg:wrk_sync}
    \begin{algorithmic}[1]
        \STATE \text{Calculate the values of $T$ \& \emph{seed}}
        \STATE \text{model $\gets$ init\_model(seed)}
        \STATE \text{r $\gets$ random\_int(1,$n_{ps}$)}
        \STATE $t \gets 0$
        \STATE \text{grad $\gets$ model.backprop()}
        \REPEAT
        \STATE \text{local\_model $\gets$ apply\_grad(model,grad)}
        \IF{$t\text{ mod }T = 0$}
            \STATE \text{models} $\gets$ \text{read\_models()}
            \STATE \text{model} $\gets$ \text{MeaMed(models)}
        \ELSE
            \STATE $i \gets 0$
            \REPEAT
                \STATE new\_model $\gets$ read\_model($(r\!+\!t\!+\!i)\text{ mod }n_{ps}$)
                \STATE \text{new\_grad $\gets$ new\_model.backprop()}
                \STATE $i \gets i+1$
            \UNTIL \text{pass\_filters(new\_model)}
            \STATE \text{model $\gets$ new\_model}
            \STATE \text{grad $\gets$ new\_grad}
        \ENDIF
        \STATE $t \gets t+1$
        \UNTIL $t > $ \text{max\_steps}
    \end{algorithmic}
\end{algorithm}
\end{minipage}
\vspace{-3mm}
\end{figure}

\subsection{\systemS{}: The Synchronous Version}
\label{subsec:alg}
Keeping the parameter server algorithm as is (Algorithm~\ref{alg:ps}), Algorithm~\ref{alg:wrk_sync} presents the workers' training loop in the synchronous case.
We focus here on the changes in the \systemS{} algorithm, compared to the asynchronous case (Section~\ref{sec:system-algorithm}).

In the initialization phase, each worker $j$ chooses a random integer $r_j$ with $1 \le r_j \le n_{ps}$ before doing one backpropagation computation to estimate the gradient at the initial model.

While parameter servers are updating the model (line~7 in Algorithm~\ref{alg:ps}), each worker $j$ speculates the updated model by computing a local view of it, using its local computed gradient and its latest local model (line 7 in Algorithm~\ref{alg:wrk_sync}).
Then, each worker $j$ pulls \emph{one} parameter vector from server $i$ where, $i = (r_j+t+1)\text{ mod }n_{ps}$.
Such a worker computes the new gradient, using backpropagation, based on the pulled model.
Based on this computation and the local estimate of the updated model, the worker applies the \lipschitzf{} and the \modelsf{} filters to check the legitimacy of the pulled model.
If the model fails to pass the filters, the worker $j$ pulls a new model from the parameter server $i_{++}$, where $i_{++} = (r_j+t+2)\text{ mod }n_{ps}$.
This process is repeated until a pulled model passes both filters.
Every $T$ steps (i.e.,\ in the \emph{gather} phase), each worker $j$ pulls models from \emph{all} servers and aggregates them using \median{}, completing the DMC computation.
\section{Experimental Evaluation}
\label{sec:eval}
We implemented our algorithms on top of TensorFlow~\cite{abadi2016tensorflow}, and we report here on some of our empirical results.
Appendix~E presents additional experiments, highlighting the effect of changing $n_{ps}$ and assessing the efficiency of our filtering mechanism.

\subsection{Evaluation Setting}
\paragraph{Testbed.}
We use Grid5000~\cite{g5k} as an experimental platform.
We employ up to 20 worker nodes and up to 6 parameter servers.
Each node has 2 CPUs (Intel Xeon E5-2630 v4) with 10 cores, 256~GiB RAM, and 2$\times$10~Gbps Ethernet.

\begin{table}
\caption{Models used throughout the evaluation.}
\vspace{-4mm}
\label{table:models}
\centering
\begin{tabular}{|c|c|c|}
\hline
\textbf{NN architecture} & \# \textbf{parameters} & \textbf{Size (MB)} \\ \hline
MNIST\_CNN & 79510    & 0.3       \\ \hline
CifarNet   & 1756426  & 6.7       \\ \hline
Inception  & 5602874  & 21.4      \\ \hline
ResNet-50  & 23539850 & 89.8      \\ \hline
ResNet-200 & 62697610 & 239.2     \\ \hline
\end{tabular}
\vspace{-4mm}
\end{table}

\paragraph{Experiments.}
We consider an image classification task due to its wide adoption as a benchmark for distributed ML systems, e.g.,~\cite{chilimbi2014project}. We use MNIST~\cite{mnist} and CIFAR-10~\cite{cifar} datasets.
MNIST consists of handwritten digits. It has 70,000 $28\times28$ images in 10 classes. CIFAR-10 is a widely--used dataset in image classification~\cite{srivastava2014dropout,poseidon}. It consists of 60,000 $32\times32$ colour images in 10 classes.

We employ several NN architectures with different sizes ranging from a small convolutional neural network (CNN) for MNIST, training <100k parameters, to big architectures like ResNet-200 with around 63M parameters (see Table~\ref{table:models}).
We use \emph{CIFAR10} (as a dataset) and \emph{CifarNet} (as a model) as our default experiment.

\paragraph{Metrics.}
We evaluate the performance of \systemA{} using the following standard metrics.
{\it 1. Accuracy (top--1 cross--accuracy).} The fraction of correct predictions among all predictions, using the \emph{test} dataset.
We measure accuracy with respect to time and model updates.
{\it 2. Throughput.} The total number of updates that the deployed system can do per second.

\paragraph{Baseline.} We consider vanilla TensorFlow (vanilla TF) as our baseline.
Given that such a baseline does not converge in Byzantine environments~\cite{aggregathor}, we use it only to quantify the overhead in non--Byzantine environments.

\subsection{Evaluation Results}
\label{sec:nonByz}
First, we show \systemS{}'s performance, highlighting the overhead, in a non--Byzantine environment.
Then, we compare the throughput of the synchronous variant to that of the asynchronous variant in a Byzantine--free environment.
Finally, we report on the performance of \systemA{} in a Byzantine environment, i.e.,\ with Byzantine workers and Byzantine servers.
For the Byzantine workers, we show the effect of a recent attack~\cite{baruch2019little} on \systemA{}, and then we show the results of 4 different attacks in the case of Byzantine servers.
In all experiments, and based on our setup, we use $T=333$. We discuss the effect of changing $T$ on \systemA{}'s performance in Appendix~E.2.

\begin{figure}[t]
\centering
\subfigure[]{\includegraphics[width=0.48\linewidth,keepaspectratio]{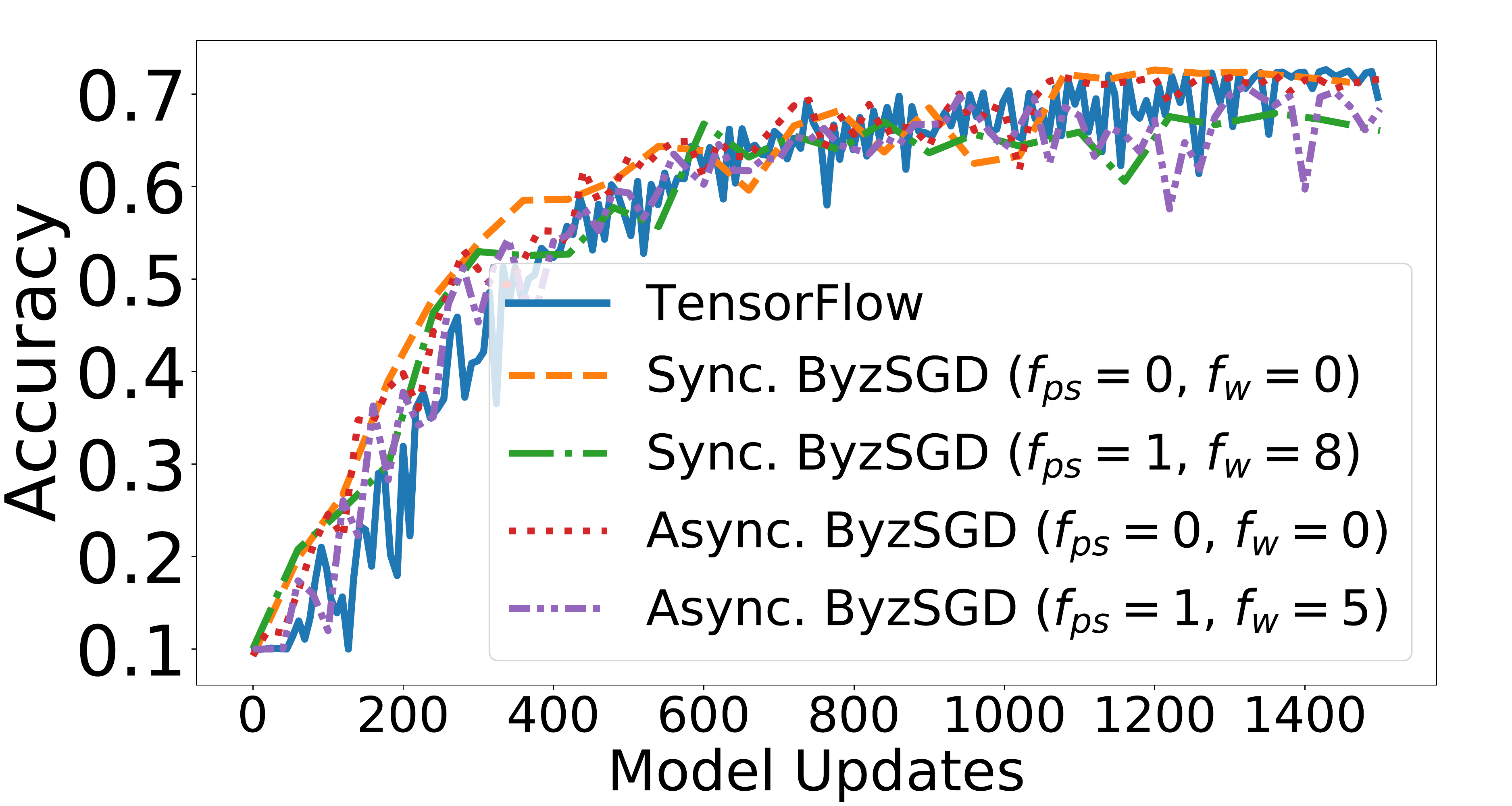}
\label{subfig:conv_b250step}}
\subfigure[]{\includegraphics[width=0.48\linewidth,keepaspectratio]{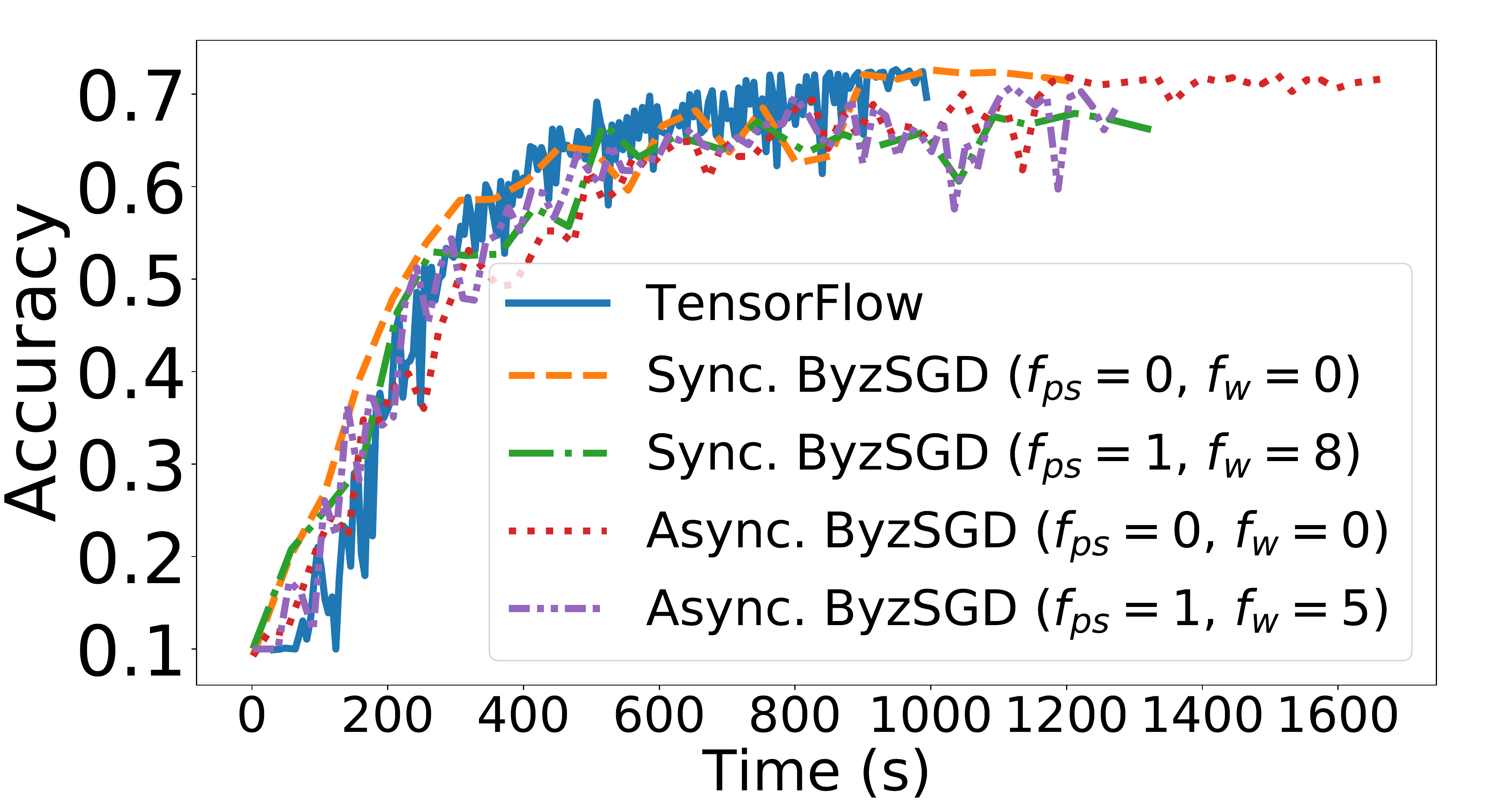}
\label{subfig:conv_b250time}}
\\[-4mm]
mini-batch size = 250
\\
\subfigure[]{\includegraphics[width=0.48\linewidth,keepaspectratio]{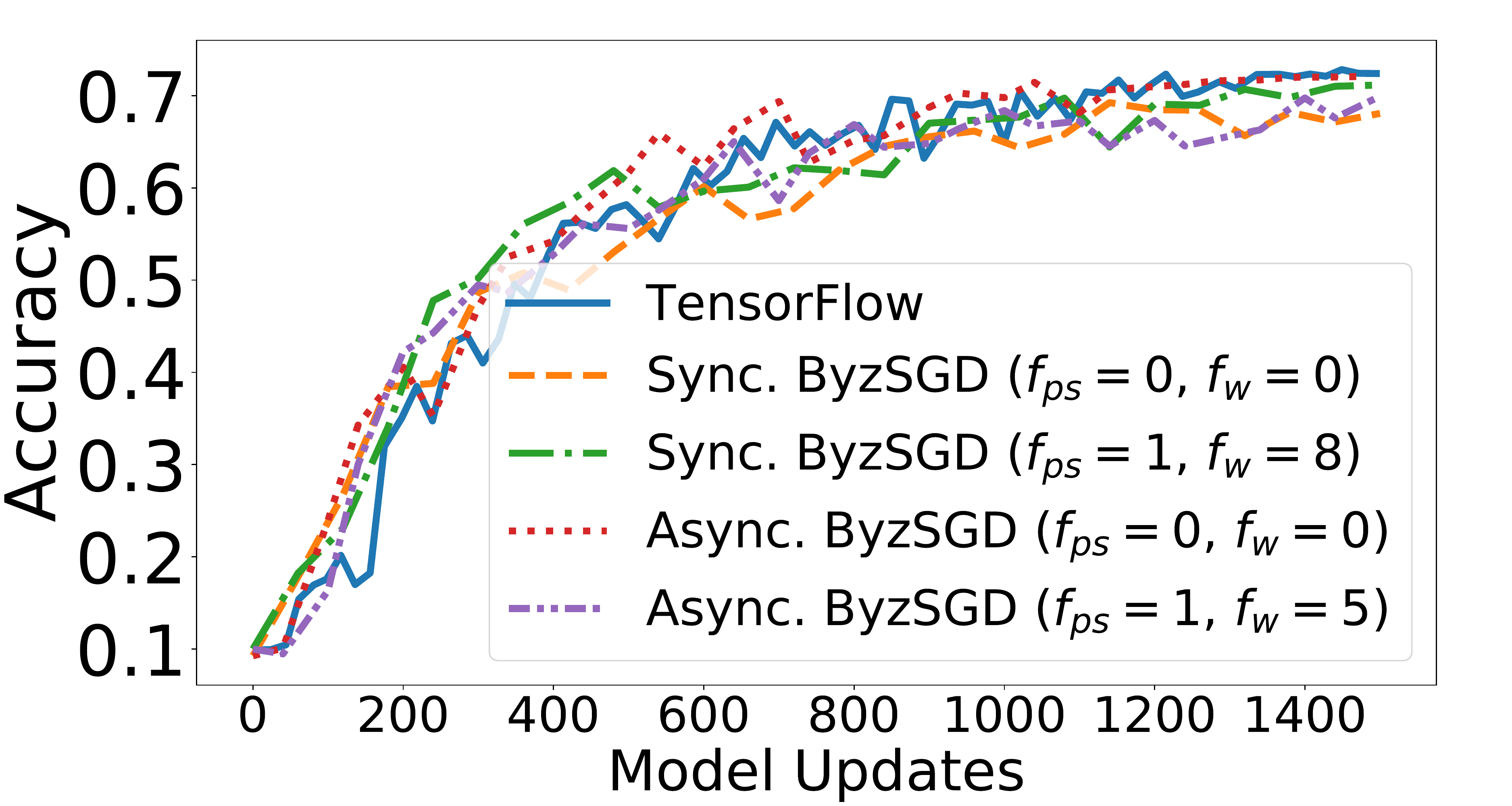}
\label{subfig:conv_b100step}}
\subfigure[]{\includegraphics[width=0.48\linewidth,keepaspectratio]{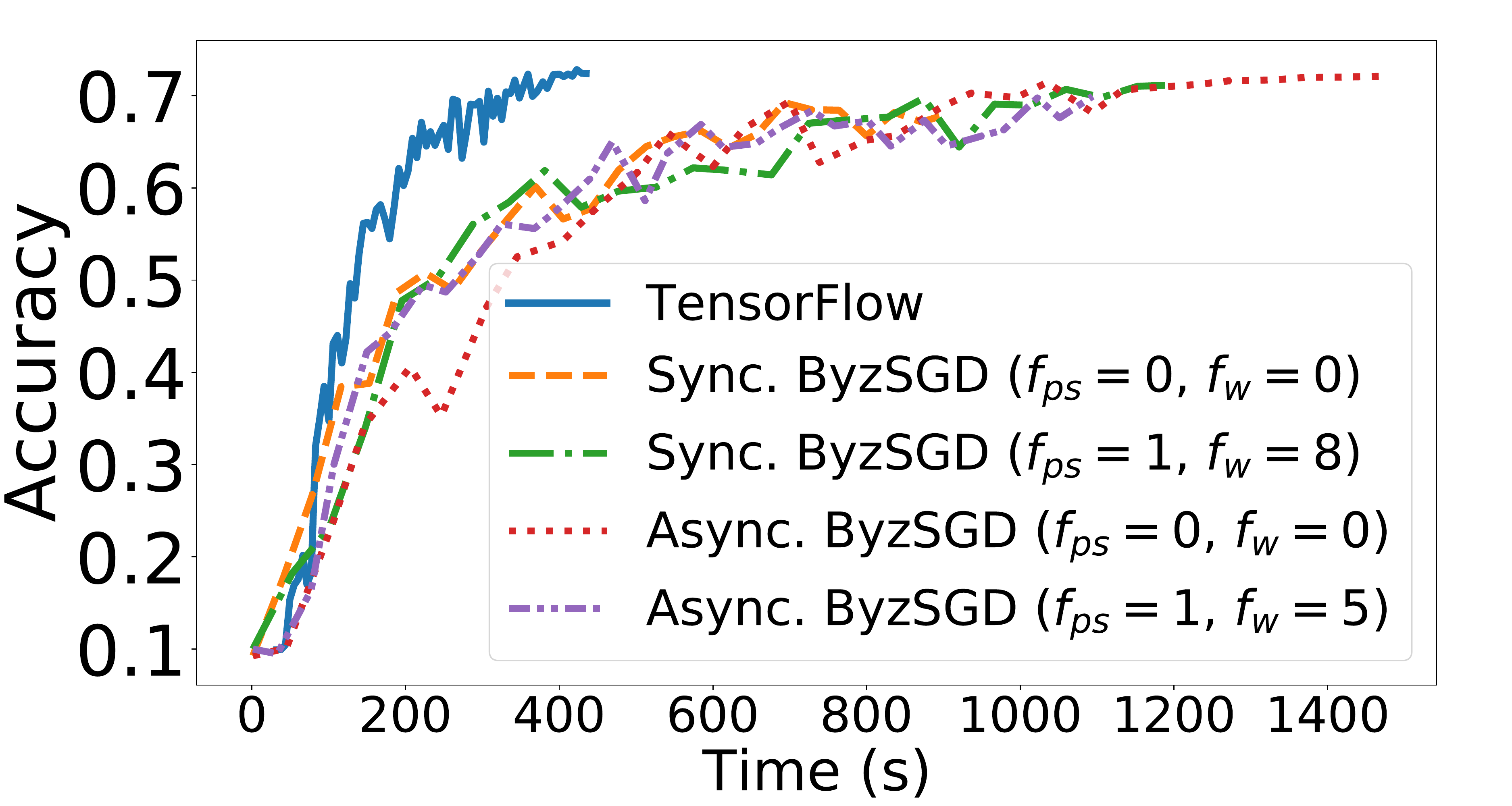}
\label{subfig:conv_b100time}}
\\[-4mm]
mini-batch size = 100
\vspace{-2mm}
\caption{Convergence in a non-Byzantine environment.}
\label{fig:conv}
\end{figure}

\begin{figure}[t]
\centering
\begin{minipage}{0.48\linewidth}
\includegraphics[width=\linewidth]{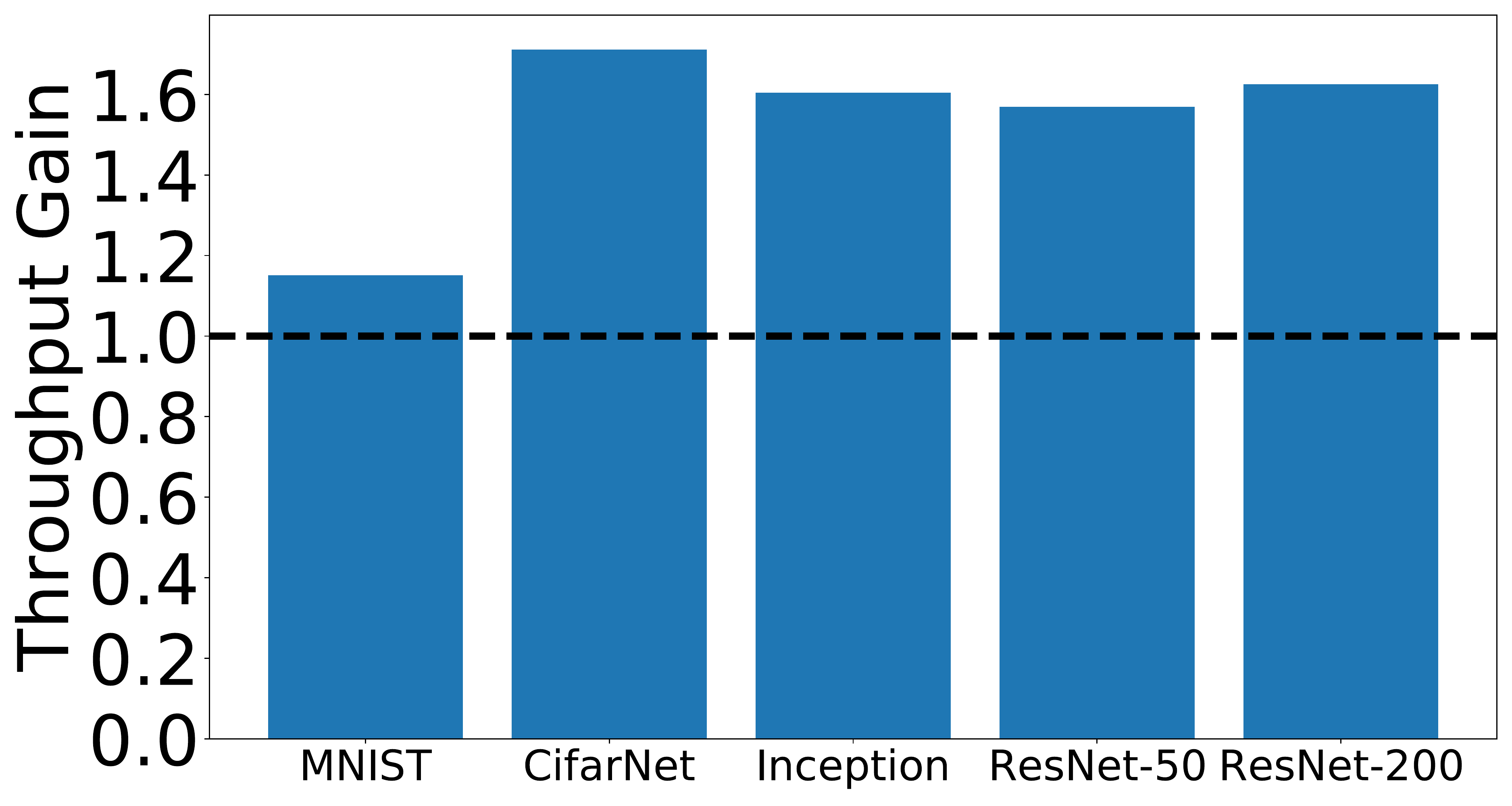}
\vspace{-4mm}
\caption{Throughput gain: sync.\ relative to async.}
\label{fig:throughput}
\end{minipage}
\hfill
\begin{minipage}{0.48\linewidth}
\includegraphics[width=\linewidth]{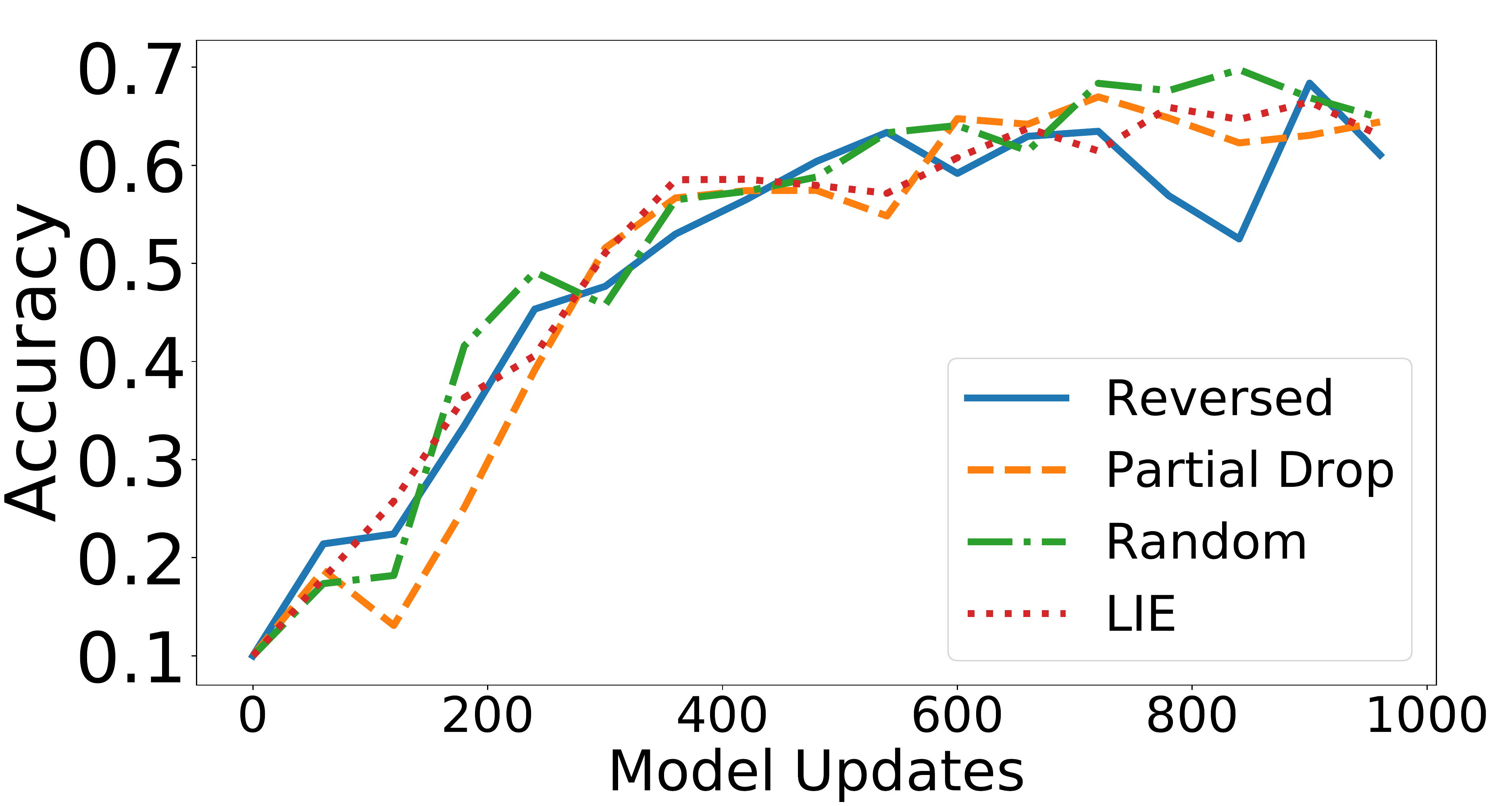}
\vspace{-4mm}
\caption{Convergence with one Byzantine server.}
\label{fig:byzps}
\end{minipage}
\vspace{-2mm}
\end{figure}

\begin{figure}[t]
\centering
\subfigure[The ratio $\frac{f_w}{n_w}$]{\includegraphics[width=0.48\linewidth,keepaspectratio]{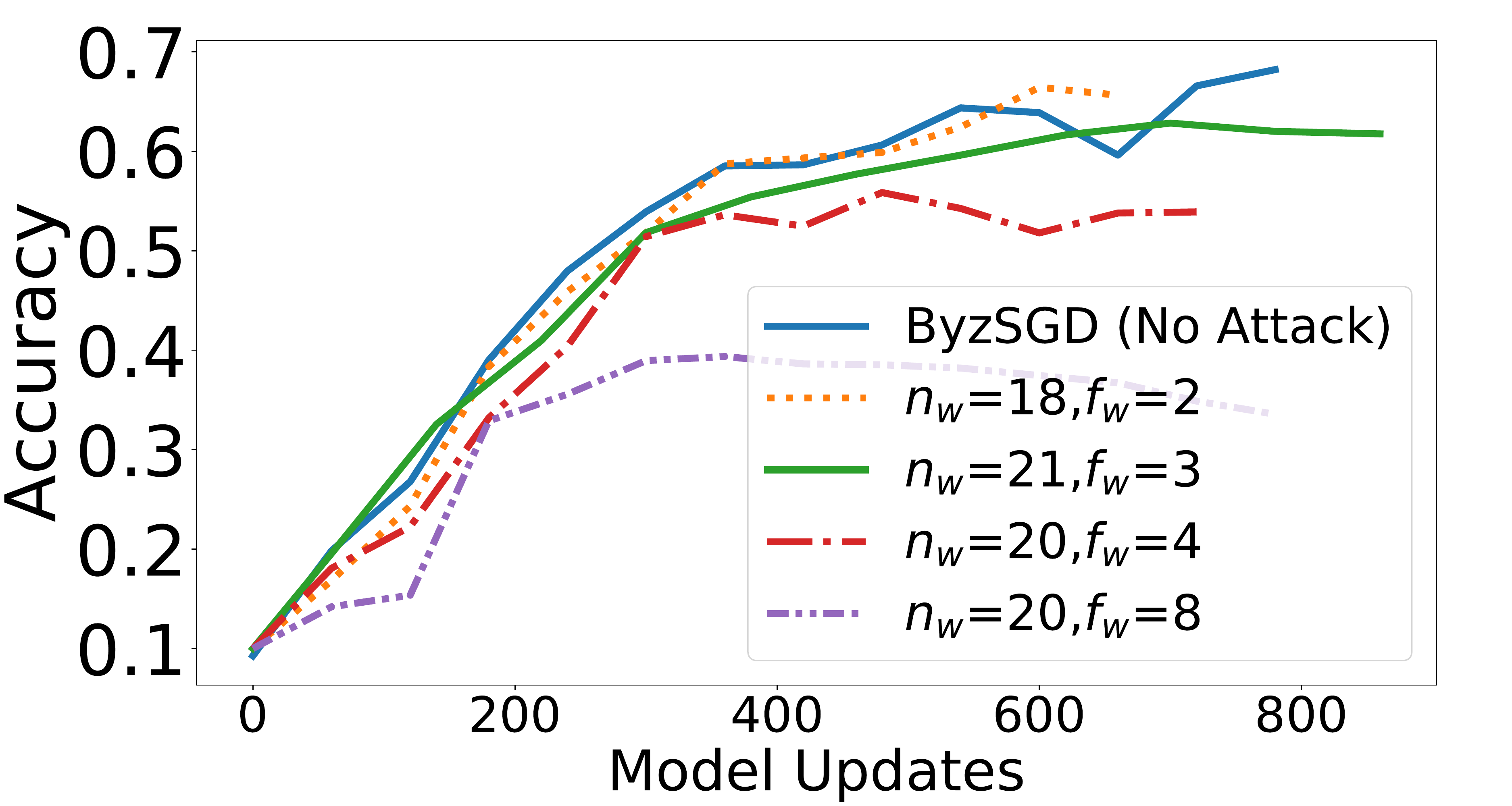}
\label{subfig:lie_ntof}}
\subfigure[Batch size]{\includegraphics[width=0.48\linewidth,keepaspectratio]{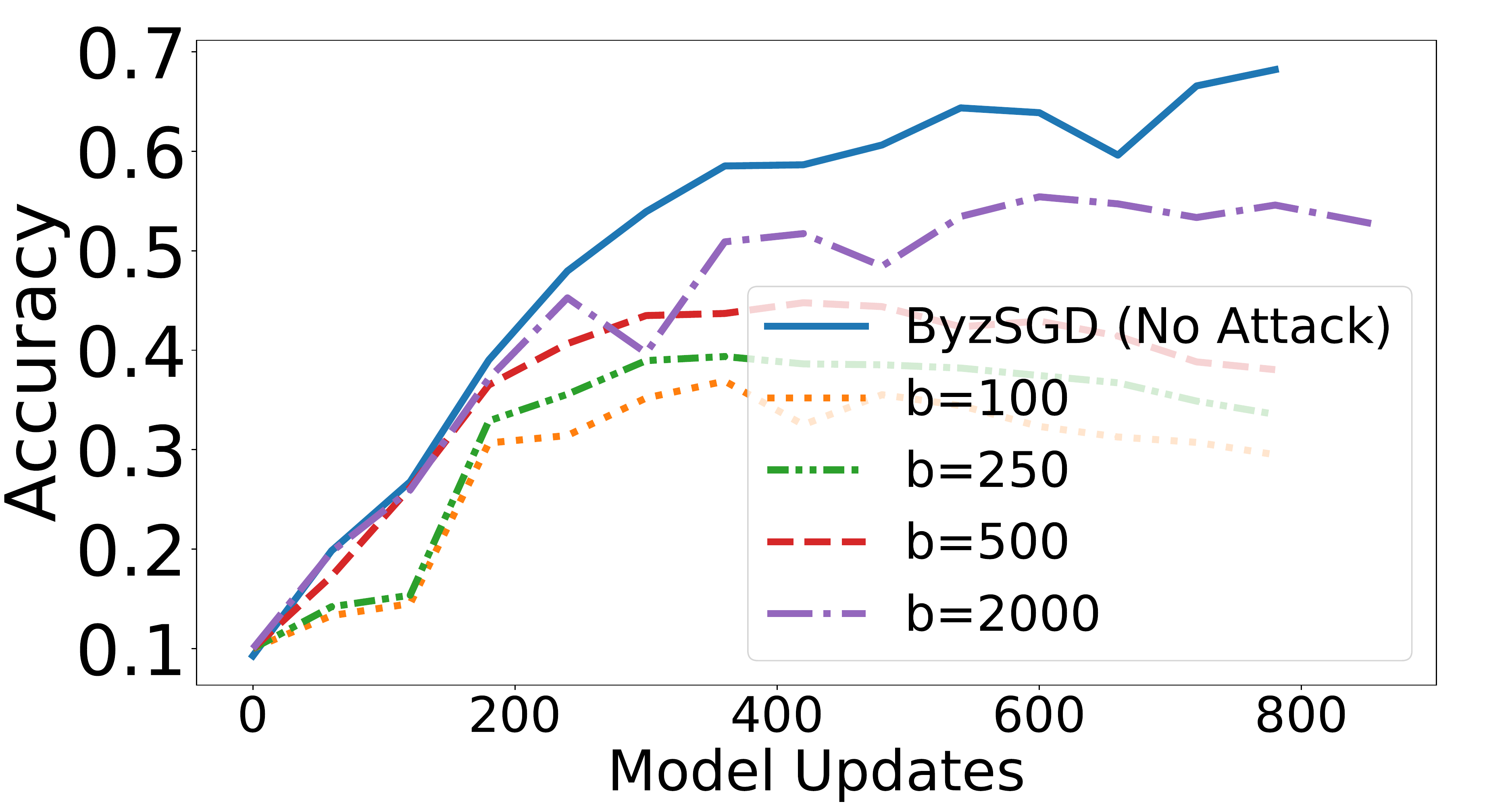}
\label{subfig:lie_batch}}
\caption{Convergence in the presence of Byzantine workers.}
\vspace{-2mm}
\label{fig:lie}
\end{figure}

\paragraph{Non--Byzantine environment.} Figure~\ref{fig:conv} shows the convergence (i.e.,\ progress of accuracy) of all experimented systems with both time and model updates (i.e.,\ training steps). We experiment with two batch sizes and different values for declared Byzantine servers and workers (only for the Byzantine--tolerant deployments).
Figure~\ref{subfig:conv_b250step} shows that all deployments have almost the same convergence behavior, with a slight loss in final accuracy for the Byzantine--tolerant deployments, which we quantify to around 5\%.
Such a loss is emphasized with the smaller batch size (Figure~\ref{subfig:conv_b100step}).
This accuracy loss is admitted in previous work~\cite{xie2018generalized} and is inherited from using statistical methods (basically, \brute{} in our case) for Byzantine resilience. In particular, \brute{} ensures convergence only to a ball around the optimal solution, i.e.,\ local minimum~\cite{bulyanPaper}.
Moreover, the figures confirm that using a higher batch size gives a more stable convergence for \systemS{}.
Figures~\ref{subfig:conv_b250step} and~\ref{subfig:conv_b100step} show that both variants of
\systemS{} almost achieve the same convergence.

The cost of Byzantine resilience is more clear when convergence is observed over time (Figure~\ref{subfig:conv_b250time}), especially with the lower batch size (Figure~\ref{subfig:conv_b100time}).
We define the convergence overhead by the ratio of the time taken by \systemS{} to reach some accuracy level compared to that taken by TensorFlow to reach the same accuracy level.
For example, in Figure~\ref{subfig:conv_b250time}, TensorFlow reaches 60\% accuracy in 268 seconds which is around 32\% better than the slowest deployment of \systemS{}.
We draw two main observations from these figures.
First, changing the number of declared Byzantine machines affects the progress of accuracy, especially with the asynchronous deployment of \systemS{}.
This is because servers and workers in such case wait for replies from only $n-f$ machines.
Hence, decreasing $f$ forces the receiver to wait for more replies, slowing down convergence.
Second, the synchronous variant always outperforms the asynchronous one, especially with non--zero values for declared Byzantine machines, be they servers and workers. Such a result is expected as the synchronous algorithm uses less number of messages per round compared to the asynchronous one.
Given that distributed ML systems are network--bound~\cite{hsieh2017gaia,poseidon}, reducing the communication overhead significantly boosts the performance (measured by convergence speed in this case) and the scalability of such systems.

\paragraph{Throughput.}
We do the same experiment again, yet with different state--of--the--art models so as to quantify the throughput gain of the synchronous variant of \systemS{}.
Figure~\ref{fig:throughput} shows the throughput of synchronous \systemS{} divided by the throughput of the asynchronous \systemA{} in each case.
From this figure, we see that synchrony helps \systemS{} achieve throughput boost (up to 70\%) in all cases, where such a boost is emphasized more with large models.
This is expected because the main advantage of synchronous \systemS{} is to decrease the number of communication messages, where bigger messages are transmitted with bigger models.

\paragraph{Byzantine workers.}
We study here the convergence of \systemS{} in the presence of Byzantine workers. Simple misbehavior like message drops, unresponsive machines, or reversed gradients are well-studied and have been shown to be tolerated by Byzantine--resilient GARs, e.g.,~\cite{xie2018generalized}, which \systemS{} also uses.
Thus, here we focus on a more recent attack that is coined as \emph{A little is enough attack}~\cite{baruch2019little}.
This attack focuses on changing each dimension in gradients of Byzantine workers to trick some of Byzantine--resilient GARs, e.g.,~\cite{krum,bulyanPaper}.

We apply this attack to multiple deployments of \systemS{}. In each scenario, we apply the strongest possible change in gradients' coordinates so as to hamper the convergence the most.
We study the effect of this attack on the convergence of \systemS{} with both the ratio of Byzantine workers to the total number of workers (Figure~\ref{subfig:lie_ntof}) and the batch size (Figure~\ref{subfig:lie_batch}).
We use the deployment with no Byzantine behavior (\emph{No Attack}) as a baseline.

Figure~\ref{subfig:lie_ntof} shows that the effect of the attack starts to appear clearly when the number of Byzantine workers is a significant fraction (more than 20\%) of the total number of workers. This is intuitive as the attack tries to increase the variance between the submitted gradients to the parameter servers and hence, increases the ball (around the local minimum) to which the used GAR converges (see e.g., \cite{krum, bulyanPaper} for a theoretical analysis of the interplay between the variance and the Byzantine resilience).
Stretching the number of Byzantine workers to the maximum ($f_w = 8$) downgrades the accuracy to around 40\% (compared to 67\% in ``No Attack" case).
This can be explained by the large variance between honest gradients, above what \brute{} requires, as we discuss in Appendix~D.

Increasing the batch size not only improves the accuracy per training step but also the robustness of \systemS{} (by narrowing down the radius of the ball around the convergence point, where the model will fluctuate as proven in~\cite{krum,bulyanPaper}).
Figure~\ref{subfig:lie_batch} fixes the ratio of $f_w$ to $n_w$ to the biggest allowed value to see the effect of using a bigger batch size on the convergence behavior.
This figure confirms that increasing the batch size increases the robustness of \systemS{}.
Moreover, based on our experiments, setting 25\% of workers to be Byzantine while using a batch size of (up to) 256 does not experimentally satisfy the assumption on the variance of \brute{} in this deployment, which leads to a lower accuracy after convergence (see Appendix~D).

\paragraph{Byzantine servers.}
Figure~\ref{fig:byzps} shows the convergence of \systemS{} in the presence of 1 Byzantine server.
We experimented with 4 different adversarial behaviors: \textbf{1.} \emph{Reversed:} the server sends a correct model multiplied by a negative number,
\textbf{2.} \emph{Partial Drop:} the server randomly chooses 10\% of the weights and set them to zero (this simulates using unreliable transport protocol in the communication layer, which was proven beneficial in some cases, e.g.,\ \cite{aggregathor}),
\textbf{3.} \emph{Random:} the server replaces the learned weights by random numbers, and
\textbf{4.} \emph{LIE}, an attack inspired from the \emph{little is enough} attack~\cite{baruch2019little}, in which the server multiplies each of the individual weights by a small number $z$, where $\lvert z - 1\rvert < \delta$ with $\delta$ close to zero;
$z=1.035$ in our experiments.
Such a figure shows that \systemS{} can tolerate the experimented Byzantine behavior and guarantee the learning convergence to a high accuracy.
\section{Concluding Remarks}
\label{sec:conc}

\paragraph{Summary.}
This paper is a first step towards genuinely distributed Byzantine--resilient Machine Learning (ML) solutions that do not trust any network component.
We present \systemS{} that guarantees learning convergence despite the presence of Byzantine machines.
Through the introduction of a series of novel ideas, the \emph{Scatter/Gather} protocol, the \emph{Distributed Median-based Contraction (DMC)} module, and the filtering mechanisms, we show that \systemS{} works in an asynchronous setting,
and we show how we can leverage synchrony to boost performance.
We built \systemS{} on top of TensorFlow, and we show
that it tolerates Byzantine behavior with $32\%$ overhead compared to vanilla TensorFlow.

\paragraph{Related work.}
With the impracticality (and sometimes impossibility~\cite{fischer1985impossibility}) of applying exact consensus to ML applications, the approximate consensus~\cite{Fekete87} seems to be a good candidate.
In approximate consensus, all nodes try to decide values that are arbitrarily close to each other and that are within the range of values proposed by correct nodes.
Several approximate consensus algorithms were proposed with different convergence rates, communication/computation costs, and supported number of tolerable Byzantine nodes, e.g.,~\cite{dolev1986reaching,fekete1986asymptotically,abraham2004optimal,MHVG15}.

Inspired by approximate consensus, several Byzantine--resilient ML algorithms were proposed yet, all assumed a single correct parameter server: only workers could be Byzantine.
Three Median-based aggregation rules were proposed to resist Byzantine attacks~\cite{xie2018generalized}.
Krum~\cite{krum} and Multi-Krum~\cite{aggregathor} used a distance-based algorithm to eliminate Byzantine inputs and average the correct ones.
Bulyan~\cite{bulyanPaper} proposed a meta-algorithm to guarantee Byzantine resilience
against a strong adversary that can fool the aforementioned aggregation rules.
Draco~\cite{chen2018draco} used coding schemes and redundant gradient computation for Byzantine resilience, where Detox~\cite{rajput2019detox} combined coding schemes with Byzantine--resilient aggregation for better resilience and overhead guarantees.
Kardam~\cite{kardam} is the only proposal to tolerate Byzantine workers in asynchronous learning setup.
\systemS{} augments these efforts by tolerating Byzantine servers in addition to Byzantine workers.

\paragraph{Open questions.}
This paper opens interesting questions.

First, the relation between the frequency of entering the \emph{gather} phase (i.e.,\ the value of $T$) and the variance between models on correct servers is both data and model dependent.
In our analysis, we provide safety guarantees on this relation that always ensure Byzantine resilience and convergence.
However, we believe that in some cases, entering the \emph{gather} phase more frequently may lead to a noticeable improvement in the convergence speed (see Appendix~E.2). The trade-off between this gain and the corresponding
communication overhead is an interesting open question.

Second, the Lipschitz filter requires $n_{ps} > 3f_{ps}$ (see Appendix~C.2.3).
There is another tradeoff here between the communication overhead and the required number of parameter servers.
One can use Byzantine--resilient aggregation of models, which requires only $n_{ps} > 2f_{ps}$, yet requires communicating with \emph{all} servers in each step.
In our design, we strive for reducing the communication overhead, given that communication is the bottleneck~\cite{poseidon,hsieh2017gaia}.

Third, it is interesting to explore similar Byzantine--resilient solutions in the context of decentralized settings, i.e.,\ in which all machines are considered as workers with no central servers and also with non-iid data.
Although it seems that we can directly apply the same algorithms presented in this paper to such settings, designing better algorithms with low communication overhead with provable resilience guarantees remains an open question.

\section*{Acknowledgments}
This work has been supported in part by the Swiss National
Science Foundation (FNS grant 200021\_182542/1).

Most experiments presented in this paper were carried out using the Grid'5000 testbed, supported by a scientific interest group hosted by Inria and including CNRS, RENATER and several Universities as well as other organizations (see \url{https://www.grid5000.fr}).

\bibliography{bib}
\bibliographystyle{ACM-Reference-Format}

\appendix

\section{Preliminary Material}

\subsection{Byzantine--resilient aggregation}

Byzantine--resilient aggregation of gradients is the key for Byzantine workers' resilience. To this end, gradients are processed by a gradient aggregation rule (GAR), which purpose is to ensure that output of aggregation is as close as possible to the real gradient of the loss function.

In the general theory of stochastic gradient descent (SGD) convergence, a typical validity assumption is that the gradient estimator is unbiased~\cite{bottou1998online}. The role of a GAR is to ensure a relaxed version of this assumption in order to accommodate for the presence of malicious workers (whose gradients are potentially biased).

Definition~\ref{gar_def} gives such a relaxation, which we adapt from~\cite{krum, bulyanPaper} and which was used as a standard for Byzantine resilience in, e.g. ~\cite{xie2018phocas, xie2018zeno, xie2018generalized}.

\begin{definition}
Let $\alpha \le 0 < \frac{\pi}{2}$ be any angular value and $0 \le f \le n$ with $n$ the total number of input vectors to the GAR and $f$ the maximum number of Byzantine vectors.
Let $g$ be an unbiased estimate of the true gradient $G$, i.e.,\ $\expect G=g$.

A GAR (which output noted as $\mathcal{F}$) is robust (said to be \alphaf{}--Byzanitne resilient) if

$\left\langle\expect\mathcal{F},g \right\rangle \ge (1 - \sin\alpha) . \norm{g}^2 > 0$
\label{gar_def}
\end{definition}

\systemS{} uses two GARs: \brutelong{}~\cite{bulyanPaper} (which can be safely replaced with any other GAR that gives similar guarantees such as Krum~\cite{krum} or Bulyan~\cite{xie2018generalized}) and \median{}~\cite{xie2018generalized}.

\subsection{\brutelong{} (\brute)}
\label{sec:mda}
\brute{} is a gradient aggregation rule (GAR) that ensures resilience against a minority of Byzantine input gradients. Mathematically, this function was introduced in~\cite{rousseeuw1985multivariate} and its Byzantine resilience proof was given in~\cite{bulyanPaper}. \brute{} satisfies the \alphaf{} Byzantine resilience guarantees\footnote{Basically, any GAR that satisfies such a form of resilience~\cite{krum,bulyanPaper,xie2018generalized} can be used with \systemS{}; \brute{} is just an instance.} introduced in~\cite{krum}. Formally, let $\mathcal{X}$ be the set of input gradients (all follow the same distribution), out of them $f$ are Byzantine, and $y$ be the output of the Byzantine resilient GAR. Then, the following properties hold:
\begin{enumerate}[nolistsep,noitemsep]
    \item{$\expect{}y$ is in the same half--space as $\expect\mathcal{X}$}.
    \item{the first $4$ statistical moments of $y$ are bounded above by a linear combination of the first $4$ statistical moments of $x \sim \mathcal{X}$}.
\end{enumerate}

Such conditions are sufficient to show that the output of this GAR guarantees convergence of the learning procedure. More formally, these conditions enable the GAR to have a proof that follows from the \emph{global confinement} proof of Stochastic Gradient Descent (SGD)~\cite{bottou1998online}.

In order to work, \brute{} assumes the following (any statistically-robust GAR depends on a similar condition):
\begin{align}
    &\exists \kappa \in \left] 1, +\infty \right[ \mathsep
    \forall \left( i, t, \params{}{} \right) \in \range{1}{n - f} \times \setn \times \setr^d \mathsep \nonumber \\
    &\kappa \, \frac{2 f}{n - f} \, \sqrt{\expect\left( \norm{\gradvar{i}{t} - \expect\gradvar{i}{t}}^2 \right)} \le \norm{\realgrad{\params{}{}}},
    \label{itm:assume-kappa}
\end{align}
where $\theta$ is the model state at the training step $t$, $n$ is the total number of input gradients, $f$ is the maximum number of Byzantine gradients, $g_t$ is an unbiased estimate of the gradient at step $t$, and $L$ is the loss function.

The \brute{} function works as follows. Consider that the total number of gradients is $n$ and the maximum number of Byzantine gradients is $f$ with $n \ge 2f+1$. \brute{} enumerates all subsets of size $n - f$ from the input gradients and finds the subset with the \emph{minimum diameter} among all subsets of this size, i.e.,\ $n - f$. The \emph{diameter} of a subset is defined as the maximum distance between any two elements of this subset. The output of the \brute{} function is the average of gradients in such a subset. More formally, the \brute{} function is defined as follows~\cite{bulyanPaper}:

\paragraphspace
Let $\left( g_1 \ldotexp g_{n} \right) \in \left( \mathbb{R}^d \right)^n$,
and $\mathcal{X} \triangleq \left\lbrace\, g_1 \ldotexp g_n \,\right\rbrace$ the set containing all the input gradients.

Let $\mathcal{R} \triangleq \left\lbrace\, \mathcal{Q} \mid \mathcal{Q} \subset \mathcal{X} \mathsep \card{\mathcal{Q}} = n - f \,\right\rbrace$ the set of all the subsets of $\mathcal{X}$ with a cardinality of $n - f$,
and let:
\begin{equation*}
    \mathcal{S} \triangleq \argmin\limits_{\mathcal{Q} \in \mathcal{R}}\left( \max\limits_{\left( g_i, g_j \right) \in \mathcal{Q}^2}\left( \norm{g_i - g_j} \right) \right).
\end{equation*}

Then, the aggregated gradient is given by:
\begin{equation*}
    \brute{}\left( g_1 \ldotexp g_n \right) \triangleq \frac{1}{n - f} \sum\limits_{g \in \mathcal{S}}{g}.
\end{equation*}

\subsection{Coordinate--wise \median{}}

We formally define the \emph{median} as follows:
\begin{align*}
    \forall q \in \setn - \left\lbrace 0 \right\rbrace \mathsep
    \forall \left( x_1 \ldotexp x_q \right) \in \setr^q \mathsep
    \emph{median}\left( x_1 \ldotexp x_q \right) \triangleq x_s \in \setr
\end{align*}
such that:
\begin{align*}
&   \exists \left( L, H \right) \subset \range{1}{q}^2 \mathsep
    L \cap H = \emptyset \mathsep
    \card{L} = \card{H} = \floor{\frac{q}{2}} \mathsep \\
&   \forall l \in L \mathsep x_l \le x_s \mathsep
    \forall h \in H \mathsep x_h \ge x_s
\end{align*}
and:
\begin{equation*}
    \left\lbrace\begin{array}{l l}
        \exists \left( a, b \right) \in \range{1}{q}^2 \mathsep x_s = \frac{x_a + x_b}{2} & \text{ if }q\text{ is even} \\
        x_s \in \left\lbrace x_1 \ldotexp x_q \right\rbrace & \text{ if }q\text{ is odd}
    \end{array}\right.
\end{equation*}

\paragraphspace
We formally define the coordinate--wise median as follows:
\begin{align*}
    &\forall \left( d, q \right) \in \left( \setn - \left\lbrace 0 \right\rbrace \right)^2 \mathsep
    \forall \left( x_1 \ldotexp x_q \right) \in \left( \setr^d \right)^q \mathsep \\
    &\emph{median} \left( x_1 \ldotexp x_q \right) \triangleq x_s \in \setr^d
\end{align*}
such that:
\begin{equation*}
    \forall i \in \range{1}{d} \mathsep
    x_s\indexed{i} = \emph{median}\left( x_1\indexed{i} \ldotexp x_q\indexed{i} \right)
\end{equation*}
\systemS{} uses \emph{coordinate-wise median}, hereafter simply: \median{}.

\section{\systemS{}'s Notations and Algorithm}
\label{app:alg}
\subsection{Notations}
\label{sec:not}
Let $\left( n_w, n_{ps}, f_w, f_{ps}, d \right) \in \setn ^5$, each representing:
\begin{itemize}[noitemsep,nolistsep]
    \item{$n_{ps} \ge 3 f_{ps} + 1$ the total number of parameter servers, among which $f_{ps}$ are Byzantine}
    \item{$n_w \ge 3 f_w + 1$ the total number of workers, among which $f_w$ are Byzantine}
    \item{$2 f_{ps} + 2 \le q_{ps} \le n_{ps} - f_{ps}$ the \emph{quorum} (see Section \ref{sec:algo}) used for aggregating servers' replies \\
    $2 f_w + 1 \le q_w \le n_w - f_w$ the \emph{quorum} used for aggregating workers' replies}
    \item{$d$ the dimension of the parameter space $\setr^d$}
\end{itemize}

Let (without loss of generality):
\begin{itemize}[noitemsep,nolistsep]
    \item{$\range{1}{n_{ps} - f_{ps}}$ be indexes of correct parameter servers \\
    $\range{n_{ps} - f_{ps} + 1}{n_{ps}}$ be indexes of Byzantine servers}
    \item{$\range{1}{n_w - f_w}$ be indexes of correct workers \\
    $\range{n_w - f_w + 1}{n_w}$ be indexes of Byzantine workers}
\end{itemize}

\paragraphspace
Let $\params{i}{t}$ be a notation for the parameter vector (i.e., model) at parameter server $i \in \range{1}{n_{ps}-f_{ps}}$ for step $t \in \setn$.

Let $\gradvar{i}{t}$ be a notation for the gradient estimation at worker $i \in \range{1}{n_{w}-f_{w}}$ for step $t \in \setn$.

Let $\graddist{i}{t}$ be a notation for the gradient distribution at worker $i \in \range{1}{n_w-f_w}$ for step $t \in \setn$.

Let $L$ be the loss function we aim to minimize, let $\realgrad{\params{}{}}$ be the real gradient of the loss function at $\params{}{}$, and let $\stocgrad{\params{}{}}$ be a stochastic estimation of the gradient, following $\graddist{}{}$, of $L$ at $\params{}{}$.

\paragraphspace
Let $\xi_t \argsfor{a}{k}$ denote the subset of size $k$ of some set $\left\lbrace \xi^{(1)}_t \ldotexp \xi^{(n)}_t \right\rbrace$ delivered by node $a$ at step $t$.
To highlight the fact that such a subset can contain up to $f$ arbitrary (Byzantine) vectors, we will also denote it by $\left( \xi_t \argsfor{a}{k - f}, \xi_t \argsfor{a}{f} \right)$.
Also, the exact value of $n$ depends on the context: in the proof, it will always be $n_w$ when $\xi^{(i)}_t$ denotes a gradient, and $n_{ps}$ otherwise.

Let $\eta_t$ be the learning rate at the learning step $t \in \setn$ with the following specifications:
\begin{enumerate}[noitemsep,nolistsep]
    \item The sequence of learning rates $\eta_t$ is decreasing~\footnote{In fact, it is sufficient that the sequence is decreasing only once every $T$ steps, with $T = \frac{1}{3.l.\eta_1}$ where $l$ is the Lipschitz coefficient of assumption~\ref{itm:Lipschitz-assumption} (cf Section~\ref{sec:app-assumptions}).} with $t$, i.e.,\ if $t_a > t_b$ then, $\eta_{t_a} < \eta_{t_b}$.
    Thus, the initial learning rate $\eta_0$ is the largest value among learning rates used in subsequent steps.
    \item The sequence of learning rates $\eta_t$ satisfies $\sum_{t}{\eta_t} = \infty$ and $\sum_{t}{{\eta_t}^2} < \infty$.
\end{enumerate}

\subsection{\systemS{}'s Algorithm}
\label{sec:algo}
\paragraph{Initialization.} Each correct parameter server $i$ and worker $j$ starts (at step $t = 0$) with the same parameter vector:
\begin{align*}
    \forall i \in \range{1}{n_{ps} - f_{ps}} \mathsep \params{i}{0} \triangleq \params{}{0} \\
    \forall j \in \range{1}{n_w - f_w} \mathsep \params{j}{0} \triangleq \params{}{0}
\end{align*}

Additionally, in the synchronous case, each correct worker $j$ generates a random integer $r_j \in \range{1}{n_{ps}}$ and computes $\gradvar{j}{0}$ (via backpropagation) at the initial model $\params{}{0}$.

\paragraphspace
\paragraph{Training loop.}
Each training step $t \in \setn$, the following sub-steps are executed sequentially (unless otherwise stated).
\begin{enumerate}
    \item Each parameter server $i$ pulls gradients $\gradvar{}{t}$ from \emph{all}\footnote{In asynchronous networks, the server waits for only $n_w-f_w$ replies.} workers and then applies the \brute{} function on the received gradients, computing the aggregated gradient $\gradvar{i}{\text{agg}}$. Then, each server uses its own computed $\gradvar{i}{\text{agg}}$ to update the model as follows: $\params{i}{t+1} = \params{i}{t} - \eta_t \gradvar{i}{\text{agg}}$.
    \label{itm:ps-compute}
    \item While parameter servers are doing step~\ref{itm:ps-compute}, each worker $j$ does a speculative step as follows: a worker $j$ calculates its local view to the updated model: $\params{j(l)}{t+1} = \params{j}{t} - \eta_t \gradvar{j}{t}$.
    \item \textbf{a.} In the asynchronous algorithm, each worker $j$ asks for the updated models from \emph{all} servers; yet, it does not expect to receive more than $n_{ps}-f_{ps}$ replies.
    The worker $j$ then computes \median{} on the received models to obtain $\params{agg}{t+1}$. In normal (i.e.,\ \emph{scatter}) steps, worker $j$ sets  $\params{j}{t+1} = \params{agg}{t+1}$.\\
    \textbf{b.} In the synchronous algorithm, each worker $j$ pulls one parameter vector $\params{i}{t+1}$ from server $i$ where, $i = (r_j+t+1)\text{ mod }n_{ps}$. Each worker $j$ does the backpropagation step, computing $\gradvar{j}{t+1}$ at the pulled model $\params{i}{t+1}$.
    \label{itm:pull-pv}
    \item To complete the picture in the synchronous variant, each worker $j$ tests the legitimacy of the received model $\params{i}{t+1}$ using the \emph{Lipschitz criterion} (i.e.,\ \lipschitzf{} filter) and the \emph{difference on model norms} (i.e.,\ \modelsf{} filter) as follows. First, a worker $j$ calculates $k$, an empirical estimation of the Lipschitz coefficient, which is defined as:
    \begin{equation*}
        k = \frac{\norm{ \gradvar{j}{t+1} - \gradvar{j}{t}}}{\norm{ \params{j(l)}{t+1} - \params{j}{t}}}
    \end{equation*}
    Then, the worker tests whether this value $k$ lies in the non-Byzantine quantile of Lipschitz coefficients: $k \le K_p \triangleq \text{quantile}_\frac{n_{ps} - f_{ps}}{n_{ps}} \{K\}$ where, $K$ is the list of all previous Lipschitz coefficients $k$ (i.e.,\ with $t_{prev} < t$).
    Second, the worker $j$ computes the distance between the local and the pulled (from server $i$) models as follows: $\norm{ \params{j(l)}{t+1} - \params{i}{t+1}}$ and makes sure that such a difference remains strictly below:
    \begin{align*}
        &\eta_{T \cdot (t\text{ mod }T)} \norm{ g_{T \cdot (t\text{ mod }T)}} \\
        \times~&\bigg( \frac{(3T+2)(n_w - f_w)}{4 f_w} + 2 \big((t-1)\text{ mod }T\big) \bigg),
    \end{align*}
    with $T = \frac{1}{3l \eta_1}$, where $l$ is the Lipschitz coefficient (assumption~\ref{itm:Lipschitz-assumption}.
    If both conditions are satisfied, the received model $\params{i}{t+1}$ is approved and the algorithm continues normally. Otherwise, the parameter server $i$ is suspected and its model is ignored; worker $j$ continues by repeating step $t$ again from step~\ref{itm:pull-pv}.
    \label{itm:end-scatter}
    \item To bound the drifts between parameter vectors at correct servers, each $T = \frac{1}{3l \eta_1}$ steps, a global \emph{gather} phase is entered on both servers and workers sides by completing the \emph{Distributed Median-based Contraction} module. During this phase, the following happens:
    each server $i$ sends to all other servers its current view of the model $\params{i}{t}$.
    After gathering models from all servers (or $n_{ps}-f_{ps}$ servers in the asynchronous case), each server $i$ aggregates such models with \median{}, computing $\params{i(agg)}{t}$.
    Then, each worker $j$ pulls the model $\params{i(agg)}{t}$ from all parameter servers (or $n_{ps}-f_{ps}$ servers in the asynchronous case) and aggregates the received models using \median{}.
    Finally, each worker $j$ uses the aggregated model to compute the backpropagation step, and the algorithm continues normally from step~\ref{itm:ps-compute}.
    \label{itm:gather-step}
\end{enumerate}
We call steps~\ref{itm:ps-compute} through~\ref{itm:end-scatter} \emph{\textbf{scatter}} phase and step~\ref{itm:gather-step} \emph{\textbf{gather}} phase. During \emph{scatter} phase(s) servers do not communicate and hence, their views of the model deviate from each other. The goal of the \emph{gather} phase (through applying the DMC computation) is to bring back the models at the correct servers close to each other.

\section{\systemS{}'s Convergence}
\label{sec:byzsgd_proof}
In this section, we show that \systemS{} guarantees convergence and tolerates Byzantine workers and servers. We first state the assumptions for \systemS{} to work and then dive into the proof.

Formally, we prove that, noting $\params{}{t} \triangleq \median{} \left( \params{1}{t} \ldotexp \params{n}{t} \right)$:
\begin{equation}
    \lim\limits_{t \rightarrow +\infty}{\norm{\realgrad{\params{}{t}}}} = 0
    \label{eqn:app-convergence}
\end{equation}

\subsection{Assumptions}
\label{sec:app-assumptions}
\begin{enumerate}
    \item{$\forall t \in \setn \mathsep \gradvar{1}{t} \ldotexp \gradvar{n_w - f_w}{t}\text{ are mutually independent}$}.
    \label{itm:assume-iid}
    \item{$\exists \sigma' \in \setr_{+} \mathsep \forall \left( i, t \right) \in \range{1}{n_w - f_w} \times \setn \mathsep \expect\norm{\gradvar{i}{t} - \expect\gradvar{i}{t}} \le \sigma'$.
    \label{itm:assume-bounded-deviation}}
    \item{$L$ is positive, and $3$--times differentiable with continuous derivatives}.
    \item{$\forall r \in \range{2}{4} \mathsep \exists \left( A_r, B_r \right) \in \setr^2 \mathsep \forall \left( i, t, \params{}{} \right) \in \range{1}{n_w - f_w} \times \setn \times \setr^d \mathsep \expect\norm{\gradvar{i}{t}} \le A_r + B_r \norm{\params{}{}}^r$.%
    \label{itm:assume-bounded-moments}}
     \item $L$ is Lipschitz continuous, i.e.\ $\exists l > 0 \mathsep \forall \left( x, y \right) \in \left( \setr^d \right)^2 \mathsep \\ \norm{\realgrad{x} - \realgrad{y}} \le l \norm{x - y}$.
    \label{itm:Lipschitz-assumption}
    \item{$\exists D \in \setr \mathsep \forall \params{}{} \in \setr^d \mathsep \norm{\params{}{}}^2 \ge D \mathsep \exists \left( \varepsilon, \beta \right) \in \setr_{+} \times \left[ 0, \frac{\pi}{2} - \eta \right[ \mathsep \\ \norm{\realgrad{\params{}{}}} \ge \varepsilon \mathsep \left\langle \params{}{}, \realgrad{\params{}{}} \right\rangle \ge \cos\left( \beta \right) \, \norm{\params{}{}} \, \norm{\realgrad{\params{}{}}}$.
    \label{itm:assume-end-krum}}
    \item{Denote $S_j \triangleq \left\lbrace s\!\subset\!\range{1}{n_{ps}\!-\!f_{ps}}\!-\!\left\lbrace j \right\rbrace \suchthat \card{s}\!\in\!\range{q\!-\!f\!-\!1}{q\!-\!1} \right\rbrace$ the subset of correct parameter server indexes that parameter server $j$ can deliver at a given step.
    We then call $S = \prod_{j \in \range{1}{n_{ps}-f_{ps}}} S_j$ the set of all correct parameter server indexes the $n_{ps}-f_{ps}$ correct parameter servers can deliver.
    We assume $\exists \rho > 0 \mathsep \forall s \in S \mathsep P\left( X = s \right) \ge \rho$.%
    \label{itm:assume-any-quorum}}
\end{enumerate}

Assumptions~\ref{itm:assume-iid} to~\ref{itm:Lipschitz-assumption} (i.i.d, bounded variance, differentiability of the loss, bounded statistical moments, and Lipschitz  continuity of the gradient) are the most common ones in classical SGD analysis~\cite{bottou1998online, bousquet2008tradeoffs}.

Assumption~\ref{itm:assume-end-krum} was first adapted from~\cite{bottou1998online} by~\cite{krum, bulyanPaper, kardam} to account for Byzantine resilience. It intuitively says that beyond a certain horizon, the loss function is ``steep enough" (lower bounded gradient) and  ``convex enough" (lower bounded angle between the gradient and the parameter vector). The loss function does not need to be convex, but adding regularization terms such as $\| \theta \|^2$ ensures assumption~\ref{itm:assume-end-krum}, since close to infinity, the regularization dominates the rest of the loss function and permits the gradient $\nabla L (\theta)$ to point to the same half space as $\theta$. The original assumption of~\cite{bottou1998online} is that $\left\langle \params{}{}, \realgrad{\params{}{}} \right\rangle > 0$; in ~\cite{krum, bulyanPaper} it was argued that requiring this scalar product to be strictly positive is the same as requiring the angle between $\params{}{}$ and  $\realgrad{\params{}{}}$ to be lower bounded by an acute angle ($\beta < \pi / 2$).

\subsection{Lemmas}
\label{subsec:lemmas}
In this section, we develop lemmas that will be used in proving the convergence of \systemS{}.

\subsubsection{\brute{} bounded deviation from majority}
\label{sec:bound-effect}

Let $\left( d, f \right) \in {\setn_{>0}}^2$,
let $q \in \setn$
such that $q \ge 2 \, f + 1$.

Let note $H_1 \triangleq \range{1}{q - f}$ and $H_2 \triangleq \range{q + 1}{2 \, q - f}$.

\paragraphspace
\begin{lemma}[Guarantee of \brute{}]
We will show that:
\begin{align*}
    & \hspace{-5mm} \forall \left( x_1 \ldotexp x_{2 q} \right) \in \left( \setr^d \right)^{2 q} \mathsep \\
    & \norm{\brute{}\left( x_1 \ldotexp x_q \right) - \brute{}\left( x_{q + 1} \ldotexp x_{2 q} \right)} \\
    \le~&3 \! \max\limits_{\left( i, j \right) \in \left( H_1 \cup H_2 \right)^2}{\norm{x_i - x_j}}
\end{align*}
\end{lemma}

\begin{proof}
We will proceed by construction of $\brute{}$ (Section \ref{sec:mda}).

Reusing the formal notation from the end of \ref{sec:mda}, we recall that $\brute{}\left( x_1 \ldotexp x_q \right) \triangleq \frac{1}{q - f} \sum_{x \in \mathcal{S}_1}{x}$.

Since $\mathcal{S}_1$ is a subset of size $q - f$ of smallest diameter in $\left\lbrace x_1 \ldotexp x_q \right\rbrace$, the following holds:
\begin{equation*}
    \exists \left( i, j \right) \in {H_1}^2 \mathsep
    \forall \left( y, z \right) \in \mathcal{S}_1^2 \mathsep
    \norm{y - z} \le \norm{x_i - x_j}
\end{equation*}

Then, observing that $q - f > f \implies \exists k \in H_1 \mathsep x_k \in \mathcal{S}_1$, we can compute:
\begin{align*}
    \norm{\brute{}\left( x_1 \ldotexp x_q \right) - x_k}
    &= \norm{\left( \frac{1}{q - f} \sum\limits_{x \in \mathcal{S}_1}{x} \right) - x_k } \\
    &= \frac{1}{q - f} \norm{\sum\limits_{x \in \mathcal{S}_1}{\left( x - x_k \right)}} \\
    &\le \frac{1}{q - f} \sum\limits_{x \in \mathcal{S}_1}{\norm{x - x_k}} \\
    &\le \frac{1}{q - f} \sum\limits_{x \in \mathcal{S}_1}{\left( \max\limits_{\left( i, j \right) \in {H_1}^2}{\norm{x_i - x_j}} \right)} \\
    &\le \max\limits_{\left( i, j \right) \in {H_1}^2}{\norm{x_i - x_j}}
\end{align*}

Using the same construction, with $l \in H_2$:
\begin{equation*}
    \norm{\brute{}\left( x_{q + 1} \ldotexp x_{2 q} \right) - x_l}
    \le \max\limits_{\left( i, j \right) \in {H_2}^2}{\norm{x_i - x_j}}
\end{equation*}

Finally, reusing $k$ and $l$, we can compute and conclude:
\begin{align*}
    &\norm{\brute{}\left( x_1 \ldotexp x_q \right) - \brute{}\left( x_{q + 1} \ldotexp x_{2 q} \right)} \\
    =~&\norm{\brute{}\left( x_1 \ldotexp x_q \right) - x_k + x_k - x_l + x_l - \brute{}\left( x_{q + 1} \ldotexp x_{2 q} \right)} \\
    \le~&\norm{\brute{}\left( x_1 \ldotexp x_q \right) - x_k} \\
    &+ \norm{x_k - x_l} \\
    &+ \norm{\brute{}\left( x_{q + 1} \ldotexp x_{2 q} \right) - x_l} \\
    \le~&3 \cdot \max\limits_{\left( i, j \right) \in \left( H_1 \cup H_2 \right)^2}{\norm{x_i - x_j}}
\end{align*}
\end{proof}

Using the exact same derivation (while replacing the general vector $x$ by gradient $g$), we can show that the distance between aggregated gradients on two correct parameter servers, at any time $t$, is bounded.

Consider two correct servers $x$ and $z$, we can show that: \begin{align}
    &\norm{\brute{}\left( \gradvar{1}{t} \ldotexp \gradvar{n_w}{t} \right)_x - \brute{}\left( \gradvar{1}{t} \ldotexp \gradvar{n_w}{t} \right)_z} \nonumber \\
    \le~&3 \cdot \max\limits_{\left( i, j \right) \in \left( H_x \cup H_z \right)^2}{\norm{g_i - g_j}}
    \label{eqn:mda-bound-ps}
\end{align}

In plain text, this equation bounds the difference between aggregated gradients at two different correct servers, based on the maximum distance between two correct gradients, i.e.,\ aggregated gradients on different servers will not drift arbitrarily.

\subsubsection{Bounded distance between correct models}
\label{sec:bound-models}
To satisfy Equation~\ref{itm:assume-kappa} (the required assumption by \brute{}, Section~\ref{sec:mda}) and bound $r_{max}$, models at correct parameter servers should not go arbitrarily far from each other. Thus, a global \emph{gather} phase (step~\ref{itm:gather-step} in Section~\ref{sec:algo}) is executed once in a while to bring the correct models back close to each other.
We quantify the maximum number of steps that can be executed in one \emph{scatter} phase before executing one \emph{gather} phase. From another perspective, the goal is to find the maximum possible distance between correct models that still satisfies the requirement of \brute{} on the distance between correct gradients (Equation~\ref{itm:assume-kappa}).

Without loss of generality, assume two correct parameter servers $x$ and $z$ starting with the same initial model $\params{}{0}$. After the first step, their updated models are given by:

\begin{align*}
    \params{x}{1} = \params{}{0} - \eta_1 \brute\left({\gradvar{1}{1} \ldots \gradvar{n_w}{1}}\right)_x&\\
    \params{z}{1} = \params{}{0} - \eta_1 \brute\left({\gradvar{1}{1} \ldots \gradvar{n_w}{1}}\right)_z
\end{align*}
Thus, the difference between them is given by:
\begin{align*}
    &\norm{\params{x}{1} - \params{z}{1}} \\
    =~&\eta_1 \norm{\brute\left({\gradvar{1}{1} \ldots \gradvar{n_w}{1}}\right)_z - \brute\left({\gradvar{1}{1} \ldots \gradvar{n_w}{1}}\right)_x }
\end{align*}
In a perfect environment, with no Byzantine workers, this difference is zero, since the input gradients to the \brute{} function at both servers are the same (no worker lies about its gradient estimation), and the \brute{} function is deterministic (i.e.,\ the output of \brute{} computation on both servers is the same). However, a Byzantine worker can send different gradients to different servers while crafting these gradients carefully to trick the \brute{} function to include them in the aggregated gradient (i.e.,\ force \brute{} to select the malicious gradients in the set $\mathcal{S}$). In this case, $\norm{\params{x}{1} - \params{z}{1}}$ is not guaranteed to be zero. Based on Equation~\ref{eqn:mda-bound-ps}, the difference between the result of applying \brute{} in the same step is bounded and hence, such a difference can be given by:

\begin{equation}
\label{eqn:diff-models-1}
    \norm{\params{x}{1} - \params{z}{1}} \le 3 \cdot \eta_1 \cdot \max\limits_{\left( i, j \right) \in \left( H_x \cup H_z \right)^2}{\norm{\gradvar{i}{1} - \gradvar{j}{1}}}
\end{equation}
Following the same analysis, the updated models in the second step at our subject parameter servers are given by:
\begin{align*}
    \params{x}{2} = \params{x}{1} - \eta_2 \brute\left({\gradvar{1}{2} \ldots \gradvar{n_w}{2}}\right)_x&\\
    \params{z}{2} = \params{z}{1} - \eta_2 \brute\left({\gradvar{1}{2} \ldots \gradvar{n_w}{2}}\right)_z
\end{align*}
Thus, the difference between models now will be:
\begin{align*}
    \norm{\params{x}{2} - \params{z}{2}}
    &= \Big\lVert \left( \params{x}{1} - \eta_2 \brute\left({\gradvar{1}{2} \ldots \gradvar{n_w}{2}}\right)_x \right) \\
    &\phantom{=\,}- \left( \params{z}{1} - \eta_2 \brute\left({\gradvar{1}{2} \ldots \gradvar{n_w}{2}}\right)_z \right) \Big\rVert \\
    &\le \norm{\params{x}{1} - \params{z}{1}} + \eta_2 \Big\lVert \brute\left({\gradvar{1}{2} \ldots \gradvar{n_w}{2}}\right)_x \\
    &\phantom{\le \norm{\params{x}{1} - \params{z}{1}} + \eta_2\,}- \brute\left( \gradvar{1}{2} \ldots \gradvar{n_w}{2} \right)_z \Big\rVert
\end{align*}

The bound on the first term is given in Equation~\ref{eqn:diff-models-1} and that on the second term is given in Equation~\ref{eqn:mda-bound-ps} and hence, the difference between models in the second step is given by:
\begin{align}
    \norm{\params{x}{2} - \params{z}{2}}
    \le~& 3 \cdot \eta_1 \cdot \max\limits_{\left( i, j \right) \in \left( H_x \cup H_z \right)^2}{\norm{\gradvar{i}{1} - \gradvar{j}{1}}} \nonumber \\
    +~& 3 \cdot \eta_2 \cdot \max\limits_{\left( i, j \right) \in \left( H_x \cup H_z \right)^2}{\norm{\gradvar{i}{2} - \gradvar{j}{2}}}
    \label{eqn:diff-models-2}
\end{align}
By induction, we can write that the difference between models on two correct parameter servers at step $\tau$ is given by:
\begin{equation}
\label{eqn:prelim-diff-models}
    \norm{\params{x}{t} - \params{z}{t}} \le
    \sum\limits_{t=1}^{\tau} 3 \cdot \eta_t \cdot \max\limits_{\left( i, j \right) \in \left( H_x \cup H_z \right)^2}{\norm{\gradvar{i}{t} - \gradvar{j}{t}}}
\end{equation}
Since $\gradvar{i}{t}$ and $\gradvar{j}{t}$ are computed at different workers, they can be computed based on different models $\params{i}{t}$ and $\params{j}{t}$. Following Assumption~\ref{itm:Lipschitz-assumption} (the Lipschitzness of the loss function), $\norm{\gradvar{i}{t} - \gradvar{j}{t}}$ is bounded from above with $l \norm{\params{i}{t} - \params{j}{t}}$. Noting that the sequence $\eta_t$ is monotonically decreasing with $t \rightarrow \infty$ (Section~\ref{sec:not}), Equation~\ref{eqn:prelim-diff-models} can be written as:
\begin{equation*}
    \norm{\params{x}{t} - \params{z}{t}} \le
    3 \cdot \eta_1 \cdot l \sum\limits_{t=1}^{\tau} \max\limits_{\left( i, j \right) \in \left( H_x \cup H_z \right)^2}{\norm{\params{i}{t} - \params{j}{t}}}
\end{equation*}

Assuming that the maximum difference between any two correct models is bounded by $\mathcal{K}$ (this is enforced anyway by the algorithm through entering frequently the \emph{gather} phase), this difference can be written as:
\begin{equation*}
    \norm{\params{x}{t} - \params{z}{t}} \le
    3 \cdot \eta_1 \cdot l \cdot \mathcal{K} \cdot \tau
\end{equation*}

Now, to ensure the bound on the maximum difference between models, we need the value of $\norm{\params{x}{t} - \params{z}{t}} \le \mathcal{K}$. At this point, the number of steps $\tau=T$ should be bounded from above as follows:
\begin{equation}
    T \le \frac{1}{3 \cdot \eta_1 \cdot l}
\end{equation}

$T$ here represents the maximum number of steps that are allowed to happen in the \emph{scatter} phase, i.e.,\ before entering one \emph{gather} phase. Doing more steps than this number leads to breaking the requirement of \brute{} on the variance between input gradients, leading to breaking its Byzantine resilience guarantees.
Thus, this bound is a \emph{safety} bound that one should not pass to guarantee convergence.
One can do less number of steps (than $T$) during the \emph{scatter} phase for a better performance (as we discuss in Section~\ref{sec:conc}).
Moreover, this bound requires that the initial setup satisfies the assumptions of \brute{}.
Having a deployment that does not follow such assumptions leads to breaking guarantees of our protocol (as we show in Section~\ref{sec:eval}).

\subsubsection{Byzantine models filtering}
\label{sec:Lipschitz-filter}
This section shows that the filtering mechanism that is applied by workers (in the synchronous variant of \systemS{}) (step~\ref{itm:end-scatter} in Section~\ref{sec:algo}) accepts only legitimate models that are received from servers which follow the algorithm, i.e.,\ correct servers.

Such a filter is composed of two components: (1) a \lipschitzf{} filter, which bounds the growth of models with respect to gradients, and (2) a \modelsf{} filter, which bounds the distance between models in two consecutive steps. We first discuss the \lipschitzf{} filter then the \modelsf{} filter; we show that using either of them only does not guarantee Byzantine resilience.

\paragraph{\lipschitzf{} filter.}
The \lipschitzf{} filter runs on worker side, where it computes an empirical estimation for the Lipschitz coefficient of the pulled model and suspects it if its computed coefficient is far from Lipschitz coefficients of the previous correct models (from previous steps).

The \lipschitzf{} filter, by definition, accepts on average $n_{ps} - f_{ps}$ models each pulled ${n_{ps}}$ models. Such a bound makes sense given the round robin fashion of pulling models from servers (by workers) and the existence of (at most) $f_{ps}$ Byzantine servers. Based on this filter, each worker pulls, on average, $n_{ps} + f_{ps}$ each $n_{ps}$ steps. Due to the presence of $f_{ps}$ Byzantine servers, this is a tight lower bound on the communication between each worker and parameter servers to pull the updated model. The worst attack an adversary can do is to send a model that passes the filter (looks like a legitimate model, i.e.,\ very close to a legitimate model) that does not lead to computing a large enough gradient (i.e.,\ leads to minimal learning progress); in other words: an attack that drastically slows down progress.
For this reason, such a filter requires $n_{ps} > 3f_{ps}$. With this bound, the filter ensures the acceptance of at least $f_{ps} + 1$ models for each pulled $n_{ps}$ models, ensuring a majority of correct accepted models anyway and hence, ensuring the progress of learning.
Moreover, due to the randomness of choosing the value $r_j$ and the round robin fashion of pulling the models, progress is guaranteed in such a step, as correct and useful models are pulled by other workers, leading to computing correct gradients.

Based on Assumption~\ref{itm:Lipschitz-assumption} and the round robin fashion of pulling models, a Lipschitz coefficient that is computed based on a \emph{correct} model is always bounded between two Lipschitz coefficients resulting from \emph{correct} models.
Based on this, the \emph{global confinement} property is satisfied (based on the properties of models that passes the \lipschitzf{} filter) and hence,
we can plug \systemS{} in the confinement proof in~\cite{bottou1998online}.
Precisely, given Assumptions~\ref{itm:assume-bounded-moments} and~\ref{itm:Lipschitz-assumption}, and applying~\cite{bottou1998online} (of which these assumptions are prerequisite), the models accepted by the \lipschitzf{} filter satisfy the following property:

\paragraphspace
Let r=2,3,4, $\exists A'_r \ge 0$  and $B'_r \ge 0$ such that $\forall t \ge 0 \mathsep \expect\norm{\gradvar{}{t}} \le A'_r + B'_r \norm{\params{}{t}}^r$, where $\params{}{t}$ is the model accepted by the \lipschitzf{} filter at some worker and $\gradvar{}{t}$ is the gradient calculated based on such a model.

\paragraph{\modelsf{} filter.}
The \lipschitzf{} filter is necessary, yet not sufficient to accept models from a parameter server. A Byzantine server can craft some model that is arbitrarily far from the correct model, whose gradient satisfies the \lipschitzf{} filter condition. Thus, it is extremely important to make sure that the received model is close to the expected, correct model. Such a condition is verified by the second filter at the worker side, which we call the \emph{\modelsf{} filter}. Such a filter is based on the results given in Section~\ref{sec:bound-models} and the local model estimation on the workers side.

Without loss of generality,
consider a correct worker $j$ that pulls models from parameter servers $ \params{i}{t} \forall i \in \range{1}{n_{ps}}$ in a round robin fashion. In each step $t > 1$, the worker computes a local estimation of the next (to be received) model $\params{j(l)}{t}$ based on the latest model it has $\params{j}{t-1}$ and its own gradient estimation $\gradvar{j}{t-1}$. The local model estimation is done as follows:
\begin{equation*}
    \params{j(l)}{t} = \params{j}{t-1} - \eta_{t-1}\gradvar{j}{t-1}.
\end{equation*}
Without loss of generality, assume that worker $j$ pulls the model from some server $i$ in step $t$. If such a server is correct, it computes the new model $\params{i}{t}$ as follows:
\begin{equation*}
    \params{i}{t} = \params{i}{t-1} - \eta_{t-1}\brute{}\left(\gradvar{1}{t-1} \ldots \gradvar{n_w}{t-1}\right).
\end{equation*}
Thus, the difference between the local model estimation at worker $j$ and the received model from server $i$ (if it is correct) is given by:
\begin{align*}
    \norm{\params{j(l)}{t} - \params{i}{t}}
    =~& \Big\lVert \left(\params{j}{t-1} - \eta_{t-1}\gradvar{j}{t-1}\right) \\
    &- \left( \params{i}{t-1} - \eta_{t-1}\brute{}\left(\gradvar{1}{t-1} \ldots \gradvar{n_w}{t-1}\right) \right) \Big\rVert \\
    \le~& \norm{\params{j}{t-1} - \params{i}{t-1}} \\
    &+ \eta_{t-1} \norm{\brute{}\left(\gradvar{1}{t-1} \ldots \gradvar{n_w}{t-1}\right) - \gradvar{j}{t-1}}
\end{align*}

Based on the guarantees given by \brute{}~\cite{bulyanPaper}, the following bound holds:
\begin{equation*}
    \norm{\brute{}\left(\gradvar{1}{t} \ldots \gradvar{n_w}{t}\right) - \gradvar{j}{t}} \le \frac{n_w - f_w}{2 f_w} \norm{\gradvar{j}{t}}.
\end{equation*}

Based on the results of Section~\ref{sec:bound-models}, the maximum distance between two correct models just after a \emph{gather} phase is:
\begin{equation*}
    \frac{3}{4}\frac{n_w - f_w}{f_w} \eta_{(T \cdot (t\text{ mod }T))} \norm{\gradvar{}{(T \cdot (t\text{ mod }T))}}T.
\end{equation*}
By induction, we can find the bound on the value $\norm{\params{j(l)}{t-1} - \params{i}{t-1}}$ and hence, we can write:
\begin{align}
    \norm{\params{j(l)}{t} - \params{i}{t}}
    \le~& \eta_{(T \cdot (t\text{ mod }T))} \norm{\gradvar{}{(T \cdot (t\text{ mod }T))}} \nonumber \\
    \times~&\left( 2 \bigg( (t\text{ mod }T) - 1\bigg) + \frac{(n_w - f_w)(3T + 2)}{4f_w} \right)
    \label{eqn:app-model-bound}
\end{align}
Thus, a received model $\params{i}{t}$ that satisfies Equation~\ref{eqn:app-model-bound} is considered passing the \modelsf{} filter. Such a model is guaranteed to be in the correct cone of models in one \emph{scatter} phase.
Note that such a filter cannot be used alone without the \lipschitzf{} filter.
A Byzantine server can craft a model that satisfies such a filter (i.e.,\ the \modelsf{} filter) while being on the opposite direction of minimizing the loss function.
Such a model then will be caught by the \lipschitzf{} filter.

Combining the \lipschitzf{} filter and the \modelsf{} filter guarantees that the received model at the worker side is close to the correct one (at the current specific \emph{scatter} phase), representing a reasonable growth, compared to the latest local model at the worker.

\subsection{Convergence of \systemA{}: The Asynchronous Case}

In this section, under the assumptions laid in Section \ref{sec:app-assumptions}, we formally prove \eqref{eqn:app-convergence}.

\subsubsection{\median{} contraction effect}
\label{sec:contract-effect}

Let $\left( d, f, n, q \right) \in \left( \setn - \left\lbrace 0 \right\rbrace \right)^2 \times \setn^2$ such that $2 \, f + 2 \le q \le \floor{\frac{n - f}{2}}$, and note $h \triangleq n - f$.

Denote $S_j \triangleq \left\lbrace s \subset \range{1}{h} - \left\lbrace j \right\rbrace \suchthat \card{s} \in \range{q - f - 1}{q-1} \right\rbrace$ the subset of correct parameter server indexes that parameter server $j$ can deliver at a given step.
We then call $S = \prod_{j \in \range{1}{h}} S_j$ the set of all correct parameter server indexes the $h$ correct parameter servers $i \in \range{1}{h}$ can deliver.
Our key assumption is that, for any configuration occurs with positive probability.

Formally, consider $\mathcal S$ the probability distribution over delivering configurations, and $X \sim \mathcal S$ a random set that follows the probability distribution $\mathcal S$. We assume that
\begin{equation}\label{eqn:condense-proof-assumption}
    \forall s \in S \mathsep P(X=s) > 0.
\end{equation}
We denote $\rho = \min_{s \in S} P(X=s)$ the probability of least likely configuration. Since the set $S$ of configurations is finite, our key assumption equivalently says that $\rho >0$.

\paragraphspace
Denote $\locals{j}{t} \in \mathbb R^d$ the parameter vector held by parameter server $j$ at step $t$, $\indexvar{X_j}{t}{\overline{\Theta}} \triangleq \left\lbrace \locals{k}{t} \suchthat k \in X_j \right\rbrace$ the parameter vectors from correct parameter servers that were delivered by parameter server $j$ at step $t$, and $z_t^{(j)} = (z_t^{(1, j)}, \ldots, z_1^{(q-X_j, j)})$ the Byzantine parameter vectors\footnote{A Byzantine parameter server can send different parameter vectors to each correct parameter server at each step $t$, hence the notation.} that were delivered by parameter server $j$ at step $t$, where each $z_k^{(j)} \in \mathbb R^d$.

The updated vector of parameter server $j$ at step $t$ is then given by
\begin{equation}
    \params{j}{t} \triangleq \median{}\left( \locals{j}{t}, \indexvar{X_j}{t}{\overline{\Theta}}, z_t^{(j)} \right).
\end{equation}
In this section, we prove that the updated parameters $\params{1}{t} \ldotexp \params{h}{t}$ have been contracted compared to parameters $\locals{1}{t} \ldotexp \locals{h}{t}$. It turns out that this claim does not actually hold for naive measures of how ``contracted'' a set of parameters are. In particular, the diameter of the vectors $\params{1}{t} \ldotexp \params{h}{t}$'s measured in, say, $\ell_2$ norm, does not necessarily get contracted.

The key trick of our analysis is to focus instead on coordinate-wise diameters.
More precisely, denote:
\begin{equation*}
    \Delta_{\locals{}{t}, i} \triangleq \max\limits_{j,k \in \range{1}{h}} \absv{\locals{j}{t}\indexed{i} - \locals{j}{t}[k]}
\end{equation*}
the diameter of vectors $\locals{1}{t} \ldots \locals{h}{t}$ along coordinate $i$.
We can then add up these coordinate-wise diameters to obtain the measure $\Delta_{\locals{}{t}} = \sum_{i \in \range{1}{d}} \Delta_{\locals{}{t}, i}$ which is still a measure of how spread the vectors $\locals{j}{t}$'s are. The key result of this section is that this measure decreases exponentially, i.e.
\begin{equation}
    \expect[\Delta_{\params{}{t}}] \leq m \Delta_{\locals{}{t}},
\end{equation}
for some contraction parameter $m \in [0,1[$, and where the expectation is taken over $X \sim \mathcal S$. More precisely, in expectation over random delivering configurations, no matter how Byzantines attack the different workers, the spread of the parameter servers parameters decrease exponentially.

The following lemma shows that bounding this measure is sufficient to control the diameter of servers' parameters.

\begin{lemma}
We have the following inequalities:
\begin{align*}
    \max_{j,k \in \range{1}{h}} \normtwo{\locals{j}{t}-\locals{k}{t}} &\leq \max_{j,k \in \range{1}{h}} \normone{\locals{j}{t}-\locals{k}{t}} \leq \Delta_{\locals{}{t}} \\
    &\leq d \max_{j,k \in \range{1}{h}} \normone{\locals{j}{t}-\locals{k}{t}} \\
    &\leq d^{3/2} \max_{j,k \in \range{1}{h}} \normtwo{\locals{j}{t}-\locals{k}{t}}.
\end{align*}
\end{lemma}

\begin{proof}
Let $j^*$ and $k^*$ two servers whose parameters are most distant, i.e.\ such that
\begin{equation*}
    \max_{j,k \in \range{1}{h}} \normone{\locals{j}{t}-\locals{k}{t}} = \sum_{i=1}^d \absv{\locals{j^*}{t}\indexed{i} - \locals{k^*}{t}\indexed{i}}.
\end{equation*}
Clearly the latter term is upper-bounded by
\begin{equation*}
    \sum_{i=1}^d \max_{j,k \in \range{1}{h}} \absv{\locals{j}{t}\indexed{i}-\locals{k}{t}\indexed{i}},
\end{equation*}
which is exactly the sum $\Delta_x$ of coordinate-wise diameters.
The first inequality is derived from the well-known equality $\normtwo{z} \leq \normone{z}$.

Note that a coordinate-wise distance is smaller than the $\ell_1$ distance.
Therefore,
\begin{align*}
    \Delta_{\locals{}{t}}
    &= \sum_{i=1}^d \max_{j,k} \absv{\locals{j}{t}\indexed{i} - \locals{k}{t}\indexed{i}} \\
    &\leq \sum_{i=1}^d \max_{j,k} \normone{\locals{j}{t} - \locals{k}{t}} \\
    &= d \cdot \max_{j,k} \normone{\locals{j}{t} - \locals{k}{t}}.
\end{align*}
We then conclude by using the inequality $\normone{z} \leq \sqrt{d} \normtwo{z}$.
\end{proof}

To prove this result, we start with an easy lemma.

\begin{lemma}
With probability 1, we have $\Delta_{\params{}{t}, i} \leq \Delta_{\locals{}{t}, i}$.
In other words, there can be no dilation of the diameter along a coordinate.
\end{lemma}

\begin{proof}
Let $x_{min}\indexed{i}$ and $x_{max}\indexed{i}$ the minimum and maximum $i$-coordinate of correct vectors $\locals{1}{t} \ldotexp \locals{h}{t}$'s.
Since $q \geq 2f+2 > 2f$, we know that a strict majority of vectors delivered to $j$ are correct, and thus belong to $\left[x_{min}\indexed{i}, x_{max}\indexed{i} \right]$.
Therefore, $\params{j}{t}\indexed{i}$ must belong to this interval too.
Since this holds for any parameter server $j$, we know that $\Delta_{\params{}{t}, i} \leq x_{max}\indexed{i}- x_{min}\indexed{i} = \Delta_{\locals{}{t}, i}$.
\end{proof}

Now along each coordinate, we separate the set of workers along two subsets $left\indexed{i}$ and $right\indexed{i}$, defined by
\begin{align*}
    left\indexed{i}& =  \left\lbrace j \in \range{1}{h} \suchthat \locals{j}{t}\indexed{i} \leq \frac{1}{2} \left( x_{min}\indexed{i} + x_{max}\indexed{i} \right) \right\rbrace, \\
    right\indexed{i} &= \range{1}{h} - left\indexed{i}
\end{align*}
We then go on showing that the minority subset has a positive probability of making a significant step towards the other side. First, we quantify the size of this large possible step.

\begin{lemma}
If $X_j \subset right\indexed{i}$, then
\begin{equation*}
    \params{j}{t}\indexed{i} \geq \frac{1}{4} \left( 3 \, x_{min}\indexed{i} + x_{max}\indexed{i} \right).
\end{equation*}
Moreover, if $X_j \subset left\indexed{i}$, then $\params{j}{t}\indexed{i} \leq \frac{1}{4} \left( x_{min}\indexed{i} + 3 \, x_{max}\indexed{i} \right).$
\end{lemma}

\begin{proof}
Assume $X_j \subset right\indexed{i}$. Since $q \geq 2f+2$, we know that that $\card{j \cup X_j} \geq q-f \geq q/2+1$ contains a strict majority of the inputs to compute $\params{j}{t}\indexed{i}$. Thus the median is at least the average of two smallest coordinates $i$ of points $\card{j \cup X_j}$, which is at least
\begin{align*}
    \params{j}{t}\indexed{i}
    &\geq \frac{\locals{j}{t}\indexed{i} + \frac{1}{2} \left( x_{min}\indexed{i} + x_{max}\indexed{i} \right)}{2} \\
    &\geq \frac{1}{4} \left( 3 \, x_{min}\indexed{i} + x_{max}\indexed{i} \right)
\end{align*}
The second bound is proved similarly.
\end{proof}

\begin{lemma}
We can then show expected strict contraction along every coordinate $i$, i.e.
\begin{equation*}
\expect\left[ \Delta_{\params{}{t}, i} \right] \leq \left(1-\frac{\rho}{4} \right) \Delta_{\locals{}{t}, i}.
\end{equation*}
\label{lemma:contraction}
\end{lemma}

\begin{proof}
Note that $right\indexed{i}+left\indexed{i} = h$. Thus at least one of the two subsets has at most $\lfloor h/2 \rfloor$ elements. Since $q \leq \lfloor \frac{n-f}{2} \rfloor = \lfloor h/2 \rfloor$, we know that $q-f \leq \lfloor h/2 \rfloor$. This means that there is a subset $s_0$ of $left\indexed{i}$ or of $right\indexed{i}$ with at least $q-f$ elements.

Without loss of generality, assume $s_0 \subset left\indexed{i}$ with cardinal $q-f$, and let $s_j = s_0-\{j\}$. Then the tuple $s = (s_1, \ldots, s_h) \in S$ is a delivering configuration. By virtue of our key assumption (Equation \ref{eqn:condense-proof-assumption}), we know that $P(X=s) \geq \rho > 0$.

But then, by the previous lemma, in the event $X=s$, we have $\params{j}{t}\indexed{i} \geq \frac{1}{4} \left( 3 \, x_{min}\indexed{i} + x_{max}\indexed{i} \right)$ for all servers $j$. Since, moreover, we can easily verify that $\params{j}{t}\indexed{i} \leq x_{max}\indexed{i}$, we conclude that
\begin{align*}
    \Delta_{\params{}{t}, i}
    &\leq x_{max}\indexed{i} - \frac{1}{4} \left( 3 \, x_{min}\indexed{i} + x_{max}\indexed{i} \right) \\
    &= \frac{3}{4} \left( x_{max}\indexed{i} - x_{min}\indexed{i} \right) \\
    &= \frac{3 \Delta_{\locals{}{t}, i}}{4}.
\end{align*}
The same bound can be derived for the case $s_0 \subset right\indexed{i}$.

Now note that in the event $X \neq s$, by Lemma 1, we still have $\Delta_{\params{}{t}, i} \leq \Delta_{\locals{}{t}, i}$. We can now take the average of the two above cases, which yields
\begin{align*}
    \expect[\Delta_{\params{}{t}, i}]
    =~& \expect\left[\Delta_{\params{}{t}, i} \suchthat X=s \right] P(X=s) \\
    &+ \expect\left[\Delta_{\params{}{t}, i} \suchthat X\neq s\right] (1-P(X = s)) \\
    =~& \expect\left[\Delta_{\params{}{t}, i} \suchthat X\neq s\right] \\
    &+ \left(\expect\left[\Delta_{\params{}{t}, i} \suchthat X=s \right] - \expect\left[\Delta_{\params{}{t}, i} \suchthat X\neq s\right] \right) P(X=s) \\
    \leq~& \Delta_{\locals{}{t}, i} + \left( \frac{3 \Delta_{\locals{}{t}, i}}{4} - \Delta_{\locals{}{t}, i} \right) \rho \\
    =~& \left( 1 - \frac{\rho}{4} \right) \Delta_{\locals{}{t}, i},
\end{align*}
which concludes the proof.
\end{proof}

\begin{lemma}
Despite any Byzantine attack $z \left( X \right)$ that reacts to the random choice of delivering configuration $X \sim \mathcal S$, there is a strict contraction of the sum of coordinate-wise diameters, i.e.
\begin{equation*}
    \forall z \mathsep \expect_{X \sim S} \left[ \Delta_{\params{}{t}} \right] \leq m \Delta_{\locals{}{t}},
\end{equation*}
where $m<1$ only depends on the probability distribution $\mathcal S$.
\end{lemma}

\begin{proof}
We simply use the linearity of the expectation, which yields
\begin{equation*}
    \expect[\Delta_y] = \sum_{i=1}^d \expect[\Delta_{\params{}{t}, i}] \leq \sum_{i=1}^d \left(1-\frac{\rho}{4}\right) \Delta_{\locals{}{t}, i} = \left(1-\frac{\rho}{4}\right) \Delta_{\locals{}{t}},
\end{equation*}
which is the key lemma of this section.
\end{proof}

\subsection{Expected contraction of the correct parameter vectors}
\label{sec:proof-partA}

Consider parameters $\theta_t$ at time $t$. Notice that this parameter will undergo three operations, each of which will be attacked by Byzantines.

First, each worker $j$ will be delivered a subset of parameters. It will need to compute the median of the parameters, to obtain a model $\params{j}{t}$.
It will then compute an estimate $\gradvar{j}{t}$ of the gradient $\realgrad{\params{j}{t}}$, which will then be broadcast to parameter servers. We make the following critical observations.

\begin{lemma}
Assuming $q_w \geq 2 f_w +1$, we have $\Delta_{x_t} \leq \Delta_{\theta_t}$ with probability 1.
\end{lemma}

\begin{proof}
Since there is a majority of workers, for each coordinate $i \in \range{1}{d}$, we have $\params{j}{t}\indexed{i} \in \left[ \theta_{t,min}\indexed{i}, \theta_{t,max}\indexed{i} \right]$. Therefore,
\begin{equation*}
    \absv{\params{j}{t}\indexed{i} - \params{k}{t}\indexed{i}} \leq \theta_{t,max}\indexed{i} - \theta_{t,min}\indexed{i} = \Delta_{\theta_t}\indexed{i}.
\end{equation*}
Taking the maximum over pairs $(j,k)$, and then adding over $i$ yields the lemma.
\end{proof}

\begin{lemma}
For any two workers $j$ and $k$, the delivered gradients satisfy
\begin{equation*}
    \expect \max_{j,k \in \range{1}{h_w}} \normtwo{\gradvar{j}{t} - \gradvar{k}{t}} \leq 2 h_w \sigma' + l \Delta_{\theta_t},
\end{equation*}
where the expectation is over the random delivering configuration of parameters from servers to workers and over the random estimation of the gradients.
\end{lemma}

\begin{proof}
Note that
\begin{align*}
    \normtwo{\gradvar{j}{t} - \gradvar{k}{t}}
    \leq~& \normtwo{\gradvar{j}{t} - \realgrad{\params{j}{t}}} \\
    &+ \normtwo{\realgrad{\params{j}{t}} - \realgrad{\params{k}{t}}} \\
    &+ \normtwo{\realgrad{\params{k}{t}} - \gradvar{k}{t}}.
\end{align*}
As a result, we know that
\begin{align}
    \max_{j,k \in \range{1}{h_w}} \normtwo{\gradvar{j}{t} - \gradvar{k}{t}} \hspace{-2cm}& \nonumber \\
    \leq~& 2 \max_{j \in \range{1}{h_w}} \normtwo{\gradvar{j}{t} - \realgrad{\params{j}{t}}} \nonumber \\
    &+ \max_{j,k \in \range{1}{h_w}} \normtwo{\realgrad{\params{j}{t}} - \realgrad{\params{k}{t}}}.
    \label{eqn:stochastic_gradient_estimator}
\end{align}
We now use the fact that
\begin{align*}
    \expect \max_{j \in \range{1}{h_w}} \normtwo{\gradvar{j}{t} - \realgrad{\params{j}{t}}} \\
    &\leq \expect \sum_{j=1}^{h_w} \normtwo{\gradvar{j}{t} - \realgrad{\params{j}{t}}} \\
    &= \sum_{j=1}^{h_w} \expect\normtwo{\gradvar{j}{t} - \realgrad{\params{j}{t}}}
\end{align*}
Using our assumption of unbiased gradient estimator with uniformly bounded error $\sigma'$, we see that
\begin{equation*}
    \expect \max_{j \in \range{1}{h_w}} \normtwo{\gradvar{j}{t} - \realgrad{\params{j}{t}}} \leq h_w \sigma'.
\end{equation*}
To bound the second term of Equation \ref{eqn:stochastic_gradient_estimator}, we now use the fact that $\normtwo{\params{j}{t}-\params{k}{t}} \leq \Delta_{x_t} \leq \Delta_{\theta_t}$. Since $\nabla L$ is $l$-Lipschitz, we thus have
\begin{equation*}
    \normtwo{\realgrad{\params{j}{t}} - \realgrad{\params{k}{t}}} \leq l \Delta_{\theta_t}.
\end{equation*}
This equation holds in particular for the maximum of the left-hand side, as we vary $j$ and $k$. This concludes the proof of the lemma.
\end{proof}

We now move on to the guarantee with respective to the second attack of Byzantines, which occurs as servers aggregate workers' estimations of the gradient. This guarantee is provided by the Byzantine resilience of $\brute{}$.

\begin{lemma}
The diameter of the aggregations of gradients by $\brute{}$ is at most three times the diameter of correct gradients:
\begin{equation*}
    \max_{j,k \in \range{1}{h_{ps}}} \normtwo{\gradaggr{j}{t} - \gradaggr{k}{t}} \leq 3 \max_{r,s \in \range{1}{h_w}} \normtwo{\gradvar{r}{t} - \gradvar{s}{t}}.
\end{equation*}
\end{lemma}

\begin{proof}
Let us first focus on the computation of $\gradaggr{j}{t}$. Recall that it is obtained by gathering $q_w$ vectors $\gradvar{r}{t}$ from workers, including at most $f_w$ Byzantines, and computing the $\brute{}$ of the the collected vectors. Recall that $\brute{}$ averages the subset of vectors of minimal diameter. Let $\mathcal X^*$ a subset of gradients of size $q_w-f_w$ that minimizes the diameter of the gradients.

Note that the diameter of delivered correct gradients is necessarily at most the diameter of all correct gradients, which we shall denote $D(g_t) = \max_{j,k \in \range{1}{h_w}} \normtwo{\gradvar{j}{t} - \gradvar{k}{t}}$. Since $q_w \geq 2f_w +1$, there is a subset of size $q_w-f_w$ that only contains correct gradients, and whose diameter is thus at most $D(g_t)$.

Therefore the subset $\mathcal X^*$ must have diameter at most $D(g_t)$. But since $q_w \geq 2f_w +1$, we know that at least one correct gradient $\gradvar{r^*(j)}{t}$ belongs to this subset. This means that all gradients collected by $\mathcal X^*$ must be at distance at most $D(g_t)$ from $\gradvar{r^*(j)}{t}$. In other words, for any $\gradaggr{j}{t}$, there exists a correct gradient $\gradvar{r^*(j)}{t}$ such that $\norm{\gradaggr{j}{t} - \gradvar{r^*(j)}{t}} \leq D(g_t)$.

But then, we have
\begin{align*}
    \normtwo{\gradaggr{j}{t} - \gradaggr{k}{t}}
    \leq~& \normtwo{\gradaggr{j}{t} - \gradvar{r^*(j)}{t}} \\
    &+ \normtwo{\gradvar{r^*(j)}{t} - \gradvar{r^*(k)}{t}} \\
    &+ \normtwo{\gradvar{r^*(k)}{t} - \gradaggr{k}{t}},
\end{align*}
which is at most $3 D\left( g_t \right)$.
\end{proof}

Recall that each server $j$'s parameters are now updated by adding $- \eta_t \gradaggr{j}{t}$ to $\params{j}{t}$. We can bound the drift that this update causes as follows.

\begin{lemma}
We have the following inequality:
\begin{equation*}
    \expect\left[\Delta_{\theta_t - \eta_t G_t}\right] \leq 6 d \eta_t h_w \sigma' + (1+ 3 d \eta_t l) \Delta_{\theta_t},
\end{equation*}
where the average is taken over delivering configurations and stochastic gradient estimates.
\label{lemma:scather}
\end{lemma}

\begin{proof}
Note that, on each coordinate $i$, for any two servers $j$ and $k$, we have
\begin{align*}
    &\absv{\left( \params{j}{t}\indexed{i} - \eta_t \gradaggr{j}{t}\indexed{i}\right) - \left( \params{k}{t}\indexed{i} - \eta_t \gradaggr{k}{t}\indexed{i}\right)} \\
    \leq~&\absv{\params{j}{t}\indexed{i} - \params{k}{t}\indexed{i}} + \eta_t \absv{\gradaggr{j}{t}\indexed{i} - \gradaggr{k}{t}\indexed{i}} \\
    \leq~&\absv{\params{j}{t}\indexed{i} - \params{k}{t}\indexed{i}} + \eta_t \normtwo{ \gradaggr{j}{t} - \gradaggr{k}{t}} \\
    \leq~&\Delta_{\theta_t}\indexed{i} + \eta_t \max_{j',k' \in \range{1}{h_{ps}}} \normtwo{\gradaggr{j'}{t} - \gradaggr{k'}{t}}.
\end{align*}
Note that the right-hand side is now independent from $j$ and $k$, and is thus unchanged as we take the maximum over all $j$ and $k$'s. Summing over all coordinates $i$ yields
\begin{equation*}
    \Delta_{\theta_t - \eta_t G_t} \leq \Delta_{\theta_t} + d \eta_t \max_{j',k' \in \range{1}{h_{ps}}} \normtwo{\gradaggr{j'}{t} - \gradaggr{k'}{t}}.
\end{equation*}
We now take the average, and invoke the two previous lemmas to derive the result.
\end{proof}

Finally, we can combine the result with the contraction property of the \median{}.

\begin{lemma}
We have the following inequality:
\begin{equation*}
    \expect \left[ \Delta_{\theta_{t+1}} \right] \leq \left( 1+ 3 d \eta_t l - \frac{\rho}{4} \right) \Delta_{\theta_t} + 6 d \eta_t h_w \sigma'.
\end{equation*}
\end{lemma}

\begin{proof}
This is an immediate application of lemmas \ref{lemma:contraction} and \ref{lemma:scather}, and using $(1+a)(1-b) \leq 1+a-b$ for $a,b \geq 0$.
\end{proof}

We are now almost there. We will need this elementary lemma to conclude.

\begin{lemma}
Let $k \in \left[ 0, 1 \right[$ and $\delta_t > 0$ be a positive decreasing sequence such that $\delta_t \rightarrow 0$.
Then there exists constants $C>0$, which depend on $k$ and $\delta_0$, such that
\begin{equation}
    \sum\limits_{i=0}^{t}{k^{t - i}} \, \eta_i \leq C k^{t/2} + \frac{ \delta_{\lfloor t/2 \rfloor}}{1-k}.
    \label{eqn:converge-sum-factor}
\end{equation}
In particular, the left-hand side converges to zero.
\end{lemma}

\begin{proof}
We divide the sum between elements before than $s= \lfloor t/2 \rfloor$ and elements after this threshold. This yields:
\begin{align*}
    \sum_{i=0}^t k^{t-i} \delta_i &= \sum_{i=0}^{s-1} k^{t-i} \delta_i + \sum_{i=1+s}^t k^{t-i} \delta_i \\
    &\leq k^{1+t-s} \delta_0 \sum_{j=0}^{s-1} k^j + \delta_s \sum_{i=0}^{t-s} k^j \\
    &\leq \frac{\delta_0 k^{1+t-s}}{1-k} + \frac{\delta_s}{1-k}.
\end{align*}
Since $1+t-s \leq 1+t-(t/2-1) = t/2$, defining $C = \delta_0 / (1-k)$ allows to conclude.
\end{proof}

Finally, we can derive the gathering of \systemA{}.

\begin{lemma}[Gathering of \systemA{}]
If learning rates go to zero ($\eta_t \rightarrow 0$), then servers converge to the same state, and their diameter is of the order of the learning rate, i.e. there exists a constant $C>0$ such that
\begin{equation*}
    \expect \max_{j,k \in \range{1}{h_{ps}}} \normtwo{\params{j}{t}-\params{k}{t}} \leq C \left( 1-\frac{\rho}{8} \right)^{t/2} + \frac{24 dl \eta_{\lfloor t/2\rfloor}}{\rho}.
\end{equation*}
\end{lemma}

\begin{proof}
Assume $0 \leq u_{t+1} \leq (1+C\delta_t-2\varepsilon) u_t + \delta_t$, where $\varepsilon = \rho/8$, $\delta_t = 3 d\eta_t l$ and $A = \frac{l}{2 h_w \sigma'}$. Since $\delta_t \rightarrow 0$, we know that there is a time $t_0$ such that for $t \geq t_0$, we have $A \delta_t \leq \varepsilon$. For $t \geq t_0$, we then have $u_{t+1} \leq k u_t +\delta_t$, where $k<1$ and $\delta_t \rightarrow 0$. But then, we observe that
\begin{equation*}
    u_{t+t_0} \leq k^t u_{t_0} + \sum_{i=1}^t k^{i-1} \delta_{t_0+t-i}.
\end{equation*}
Using Lemma (\ref{eqn:converge-sum-factor}) (and replacing $k$ by $\sqrt{k}$), we conclude that $u_t \leq C k^t + \frac{\delta_{\lfloor t/2\rfloor}}{1-k}$. Applying this to $u_t = \expect[\Delta_{\theta_t}]$ implies the bound of the lemma.

We can finally conclude by noting that
\begin{equation*}
    \max \limits_{j,k \in \range{1}{h_{ps}}} \normtwo{\params{j}{t}-\params{k}{t}} \leq \Delta_{\theta_t}.
\end{equation*}
\end{proof}

In particular, assuming, say $\eta_t = 1/t^\alpha$, the expected diameter is of the order $\mathcal{O}\!\left( \frac{d l}{\rho{} t^\alpha} \right)$.

\subsubsection{Expected convergence of the loss function's gradient}
\label{sec:proof-partB}

Let us now use the results of the previous lemma to show that, despite gathering, the trajectory of parameters $\params{j}{t}$ is nearly a stochastic gradient descent. The trick to do so will be to determine that the actual update of the parameters after contraction $\hat G_t^{(j)} \triangleq \frac{\params{j}{t+1} - \params{j}{t}}{\eta_t}$ satisfies the conditions of convergence of~\cite{krum}. In particular, we need to prove that
\begin{equation*}
    \exists \alpha > 0 \mathsep \hat \expect{}G_t^{(j)} \cdot \nabla L\left(\params{j}{t}\right) \geq \alpha.
\end{equation*}
To prove this, we first show that for $t$ large enough, all servers receive roughly the same gradients, using the previous lemmas.

\begin{lemma}
For any two workers, we have
\begin{equation*}
    \expect{\max_{j,k \in \range{1}{h_{ps}}} \normtwo{\gradaggr{j}{t} - \realgrad{\params{k}{t}}}} \leq 3 h_w \sigma' + 3 l \Delta_{\params{}{t}}.
\end{equation*}
\end{lemma}

\begin{proof}
This results from combining the previous lemmas.
\end{proof}

\begin{lemma}
The diameters of the servers' updates after applying gradients is upper-bounded as follows:
\begin{align*}
    &\max_{j,k \in \range{1}{h_{ps}}} \normtwo{\params{j}{t} - \eta_t \gradaggr{j}{t} - \left( \params{k}{t} - \eta_t \gradaggr{k}{t} \right)} \\
    \leq~& 6 \eta_t h_w \sigma' + (3l +1) \Delta_{\params{}{t}}.
\end{align*}
\label{lemma:liveness}
\end{lemma}

\begin{proof}
This is derived from the triangle inequality and from previous lemmas.
\end{proof}

\begin{lemma}
Assuming $\eta_t = 1/t^\alpha$, after applying gradients and after contraction, the distance between the overall motion of parameters of server $j$ and the true gradient is upper-bounded as follows:
\begin{equation*}
    \expect{} \hat G^{(j)}_t \cdot \realgrad{\params{j}{t}} \geq \normtwo{\realgrad{\params{j}{t}}}^2 - \left( 9h_w \sigma' + \frac{C dl^2}{\rho} \right) \normtwo{\realgrad{\params{j}{t}}},
\end{equation*}
where C is a constant.
\end{lemma}

\begin{proof}
Note that $\params{j}{t+1}$ is obtained by taking \median{} over updated parameters $\params{k}{t} - \eta_t \gradaggr{k}{t}$. By the guarantee of \median{}, we know that the $\ell_2$ diameter of the (attacked) outputs is at most three times the diameter of the input. This implies that
\begin{equation*}
    \normtwo{\params{j}{t+1} - \left( \params{j}{t} - \eta_t \gradaggr{j}{t} \right)} \leq \Delta_{\params{}{t} - \eta_t \gradaggr{}{t}}.
\end{equation*}
From this, and using the previous lemmas, we derive the fact that
\begin{align*}
    &\expect{} \left( \params{j}{t+1} - \params{j}{t} \right) \cdot \realgrad{\params{j}{t}} \\
    \geq~& \eta_t \gradaggr{j}{t} \cdot \realgrad{\params{j}{t}} - \Delta_{\theta_t - \eta G_t} \normtwo{\realgrad{\params{j}{t}}} \\
    \geq~& \eta_t \normtwo{\realgrad{\params{j}{t}}}^2 - \left( 9 h_w \sigma' \eta_t + (6 l + 1) \Delta_{\params{}{t}} \right) \normtwo{\realgrad{\params{j}{t}}}.
\end{align*}
Dividing by $\eta_t$ and factoring in the bound on $\Delta_{\params{}{t}}$, we obtain the lemma.
\end{proof}

More precisely, we have the following, let $(a,b)$ be any two correct parameter servers. Using a triangle inequality, the second inequality above and Assumption \ref{itm:assume-bounded-moments} (bounded moments) we have:
\begin{align*}
&   \forall t> t_\varepsilon \forall r \in \range{2}{4} \mathsep
    \exists \left( A_r, B_r \right) \in \setr^2 \mathsep \\
&   \forall \left( i, t, \params{}{} \right) \in \range{1}{n_w - f_w} \times \setn \times \setr^d \mathsep
    \expect\norminf{\gradvar{a}{t}} \le A'_r + B'_r \norminf{\params{b}{t}}^r
\end{align*}
where $A'_r$ and $B'_r$ are positive constants, depending polynomially (at most with degree $r$, by using the second inequality above, the triangle inequality and a binomial expansion) on $\varepsilon$, $\sigma$ and $(A_r,B_r)$ from Assumption \ref{itm:assume-bounded-moments}, allowing us to use the convergence proof of~\cite{krum} (Proposition~2) regardless of the identity of the (correct) parameter server.

\paragraph{Optimal resilience against $\frac{1}{3}$ of Byzantine nodes.}
The rationale stems from the optimal \emph{breakdown point}
of $\frac{1}{2}$ for any synchronous Byzantine--resilient aggregation scheme (e.g.\ \brute{} and \median{}), identified in \cite{rousseeuw1985multivariate}.
Network asynchrony makes it impossible to differentiate a slow correct node from a non--responding Byzantine node~\cite{fischer1985impossibility}.
This observation implies the over--provisioning of (at least) $\frac{1}{2}$ correct nodes for $\frac{1}{2}$ Byzantine ones and hence, the final optimal breakdown point of $\left(\frac{1}{2}\text{ Byzantine}~\big/~\left( 1 + \frac{1}{2} \right)\right) = \frac{1}{3}$.

\subsection{Convergence of \systemS{}: The Synchronous Case}
\label{sec:proof_sync}

Naively, if we enter the \emph{gather} phase in all learning steps, the algorithm
becomes straightforward, and the proof could be inherited directly and only from the GARs guarantees.
On the other extreme, if the \emph{gather} phase is never executed (i.e.,\ only the \emph{scatter} phase is executed), the models at different servers will become arbitrarily far form each other (Section~\ref{sec:bound-models}) and hence, the assumptions of the Byzantine-resilient GAR, i.e.,\ \brute{} are violated, leading to (possibly) divergence.

Based on this, the gist of the proof is to show that the maximum distance between models at correct servers is always small enough to satisfy the assumptions of \brute{}. Based on this and the fact that workers do SGD steps during the \emph{scatter} phase, it is trivial to show that there is a progress in learning between two \emph{gather} phases, i.e.,\ the loss function is minimized and hence, the proof is reduced to the standard proof of SGD convergence~\cite{bottou1998online}.

Formally, we follow the same proof scheme as we show:
\begin{equation*}
    \left.\begin{array}{r}
    \lim\limits_{t \rightarrow \infty}{\expect\left( \max\limits_{\left( x, z \right) \in \range{1}{n_{ps} - f_{ps}}}\norm{\params{x}{t} - \params{z}{t}} \right)} = 0 \\
    \lim\limits_{t \rightarrow \infty}{\expect\norm{\realgrad{\params{x}{t}}}} = 0
    \end{array}\right.
\end{equation*}

\paragraphspace
Proving these two equalities implies necessarily the convergence of the learning procedure:
\begin{equation*}
    \lim\limits_{t \rightarrow \infty}{\expect\norm{\realgrad{\params{}{t}}}} = 0
\end{equation*}

\begin{proof}
From Equation~\ref{eqn:app-model-bound}, we know that the difference between correct models is bounded by $\norm{\gradvar{}{T \cdot (t\text{ mod }T)}}$. Based on our assumptions (Section~\ref{sec:app-assumptions}) and the guarantees given by SGD, it holds:
\begin{equation*}
    \lim\limits_{t \rightarrow \infty} \norm{\gradvar{}{T \cdot (t\text{ mod }T)}}~=~0.
\end{equation*}
Therefore:
\hfill
$\lim\limits_{t \rightarrow \infty} \left( \max\limits_{\left( x, z \right) \in \range{1}{n_{ps} - f_{ps}}}\norm{\params{x}{t} - \params{z}{t}} \right) = 0$.
\end{proof}

A direct consequence of Section~\ref{sec:bound-models} is that beyond a certain time $t_\epsilon$ (with $\varepsilon > 0$ be any positive real number), the standard deviation of the gradient estimators as well as the \emph{drift} between (correct) parameter vectors can be bounded arbitrarily close to each other.
More formally, the following holds (and is a direct consequence of the limits stated above):
\begin{equation*}
    \exists t_\varepsilon \geq 0 \mathsep
    \forall t \geq t_\varepsilon \mathsep
    \left\lbrace\begin{array}{l}
        \expect\left( \max\limits_{i \in \range{1}{n_w - f_w}}\norm{\gradvar{i}{t} - \expect\gradvar{1}{t}} \right) \le \sigma' + \varepsilon \\
        \expect\left( \max\limits_{\left( x, z \right) \in \range{1}{n_{ps} - f_{ps}}^2}\norm{\params{x}{t} - \params{z}{t}} \right) < \frac{\varepsilon}{l}
    \end{array}\right.
\end{equation*}

The first inequality provides the bounded variance guarantee needed to plug \systemS{} into the convergence proof of~\cite{krum}, and the second inequality provides the remaining requirement, i.e. the bound on the statistical higher moments of the gradient estimator as if there was a \emph{single} parameter. More precisely, we have the following, let $(x,z)$ be any two correct parameter servers, using a triangle inequality, the second inequality above, and Assumption~\ref{itm:assume-bounded-moments} (bounded moments) we have (as given in Section~\ref{sec:Lipschitz-filter}):
\begin{align*}
&   \forall t> t_\varepsilon \forall r \in \range{2}{4} \mathsep
    \exists \left( A_r, B_r \right) \in \setr^2 \mathsep \\
&   \forall \left( i, t, \params{}{} \right) \in \range{1}{n_w - f_w} \times \setn \times \setr^d \mathsep
    \expect\norm{\gradvar{x}{t}} \le A'_r + B'_r \norm{\params{z}{t}}^r
\end{align*}
where $A'_r$ and $B'_r$ are positive constants, depending polynomially (at most with degree $r$, by using the second inequality above, the triangle inequality and a binomial expansion) on $\varepsilon$, $\sigma$ and $(A_r,B_r)$ from assumption \ref{itm:assume-bounded-moments}, allowing us to use the convergence proof of~\cite{krum} (Proposition~2) regardless of the identity of the (correct) parameter server.

\section{Validating the bounded gradients variance to norm ratio assumption}
\label{sec:batch_variance}
To make progress at every step, any state--of--the--art Byzantine--resilient gradient aggregation rule (GAR), that is based \emph{solely} on statistical robustness, requires a bound on the ratio $\frac{variance}{norm}$ of the correct gradient estimations.
Intuitively, not having such a bound would allow the correct gradients to become \emph{indistinguishable} from some random noise.
This is problematic, since these Byzantine--resilient GARs \cite{krum,su2017defending,bulyanPaper,xie2018phocas} rely on techniques analogous to \emph{voting} (i.e. median--like techniques in high--dimension): if the correct majority does not agree (appears ``random''), then the Byzantine minority controls the aggregated gradient.
For example, not satisfying these bounds makes the used GARs vulnerable against recently--proposed attacks like \emph{A little is enough} attack~\cite{baruch2019little}, which we experimented in Section~\ref{sec:eval}.
Such a bound is to ensure that, no matter the received Byzantine gradients, the expected value of the aggregated gradient does lie in the same half--space as the real gradient, leading for every step taken to more \emph{optimal} parameters (smaller loss).

Here, we try to understand when this assumed bound on the variance to norm ratio (e.g.,\ Equation~\ref{itm:assume-kappa}) holds, and when it does not.
The most straightforward way to fulfill such an assumption is to increase the batch size used for training.
The question is then what the minimum batch size (that can be used while satisfying such a bound) is, and whether it is small enough for the distribution of the training to still make sense.

\begin{figure*}[!ht]
\centering
\subfigure[$f=1$]{\includegraphics[width=0.45\linewidth,keepaspectratio]{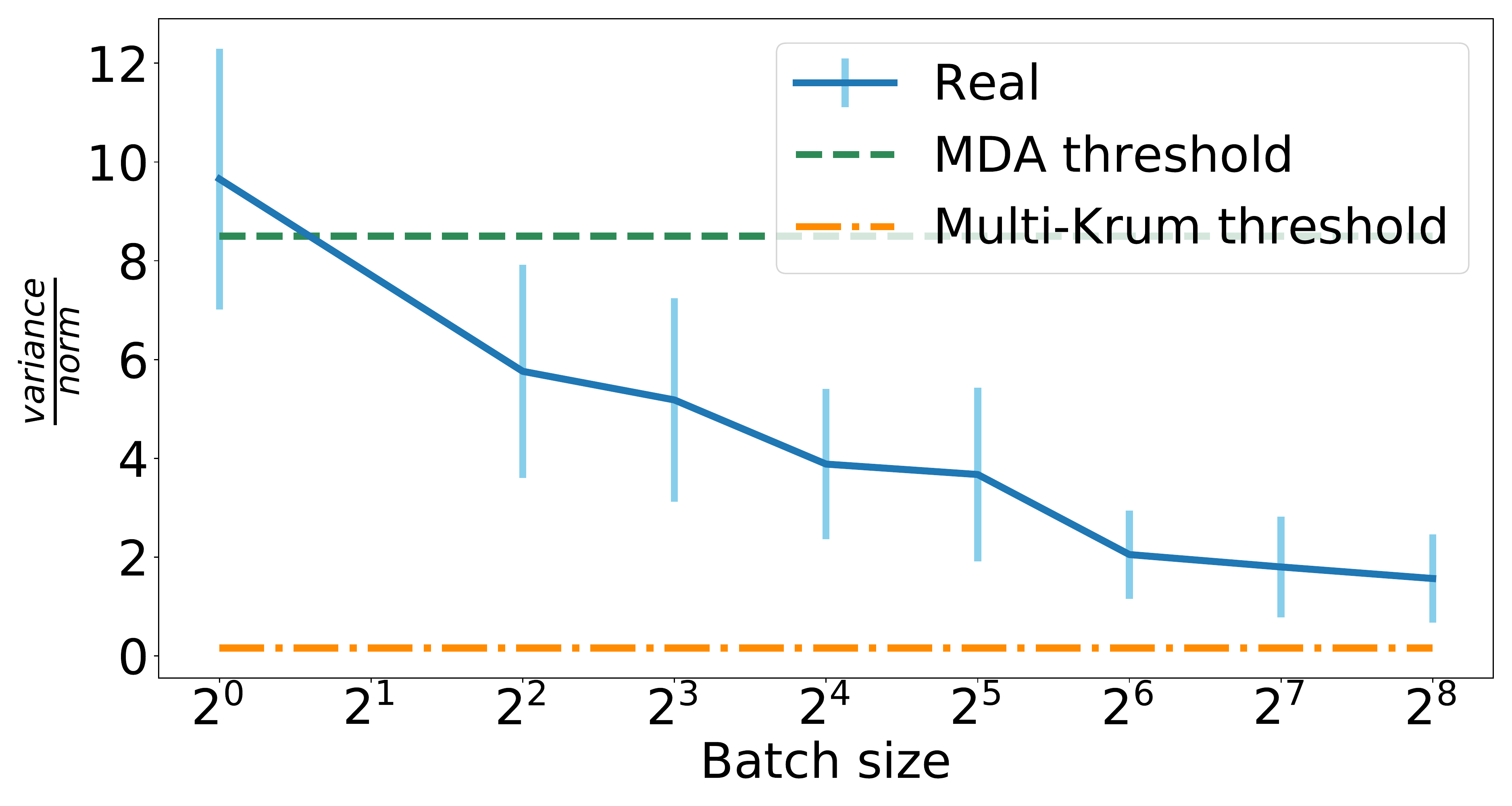}
\label{subfig:var_f1}}
\subfigure[$f=5$]{\includegraphics[width=0.45\linewidth,keepaspectratio]{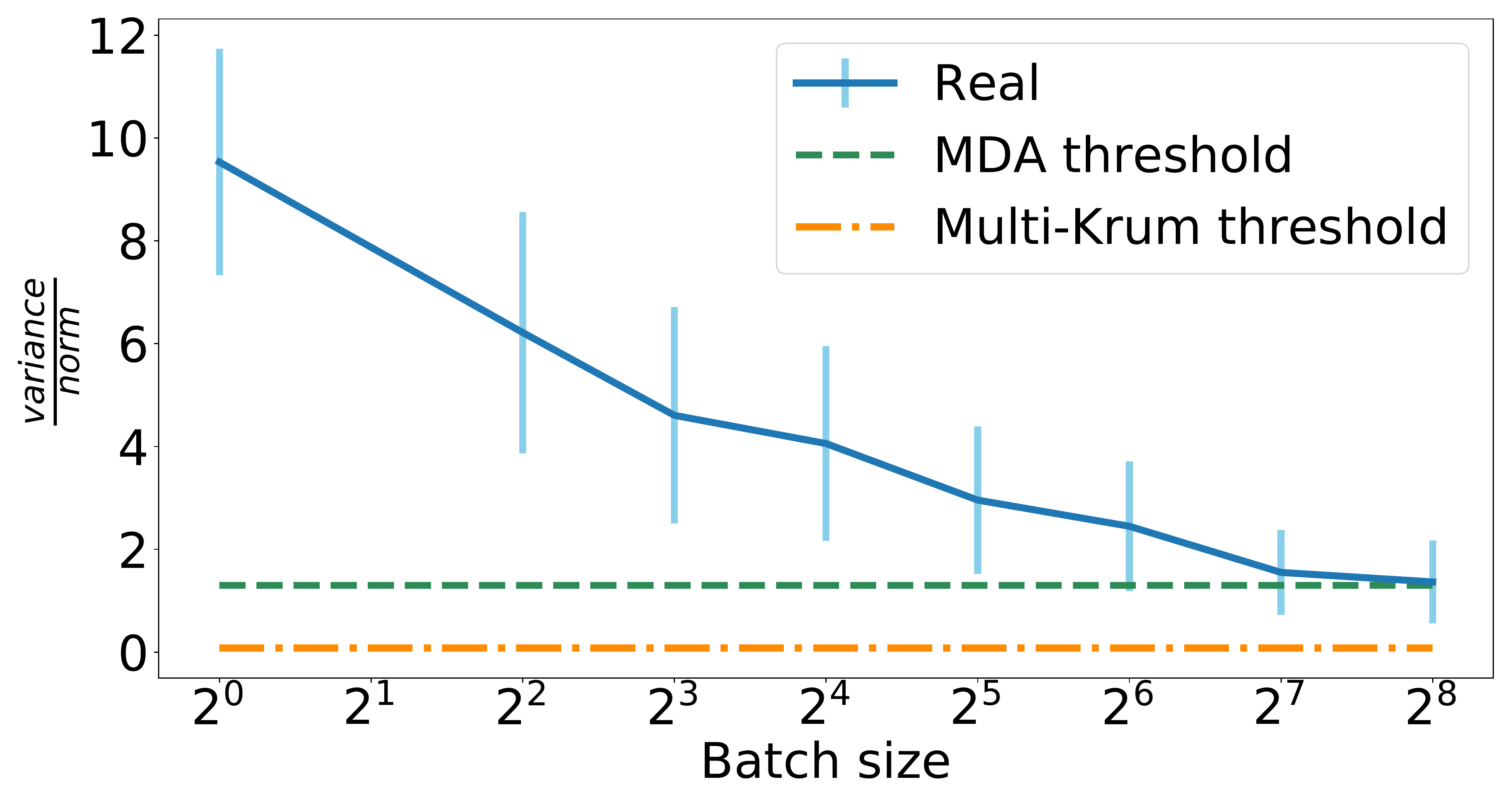}
\label{subfig:var_f5}}
\caption{Variance to norm ratio with different batch sizes, compared to the bound/threshold required by two Byzantine-resilient GARs: \brute{} and \mkrum{}. To satisfy a GAR assumption/condition, the \emph{real} $\frac{variance}{norm}$ value should be lower than the GAR bound. For instance, \brute{} can be used with batch-size=$32$ with $f=1$, but not with $f=5$ (as the \emph{real} $\frac{variance}{norm}$ value is higher than the \brute{} threshold).}
\label{fig:variance}
\end{figure*}

\paragraph{Methodology.}
We use the same setup and hyperparameters as used in our evaluation of \systemS{} (Section~7).
We estimate over the first 100 steps of training the variance to norm ratio of correct gradient estimations for several batch sizes.
We plot the average (line) and standard deviation (error bar) of these ratios over these 100 steps (Figure \ref{fig:variance}).
We show the bound required by two Byzantine-resilient GAR: \brute{}, and \mkrum{}~\cite{krum}.
We find \mkrum{} a very good example on a widely--used GAR that unfortunately does not seem to provide any practical\footnote{At least on our academic model and dataset.} guarantee, due to its unsatisfied assumption\footnote{It needs very low variance to norm ratio of correct gradient estimations, e.g. $0.08$ for $\left( n, f \right) = \left( 18, 5 \right)$.}.
We also experimented with two values of the number of declared Byzantine workers: $f=1,5$.
Increasing the value of $f$ calls for a tighter bound on the variance to norm ratio.

\paragraph{Results.} Figure~\ref{fig:variance} depicts the relation between the variance (of gradients) to norm ratio with the batch size. According to such a figure, \mkrum{} cannot be safely used even with the largest experimented batch size, i.e.,\ 256. Otherwise, the variance bound assumption such a GAR builds on is not satisfied and hence, an adversary can break its resilience guarantees~\cite{baruch2019little}.
\brute{} gives a better bound on the variance, which makes it more practical in this sense: typical batch size of 128 for example can be safely used with $f=1$.
However, \brute{} is not safe to use with $f=5$ even with the largest experimented batch size ($b=256$).
This is confirmed in Section~\ref{sec:eval}, where we show that an adversary can use such a vulnerability (due to the unsatisfied assumption) to reduce the learning accuracy.
Having the optimal bound on variance while guaranteeing Byzantine resilience and convergence remains an open question.

\section{Experimenting \systemS{} with More Scenarios}
\label{sec:add_res}
We provide here few additional results for evaluating \systemS{}, especially the efficiency of the filtering mechanism used in the synchronous setup.
First, we show the progress of accuracy over training steps and time (i.e.,\ convergence) with different values for \emph{declared} Byzantine servers.
Then, we describe the effect of changing the value of $T$ on the filters' performance and convergence.

\begin{figure*}[!ht]
\centering
\subfigure[]{\includegraphics[width=0.45\linewidth,keepaspectratio]{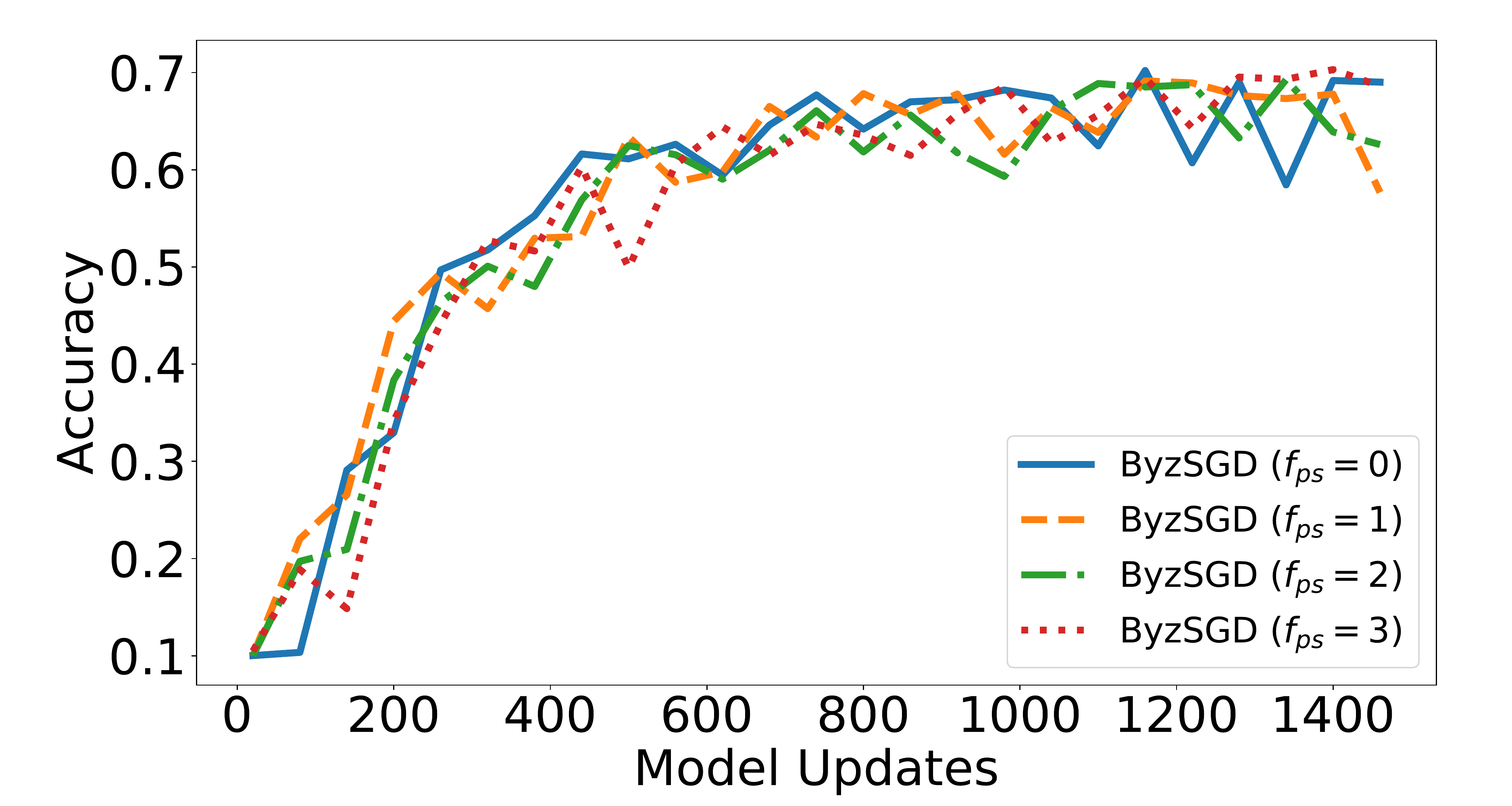}
\label{subfig:mult_byzps_step}}
\subfigure[]{\includegraphics[width=0.45\linewidth,keepaspectratio]{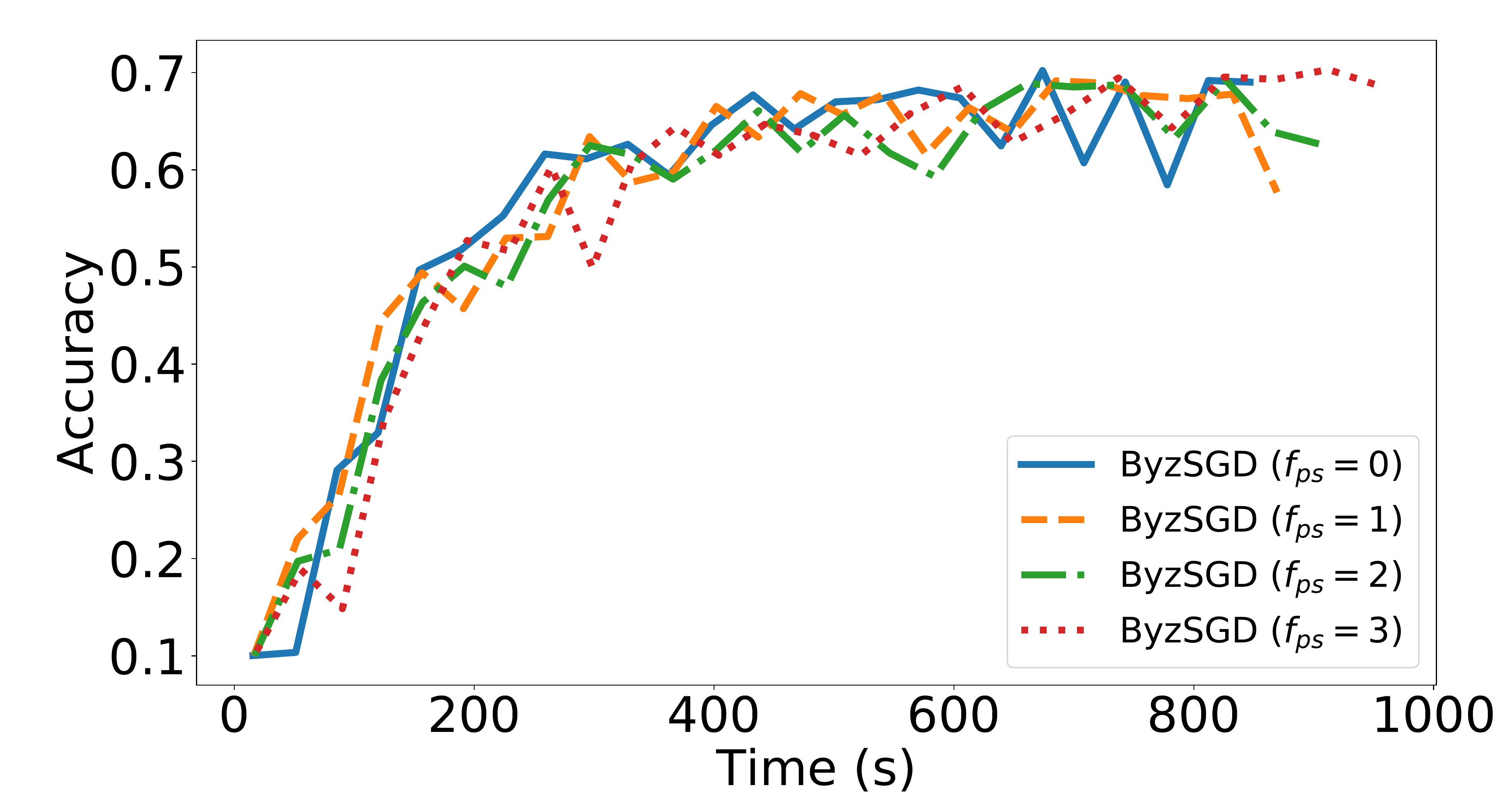}
\label{subfig:mult_byzps_time}}
\caption{The effect of changing $f_{ps}$ on performance.}
\label{fig:mult_byzps}
\end{figure*}

\subsection{Convergence with Multiple Byzantine Servers}

Figure~\ref{fig:mult_byzps} shows the convergence with different numbers of \emph{declared} Byzantine servers.
To allow for (up to) 3 Byzantine servers, we use a total of 10 servers in this experiment.
Such figures show that changing the number of Byzantine servers does not affect the the number of steps required for convergence (Figure~\ref{subfig:mult_byzps_step}).
Yet, deployments with a higher value for $f_{ps}$ require slightly more time to converge.
Note that in this experiment we have not employed real attacks nor changed the total number of machines (for both servers and workers) used.

\begin{figure*}[!ht]
\centering
\subfigure[]{\includegraphics[width=0.45\linewidth,keepaspectratio]{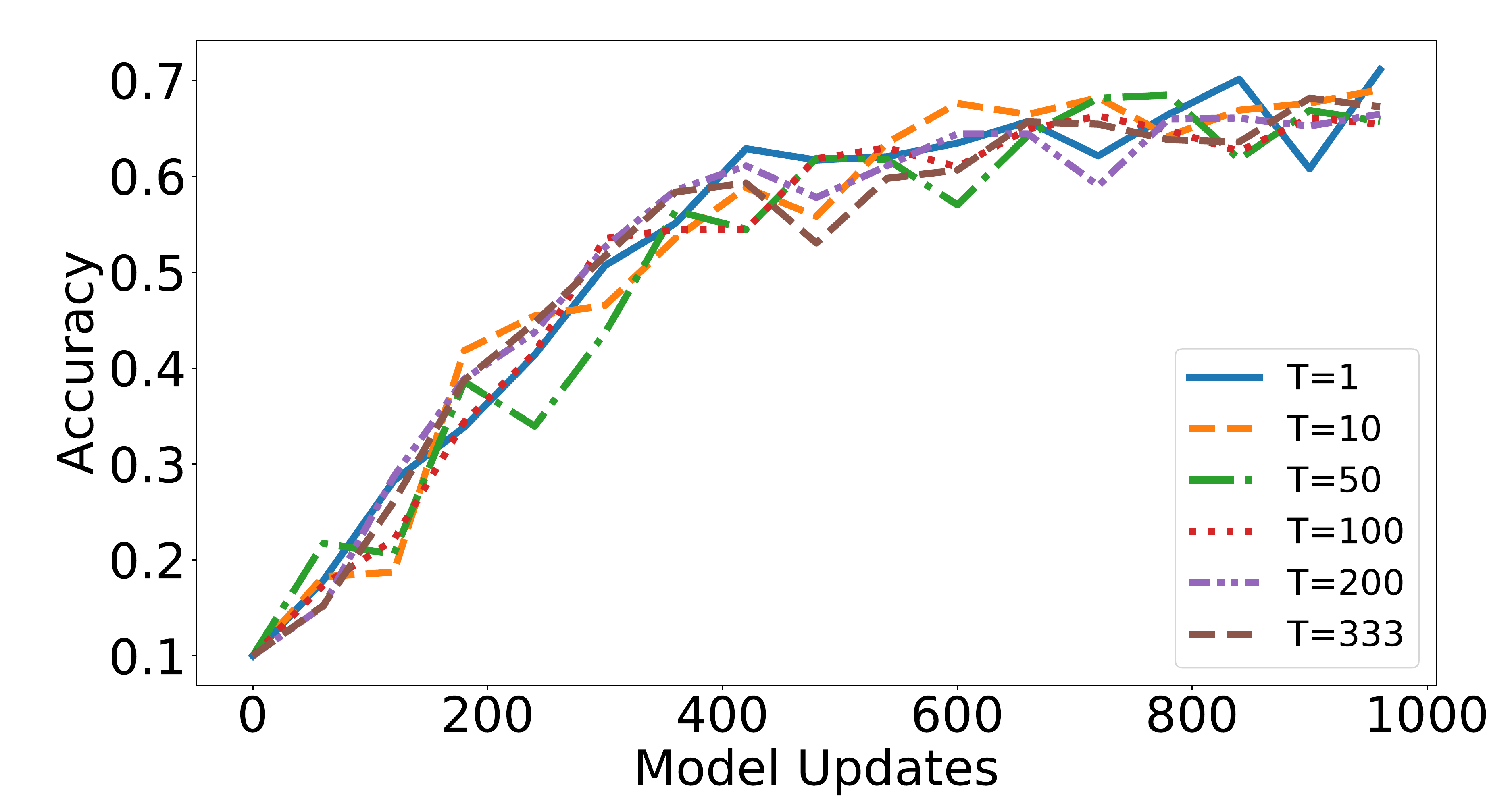}
\label{subfig:t_step}}
\subfigure[]{\includegraphics[width=0.45\linewidth,keepaspectratio]{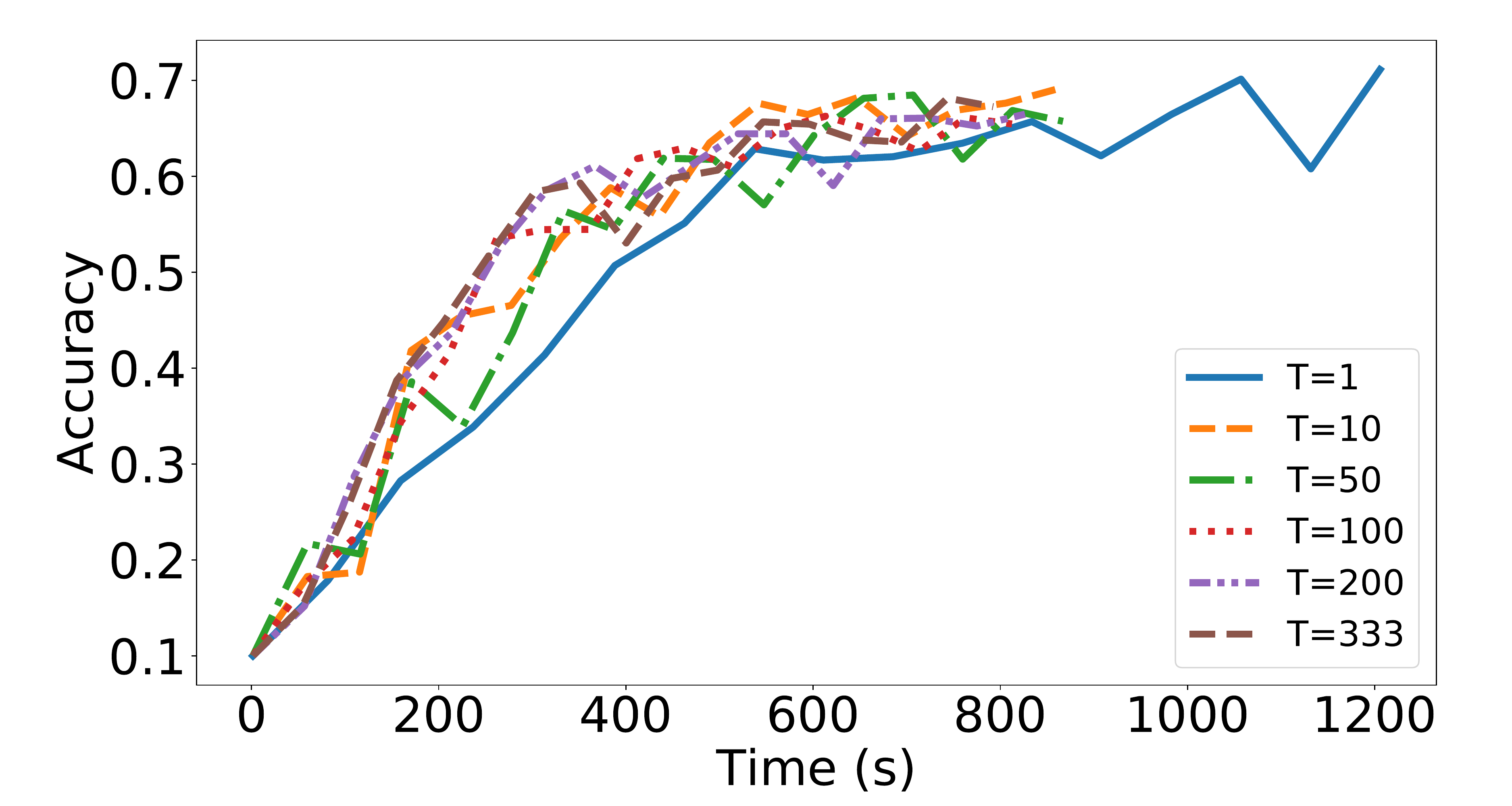}
\label{subfig:t_time}}
\\
No Attack
\\
\subfigure[]{\includegraphics[width=0.45\linewidth,keepaspectratio]{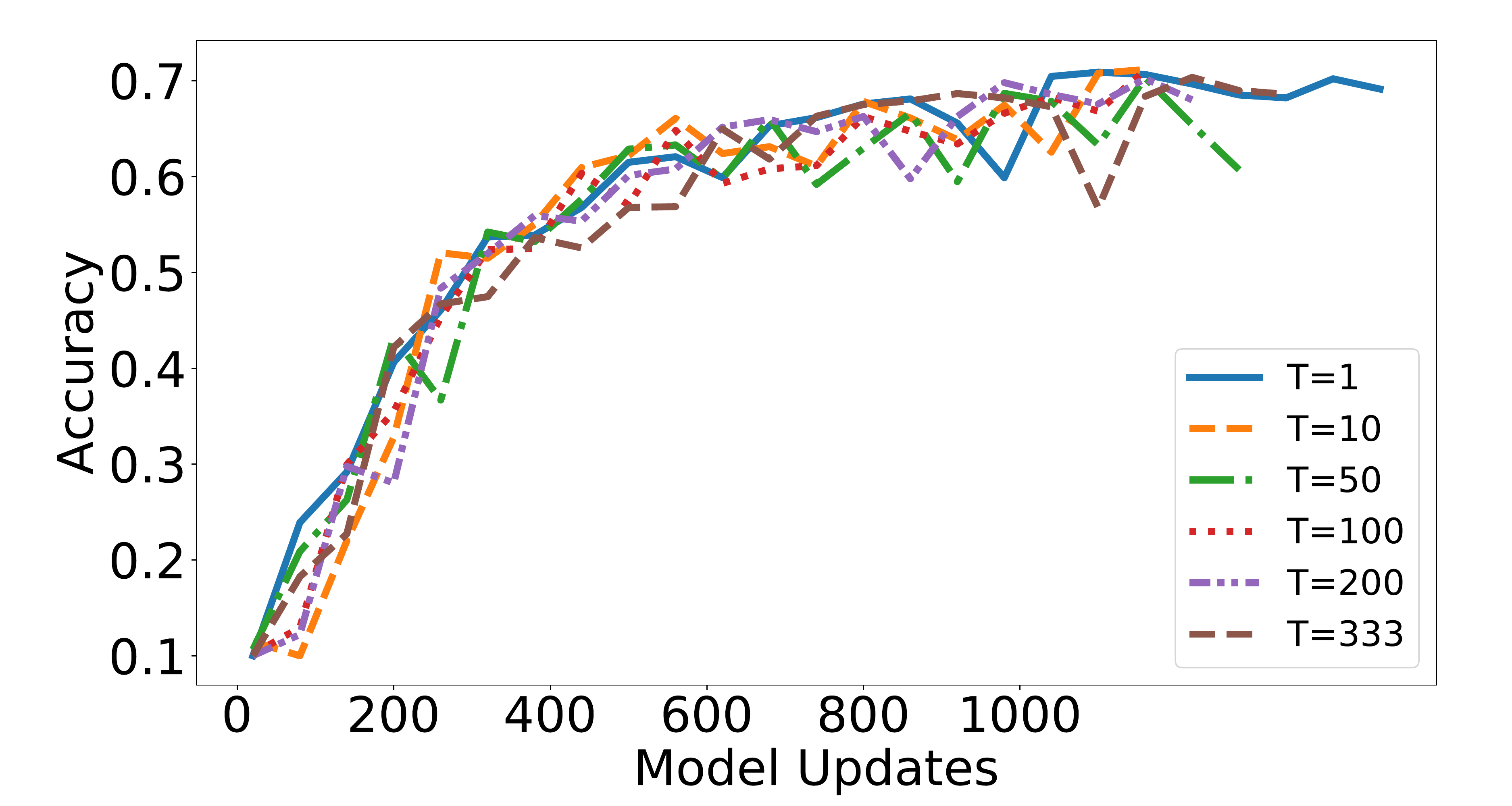}
\label{subfig:rev_t_step}}
\subfigure[]{\includegraphics[width=0.45\linewidth,keepaspectratio]{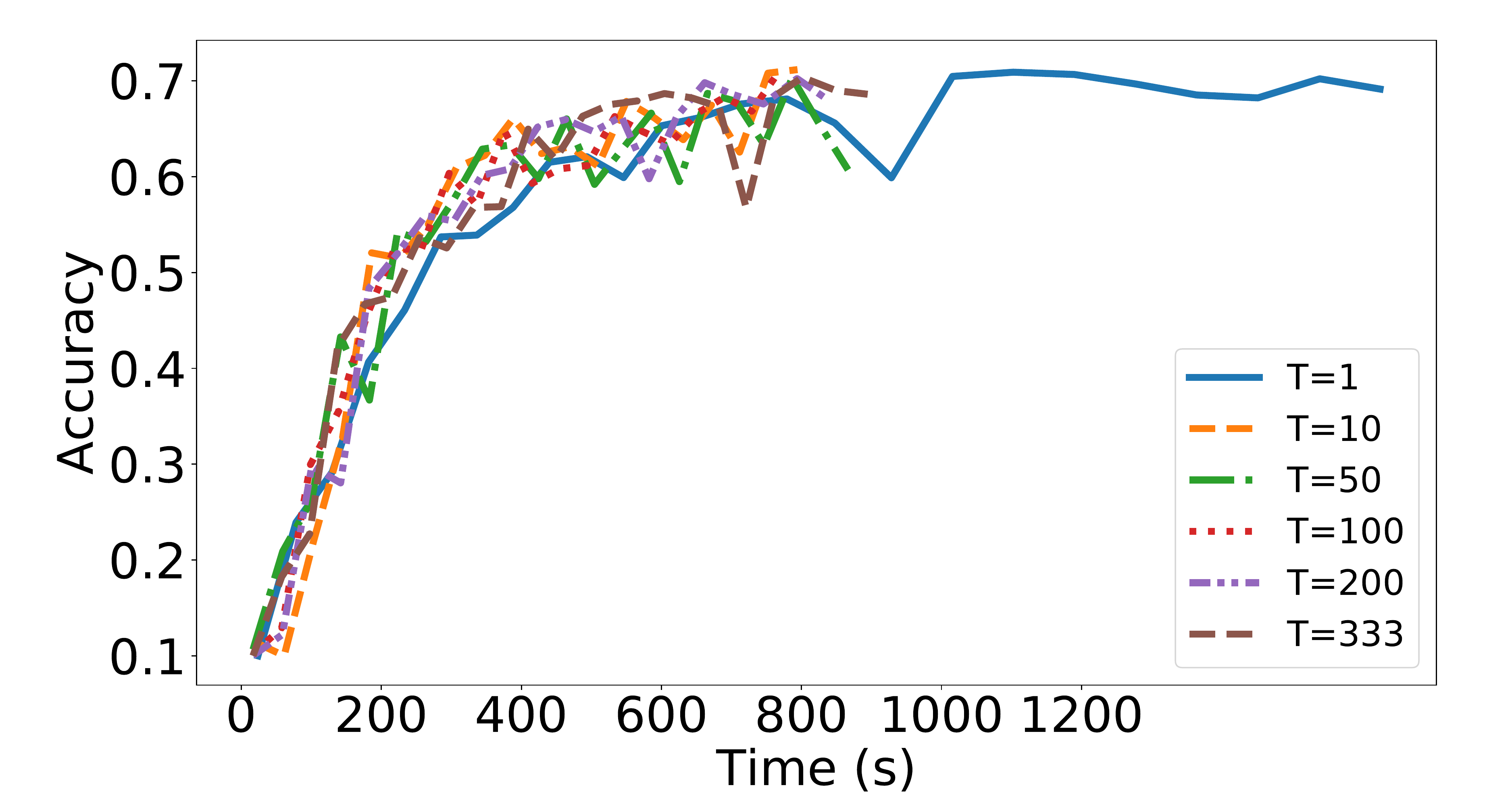}
\label{subfig:rev_t_time}}
\\
Reversed Attack
\caption{The effect of changing $T$ on performance.}
\label{fig:changing_t}
\end{figure*}

\subsection{Effect of Changing T}
\label{subsec:changing_T}
The value of $T$ denotes the number of steps done in one \emph{scatter} phase (i.e.,\ before entering one \emph{gather} phase).
Figure~\ref{fig:changing_t} shows the effect of changing the value of $T$ on convergence with both time and model updates in both a Byzantine and a Byzantine-free environments.
Figure~\ref{subfig:t_step} shows that the value of $T$ almost does not have any effect on the convergence w.r.t. the model updates.
This happens because models on correct servers almost do not drift from each other (as all the servers are correct).
Interestingly, Figure~\ref{subfig:t_time} shows that using a higher value for $T$ helps converge faster.
This is because increasing $T$ decreases the communication overhead, achieving faster updates and higher throughput.
However, it is important to note that as the value of $T$ increases, the expected drifts between models on correct servers increases, and it becomes easier for the Byzantine server to trick the workers.
Figures~\ref{subfig:rev_t_step} and~\ref{subfig:rev_t_time} shows the convergence with different values for $T$ under the \emph{Reversed} attack (in which the Byzantine machine reverses the direction of the correct vector as described in the main paper, Section~7).
Though setting $T=1$ slows down the convergence, this case shows the most stable convergence behavior (especially towards the end when approaching a local minimum; see Figure~\ref{subfig:rev_t_step}).
Yet, higher values for $T$ lead to increased noise and oscillations towards the end of convergence.
Notably, testing with $T \le 333$ is safe in this setup and that is why convergence is reached in all cases.

\begin{figure*}[!ht]
\centering
\subfigure[No Attack]{\includegraphics[width=0.3\linewidth,keepaspectratio]{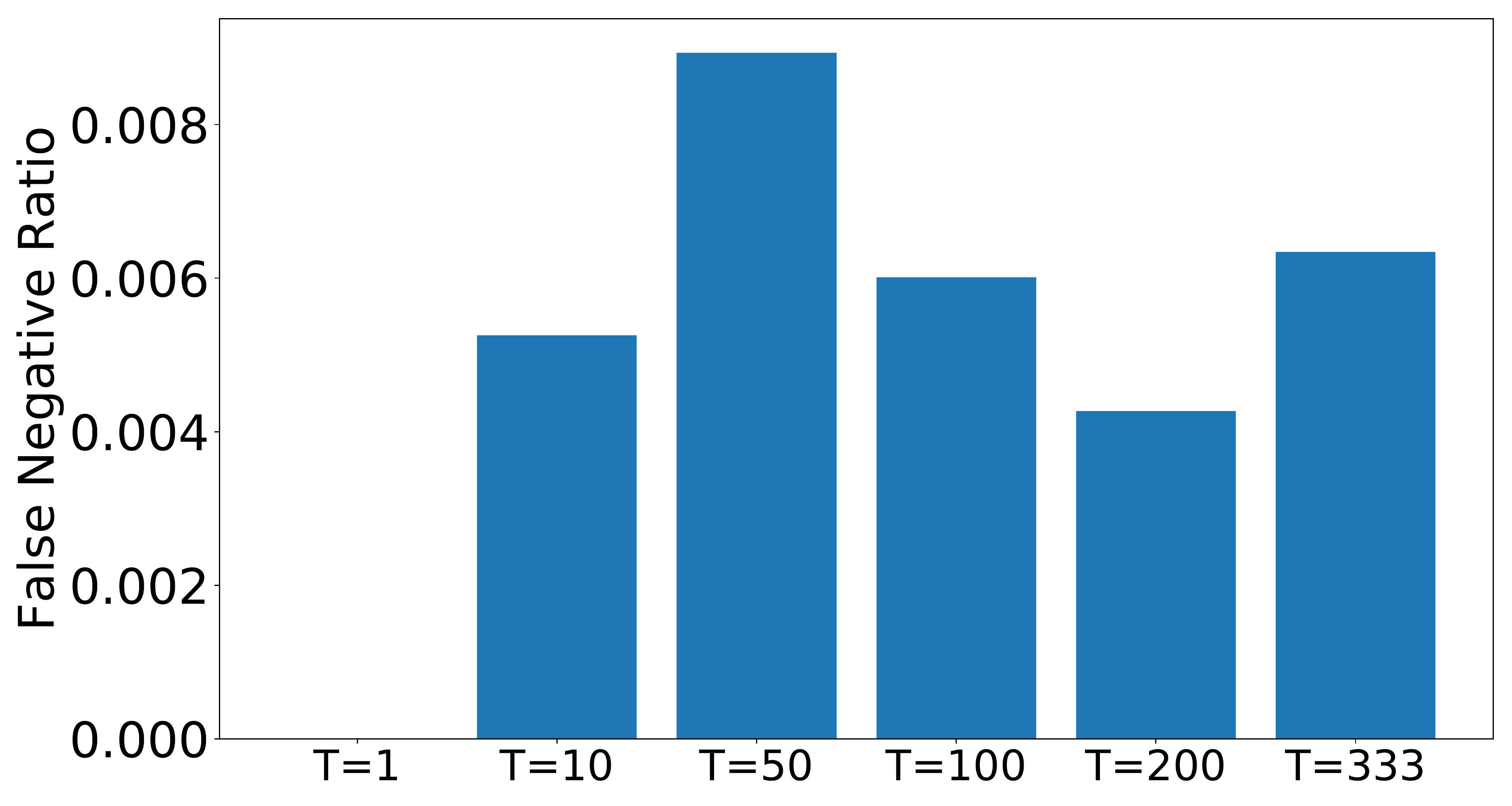}
\label{subfig:fn_noattack}}
\subfigure[Reversed Attack]{\includegraphics[width=0.3\linewidth,keepaspectratio]{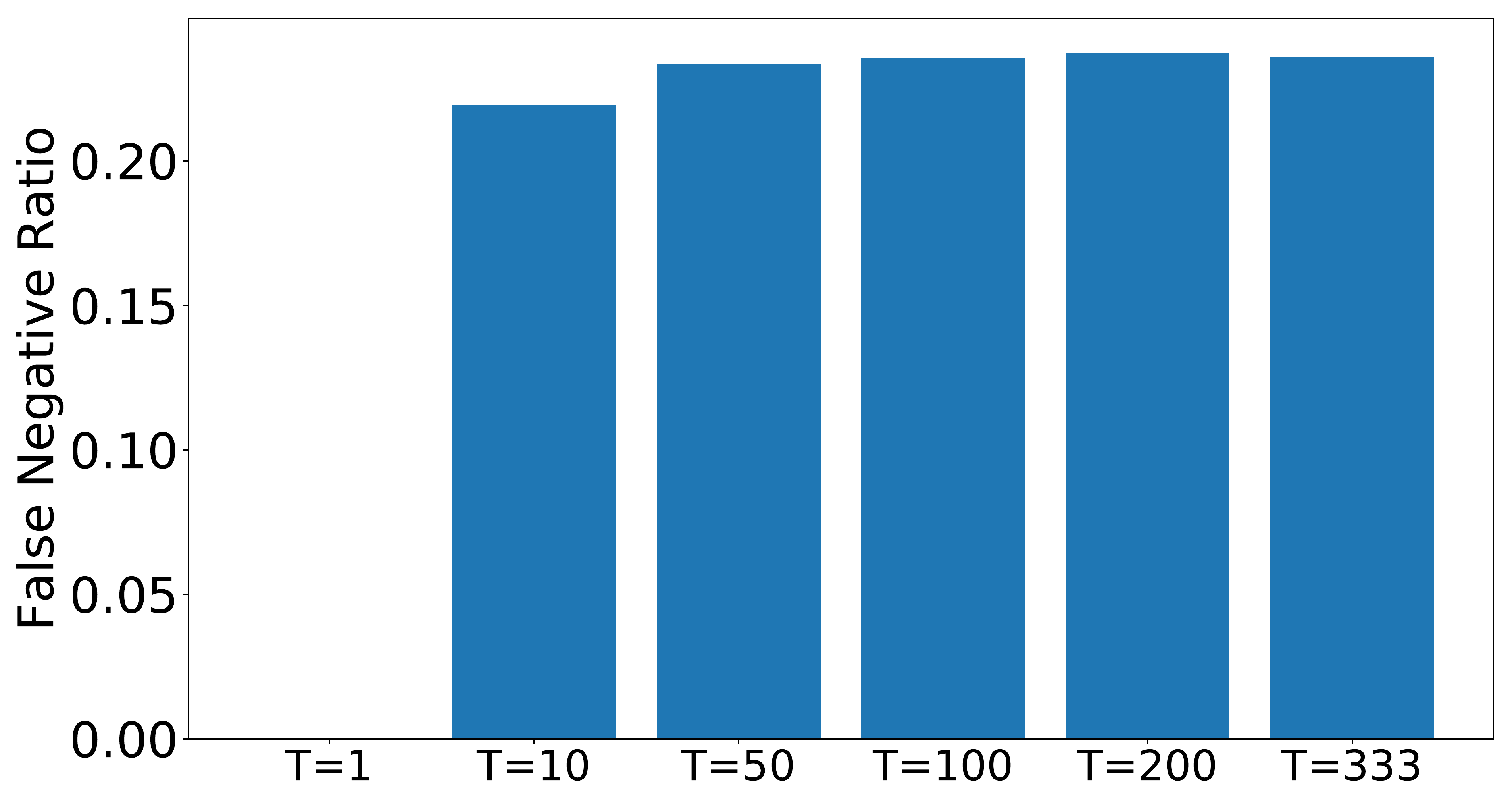}
\label{subfig:fn_attack}}
\subfigure[Multiple Attacks]{\includegraphics[width=0.3\linewidth,keepaspectratio]{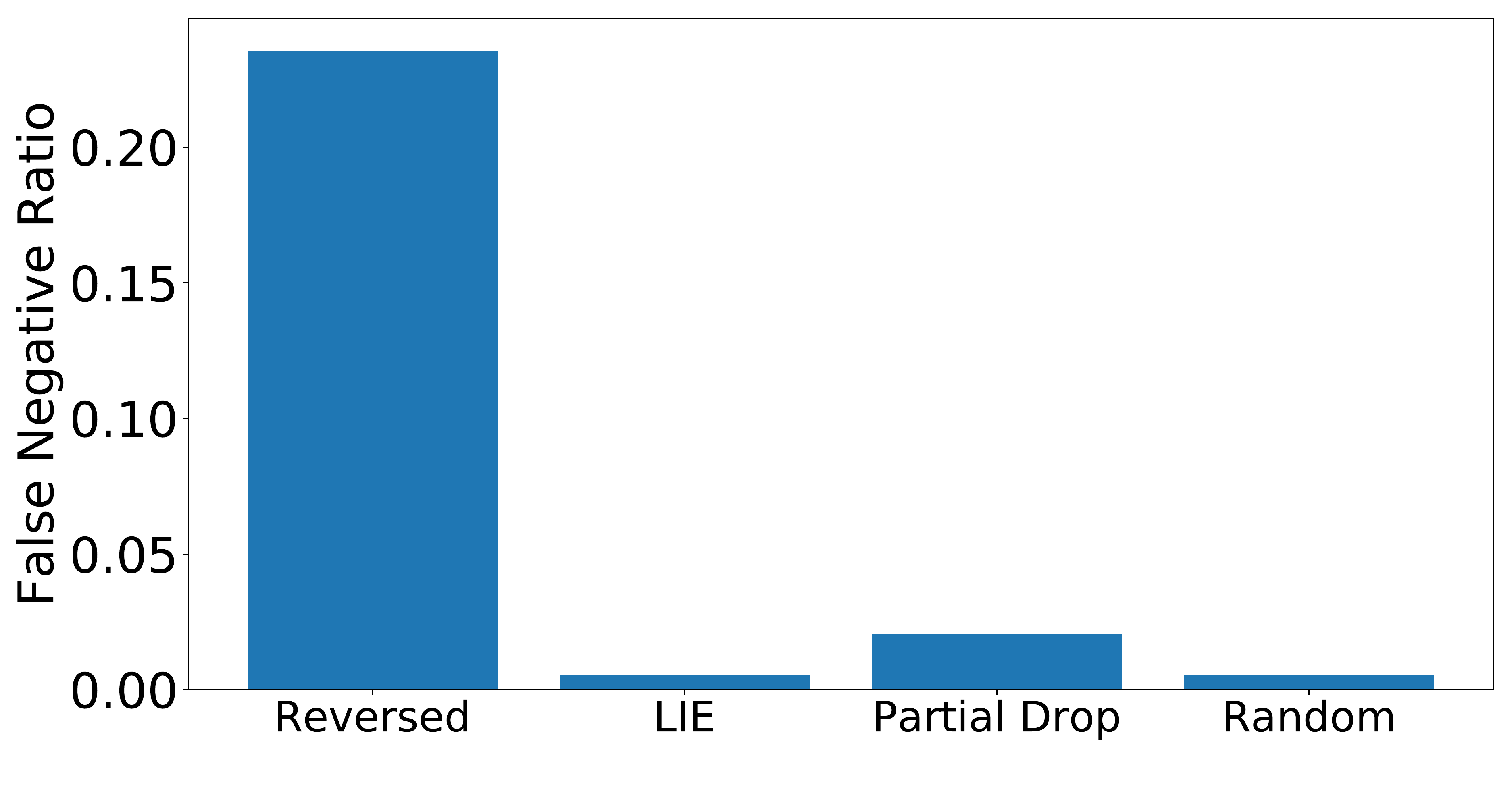}
\label{subfig:fn_attacks}}
\caption{False negative percentage with different scenarios.}
\label{fig:fnp}
\end{figure*}

\subsection{Filters Microbenchmark: False Negatives}
\systemS{}'s filters (in the synchronous case) may have false negatives, i.e.,\ falsely-reject correct models (at workers) from correct servers.
This leads to wasted communication bandwidth and possibly slows down the convergence.
We observe the number of rejected models on workers after 500 learning steps.
Figure~\ref{fig:fnp} shows the ratio of false negatives to the total number of submitted models in different scenarios.

Figure~\ref{subfig:fn_noattack} shows such a ratio with different values for $T$ while no attack is employed yet, with $f_{ps}=1$.
In general, the ratio of false negatives never exceeds 1\% in this experiment, and it is almost stable with increasing $T$.
Note that 333 is the maximum value allowed for $T$ in this setup (to follow the safety rules of \systemS{}).
With $T=1$, the false negatives are always zero, simply because the filters do not work in this setup (i.e.,\ the \emph{gather} phase is entered in every step).
Such a figure shows that our filtering mechanism is effective in not producing many false negatives and hence, do not waste communication rounds (when a model is rejected, the worker asks for another model from a different server).

We repeated the same experiment yet with employing the \emph{Reversed} attack from the Byzantine server; results in Figure~\ref{subfig:fn_attack}.
Such an attack is effective (in terms of bandwidth waste), especially with $T \ge 50$.
Yet, the wasted bandwidth is upper bounded by $25\%$ in all cases, which is the ratio of the number of Byzantine servers to the total number of servers $1:4$.
Figure~\ref{subfig:fn_attacks} show that other attacks are not that effective in wasting the bandwidth, as the filters can successfully filter out only the Byzantine models and accept the correct ones.
Other than the \emph{Reversed} attack, the false negatives ratio in this experiment do not exceed $3.5\%$.
\end{document}